\documentclass{HWart}

\usepackage{epsfig,cite}

\begin{document}
\thispagestyle{empty}
\hbox{}
\nopagebreak\vspace{-3cm}

\begin{flushright}
{\bf hep-ph/9709432}\\
{\sc  TPI--MINN--97/26} \\
{\sc    NUC--MINN--97/11--T} \\
{\sc    HEP--MINN--1607} \\
{\sc OUTP-97-45-P}\\
{\sc Cavendish-HEP 97/15}\\
\end{flushright}

\begin{center}
  {\Large \bf The Wilson renormalization group for low x physics: \\
    Gluon evolution at finite parton density.}\\
  \vspace{0.6in} {\large Jamal Jalilian-Marian$^1$, Alex
    Kovner$^{1,2}$
    and Heribert Weigert$^3$}\\
  $^1${\it Physics Department, University of Minnesota\\
    Church st. S.E., Minneapolis, MN 55455, USA}\\
  $^2${\it Theoretical Physics, Oxford University, 1 Keble Road,
    Oxford OX13NP,
    UK}\\
  $^3$ {\it University of Cambridge, Cavendish Laboratory, HEP,
    Madingley Road,CB3 0HE UK}\\
  \vspace{1.1in}{\sc Abstract}
\end{center}

\noindent

In this paper we derive the complete Wilson renormalization group
equation which governs the evolution of the gluon distribution and
other gluonic observables at low $x$ and arbitrary density.

\vfill

\newpage

\section {Introduction}
Recent years have seen a surge of activity in the area of the low $x$
physics.  Primarily this has been motivated by the new HERA data
\cite{HERA}, which has greatly extended the available kinematic region
for Deeply Inelastic Scattering (DIS).  The data now exists at Bjorken
$x$ as low as $10^{-5}$ and it was hoped initially that at such low
$x$ and relatively large $Q^2$ one would see clearly the new class of
perturbative phenomena, those that go under the general name of
"semihard physics".  First, one was hoping to see the unambiguous
signs of the perturbative BFKL pomeron \cite{BFKL}, which predicts
steeply rising gluon distribution (and consequently DIS cross section)
as a function of $x$ at fixed $Q^2$: $g(x)\propto
(1/x)^{4N_c\alpha_s/\pi}$.  Second, it was expected that in this
kinematical region the gluon densities will be large enough so that
the semihard shadowing effects due to gluon recombination \cite{GLR}
will become sizable.

In actual fact the situation turned out to be not quite so clear cut.
The DIS cross section does indeed rise quite steeply with $1/x$. It
can be fit by a power of $1/x$, although apparently not as large a
power as predicted by the BFKL formula\cite{MRS}.  The GLR parameter
$k$ which is the physical parameter for the onset of shadowing was
estimated and was indeed found to be close to 1 within the HERA regime
\cite{AGL}, which would suggest observable shadowing corrections.
Nevertheless, surprisingly enough all the data is described very well
by a simple straightforward DGLAP linear evolution \cite{MRS}. The
rise at low $x$ is then a consequence of the standard perturbative
evolution of ``flat'' partonic distributions from a low initial scale
$Q^2_0$ to $Q^2$. The shadowing corrections to this perturbative
evolution seem to be practically absent.  In this sense the higher
twist corrections seem to be irrelevant in the HERA regime.

Still it is widely believed that these nonlinear effects must make
their presence felt when the partonic densities are high enough. Even
if it does not happen in $ep$ collisions at HERA they have a very good
chance of being observed in the DIS experiments on nuclei at low $x$
if and when this program takes off at HERA and even more so in the
heavy ion collision experiments at LHC. The curious situation with the
present HERA data only adds motivation for studying the onset of the
shadowing physics.

The shadowing regime can be approached in two ways. Decreasing $Q^2$
at fixed (small) $x$ leads to the initial growth of the cross section
as $Q^{-2}$. This growth gradually slows down and eventually (almost)
stops when the unitarity bound is reached. We will refer to this slow
down and the stop of the growth as the shadowing and the saturation
respectively.  Alternatively one can decrease $x$ at a fixed value of
$Q^2$. Due to the growth of gluonic distributions the cross section
first should exhibit fast growth (powerwise according to the BFKL
prediction), which again should slow down and saturate. Our
perspective in this paper will be the second one, so that we will be
dealing with the evolution of the unintegrated gluon density with
$1/x$.

The physics of the nonlinear effects in DIS is basically the physics
of dense partonic systems. This statement perhaps needs some
clarification.  The physical picture of shadowing depends in large
measure on the Lorentz frame used to describe the DIS process. In the
rest frame of the nucleon the onset of shadowing corrections is due to
the multiple scattering processes of the hadronic component of the
photon (a quark - antiquark pair) on the nucleon.  This can be
described by the extension of Glauber multiple scattering formalism to
the context of QCD \cite{AGL}.  For multiple scattering to become
effective the partonic system does not have to be particularly dense.
In this frame the onset of the nonlinearities (multiple scattering) is
rather determined by the cross section of the scattering of the
hadronic fluctuation on a parton in the nucleon \cite{AGL}. The
authors of Ref.\cite{AGL} analyzed the corrections to the Glauber
formula and concluded that once the rescattering becomes important one
also has to take into account corrections due to rescattering of
gluons in the cascade produced by the quark - antiquark pair, into
which the photon initially fluctuates. The saturation happens in the
regime where the gluon density in the photon is high.

In the infinite momentum frame (IMF), where the nucleon carries all the energy
prior to collision, the picture is different. Low $x$ here means that one
probes low longitudinal momentum - large wavelength
fluctuations in the nucleon wave function. These long wavelength
gluons are emitted from the valence quarks and gluons in the nucleon. In the
standard linear evolution equations (DGLAP or BFKL) the interaction between
 these wee gluons is disregarded. However when their wave length is large
enough, the gluons emitted by different valence partons overlap in space
and interact. The first nontrivial effect of this interaction is the
recombination process, which slows down the growth of the gluon density and
thereby leads to shadowing \cite{GLR}. The shadowing and saturation 
in this frame are both 
clearly effects of large partonic density\footnote{We mention here that
the same conclusion emerges from the analysis carried out in the
recent paper \cite{KMW}. It was shown there, using the explicit BFKL 
expressions for the gluon density, that for collision of two hadrons
the shadowing first appears at low density in the frame where the two
hadrons share the energy equally before the collision and at high density
in the analog of IMF, where one hadron carries all the energy. The saturation
again is the high density effect in both frames.}.

Although the qualitative picture of saturation and unitarization based on
the GLR type recombination effects is very
appealing, reliable 
theoretical tools of dealing quantitatively with finite density
partonic systems are yet to be developed.
The original GLR equation\cite{GLR,MQ} 
truncates the series in the expansion in powers
of density at the first nonlinear term. As all such truncations it
has intrinsically a very limited range of validity, since in general
one expects that when the first nonlinear term becomes important
the higher order terms will be comparable to it in magnitude.
The saturation of partonic distributions and
 restoration of unitarity in high energy
(density) limit of QCD is an outstanding problem which remains
unsolved although several approaches are being explored in the
literature \cite{Li,BKP,Mueller}.

In this paper we continue to develop a theoretical approach to finite
density partonic systems at low $x$. The main goal of this program is
to derive the evolution equation for the gluon density at small $x$
without assuming that the density is in any sense small.  In the
previous papers \cite{JKMW,linear,nonlinear} we have described the
main framework of our approach and have discussed several aspects of
this evolution. In this work we complete the derivation of the full
nonlinear evolution equation.

The present approach is inspired by an idea of McLerran and
Venugopalan \cite{MV} first formulated in the context of
ultrarelativistic heavy ion collisions.  The observation in \cite{MV}
is that there is a regime of high density and weak coupling in which
semiclassical methods should apply.  It was therefore suggested that
the leading small $x$ glue structure of the nucleus is due to the
classical gluon field which is created by the random color charges of
energetic on-shell partons.  The nonlinearities of the Yang - Mills
equations exhibit themselves already on this classical level and it is
therefore possible that they provide the necessary saturation
mechanism at low $x$.  This approach assumes that the interaction of
the fluctuations of the gluon field is weak. In this sense it is a
weak coupling expansion and its validity is therefore restricted to
the perturbative ``semihard'' shadowing effects. It is however
nonperturbative in a different sense.  In the standard perturbation
theory, the charge density is small and the standard perturbative
expansion is simultaneously expansion in powers of the charge density.
In this language, higher powers of charge density appear as higher
twist contributions (although there is no one - to - one
correspondence between the twist expansion and expansion in powers of
the charge density).  The McLerran - Venugopalan (MV) method does not
assume expansion in powers of color charge density and corresponds to
resummation of a particular type of higher twist terms. In fact the
interesting saturation effects are expected when the density is order
$\alpha_s^{-1}$.  This formulation is therefore naturally suited for
discussion of the type of problems we are interested in.

Based on this idea the approach was developed which combines the
concept of the effective Lagrangian for the low $x$ DIS with the
Wilson renormalization group resummation of leading log corrections to
the MV approximation \cite{JKMW,linear}.  The main effect of
the renormalization group procedure is the change in the color charge
density distribution in the effective Lagrangian with $1/x$. The RG
equation that governs the evolution of this distribution is the
subject of the present paper. It was shown in \cite{linear} that in
the limit of small color charge densities this equation reduces to the
celebrated BFKL equation. In \cite{nonlinear} we have derived the
general form of this evolution equation at finite color charge
density. In the present work we calculate the ``coefficients'' in this
renormalization group equation, which are in fact rather ``coefficient
functions'', thereby providing the last ingredient in the derivation
of the full nonlinear evolution equation valid to leading log
approximation at finite color charge density.  We should stress, that
the calculations presented here are only valid to leading order in
$\alpha_s$. The scale of $\alpha_s$ is therefore left undetermined in
this framework and the strong interaction coupling constant is treated
as a momentum independent constant, just like in the standard BFKL
equation. Higher order perturbative calculations of the type of
\cite{fadin} are needed to determine the appropriate scale.

This paper is organized as follows. In Section~\ref{sec:effact} we
motivate and describe the form of the effective Lagrangian for the
physics of low $x$ gluons in DIS. In Section~\ref{sec:pert} we
describe in some detail the classical approximation to this effective
Lagrangian. It turns out that the proper treatment of this Lagrangian
requires careful specification of the complete gauge fixing condition,
and this is also done in Section~\ref{sec:pert}.  We then discuss the
first quantum corrections to the classical approximation which obviate
the need for a renormalization group resummation, and the physical
interpretation of the change of the color charge density distribution
with the RG flow.  In Section~\ref{sec:wilsonRG} we describe in detail
the Wilson renormalization group procedure as applied to our effective
Lagrangian and derive the general form of the RG equation.
Section~\ref{sec:sigmachi} is the central section of this paper. It
contains the calculation of the coefficient functions that appear in
the RG equation. Finally, Section~\ref{sec:disc} is devoted to a
discussion of our results. Several Appendices contain technical
details of the calculation.

We have tried to make this paper self contained, and therefore have
included some of the material already contained in the earlier work
\cite{JKMW,linear,nonlinear}.

\section{The Effective Action for Low $x$ DIS.}\label{sec:effact}

Throughout this paper we will work in the infinite momentum frame,
where the hadron moves in the positive $z$ direction with the velocity
close to velocity of light and almost infinite longitudinal momentum
$P^+\rightarrow \infty$.  Also, we will be working in the light cone
gauge $A^+=0$.

Our task now is to understand the structure of the effective Lagrangian
for low $x$
DIS.  First, it is well known that the most important degrees of
freedom at low $x$ are gluons. In the framework of standard linear
evolution equations, the evolution of gluons in the leading
approximation is independent of quarks, and the evolution of quarks is
entirely driven by the gluonic distribution.  We will therefore retain
only gluons as our dynamical degrees of freedom and disregard quarks
entirely.  Importantly, the gluons that we treat as dynamical degrees
of freedom are only those which have low longitudinal momentum, lower
than some cutoff $\Lambda^+=xP^+$. Our effective Lagrangian therefore
has to be understood as having a built in longitudinal cutoff.

So, what is the Lagrangian that governs the interactions of the low $x$
gluons? First of all, of course it must contain the standard Yang - Mills
interaction term
\begin{equation}
-\int\! d^4 x \ {1\over 4}G^2
\end{equation}
where $G^{\mu \nu} $ is the gluon field
strength tensor
\begin{equation}
G^{\mu \nu}_{a} = \partial^{\mu} A^{\nu}_{a} 
- \partial^{\nu} A^{\mu}_{a} + 
g f_{abc} A^{\mu}_{b} A^{\nu}_{c}\nonumber
\end{equation}

The gluons with the low longitudinal momentum also interact
with the rest of the partons in the hadron, which have larger
longitudinal momentum. We will refer to those partons as ``fast'' for
notational convenience, we may think of valence partons as their
initial representatives. This interaction certainly can not be
neglected, but in the kinematics of IMF and in the light cone gauge it
is very simple. The leading interaction is the eikonal vertex
$A^-J^+_{\rm fast}$, where $J^+_{\rm fast}$ is the color charge
density due to the fast partons.

The dependence of $J^+_{\rm fast}$ on $x^-$ and $x^+$ is very simple.
First, since the wavelength of the fast fields is much shorter than
that of the dynamical soft gluons the charge they produce is
effectively concentrated at $x^-=0$. Intuitively this can be
understood in the following way.  In the rest frame the valence
partons are concentrated within the nucleon radius from the center of
the nucleon.  When boosted to the infinite momentum frame due to
Lorentz contraction they are squeezed into a very thin pancake. This
picture is a little too naive for our fast partons since some of them
have much larger wave length than the nucleon radius. However as a
basic physical picture it is still correct.  We therefore have
$J^+\propto\delta(x^-)$.

Second, we can understand the $x^+$ dependence by considering the
(light cone) time scales characteristic of the problem. The relevant
time scale for the low $x$ phenomena is the inverse of the on shell
frequency of the soft gluons: $k^-\propto 1/k^+$. The frequency of the
fast modes is much lower $p^-\propto 1/p^+$ since their longitudinal
momentum is higher, so that $k^-/p^-\propto 1/x$. Therefore, as far as
the soft glue is concerned the color charge source due to fast partons
is effectively static.  We are therefore led to consider the
interaction of the type
\begin{equation}
S_{\rm int}=A^-\rho(x_\perp)\delta(x^-)
\label{naiveint}
\end{equation}

The fast partons are represented in our effective Lagrangian by the
surface charge density $\rho(x_\perp)$.  A hadron of course is not
described by a fixed single configuration of the color charge density
$\rho(x_\perp)$.  However, the crucial point is that the structure of
the fast component of the hadron is determined on a much longer time
scale than the time scale relevant for the low $x$ physics. It is
fixed by the hadronic wave function, bremsstrahlung processes that
involve fast partons, etc.  Therefore, as far as the soft glue is
concerned, there is no interference between the different
configurations of $\rho(x_\perp)$.  In the low $x$ effective
Lagrangian the hadron thus appears as an ensemble in which different
configurations of $\rho(x_\perp)$ enter with some statistical weight
$\exp\{-F[\rho]\}$.  The partition function for calculation of the
soft glue characteristics of a hadron must therefore have the form
\begin{eqnarray}
\int D[\rho,A]\exp\{-F[\rho]-{i\over 4}\int\! d^4 x\  G^2+
iS_{\rm int}[A, \rho]\}
\end{eqnarray}

At this point we do not specify the form of the functional $F[\rho]$. In fact,
as we shall see later this functional depends on the longitudinal cutoff
$\Lambda^+$ which is imposed on the soft fields. In other words, as one 
considers regions of lower and lower $x$,  $F$ changes. The 
flow of $F$ with the cutoff $\Lambda^+\propto x$ is described by a 
renormalization
group equation of the form
\begin{eqnarray}
{d\over d \ln 1/x}F[\rho]=\alpha_s\Delta[\rho]
\end{eqnarray}
This RG equation is precisely the evolution equation for the charge
density correlators (and consequently for the soft glue observables)
which we undertake to derive in this paper.  

Of course, in order to
make quantitative statements about the $x$ dependence of $F$ we have
to specify the initial condition for the evolution.  This can be done
in the perturbative region at not too small a value of $x$, where $F$
can still be expanded in powers of $\rho$.  The initial form of $F$
can then be taken as
\begin{equation}
F=\int dx_\perp dy_\perp  \rho(x_\perp )\mu^{-1}(x_\perp ,y_\perp )\rho(y_\perp )
\label{gaussi}
\end{equation} 
with
\begin{equation}
\mu(x_\perp ,y_\perp )=S(b_\perp )\int {d^2k_\perp \over (2\pi)^2}\ 
e^{ik_\perp (x_\perp -y_\perp )}\, \phi(k_\perp ,x_0)
\label{mu}
\end{equation}
Here $b_\perp ={x_\perp +y_\perp \over 2}$ is the impact parameter,
$S(b_\perp )$ is a nucleon thickness function, $x_0$ is the value of
$x$ from which we start evolving $F$ according to the RG equation and
$\phi(k_\perp ,x)$ is the unintegrated gluon density\footnote{ A
  similar Gaussian form for the statistical weight was used in
  \cite{MV,JKMW} in description of a large nucleus limit. In
  this case the Gaussian form is valid since the charge density is
  large and the color charges that build it up are randomly
  distributed in color space.}.  The relation between the parameter in
the Gaussian, $\mu^{-1}$ and $\phi$, Eq.(\ref{mu}) will become clear
when in Section~\ref{sec:pert} we consider the perturbative
calculation of the gluon structure function based on the effective
Lagrangian in Section~\ref{sec:effact}.

One important question that we have not touched upon so far, is the
question of gauge invariance of our effective action. Although we have
partially fixed the gauge by the light cone gauge condition $A^+=0$,
the action should still preserve the residual gauge symmetry. This
residual gauge symmetry group is comprised of gauge transformations
with gauge functions which do not depend of $x^-$.  The naive
``Abelian'' eikonal interaction term Eq.(\ref{naiveint}) does not
preserve this gauge symmetry. The relevant generalization takes the
form
\begin{eqnarray}
S_{\rm int}={1\over{N_c}} \int d^2 x_\perp  dx^-
\delta (x^-)
\rho^{a}(x_\perp ) {\rm tr}T_a W_{-\infty,\infty} [A^-](x^-,x_\perp )
\label{int}
\end{eqnarray}
Here $T_a$ are the $SU(N)$ color matrices in the adjoint representation
and $W$ is the path ordered exponential along the $x^+$ direction in
the adjoint representation of the $SU(N_c)$ group
\begin{equation}
W_{-\infty,\infty}[A^-](x^-,x_\perp ) = P\exp \bigg[-ig \int dx^+
A^-_a(x^+,x^-,x_\perp )T_a \bigg] 
\end{equation}
This form is explicitly gauge invariant under the residual gauge
transformations with gauge functions which do not depend on $x^-$ and
vanish at $x^+\rightarrow\pm\infty$.  Requiring $F[\rho]$ to be gauge
invariant, we also restore gauge invariance of the action under the
gauge transformations which do not vanish at $x^+\rightarrow\pm\infty$
but rather are periodic in $x^+$.

This form of the interaction consistently leads to a source term in
the corresponding Yang-Mills equation that represents classical
colored particles moving along the light cone:
\begin{equation} 
D_\mu G^{\mu\nu}=J^+\delta^{\nu +}
\label{eom}
\end{equation}
with
\begin{equation}
J^+_a(x)={g\over{N_c}} 
\delta (x^-) 
\rho^{b}(x_\perp ) 
{\rm tr}\Bigg[T_b W_{-\infty,x^+} [A^-]T_a 
W_{x^+,\infty} [A^-]\Bigg]
\label{current}
\end{equation}
Expanding this expression for the current to lowest
order in the field $A^-$ yields
\begin{equation}
J^+_a(x)=g\delta (x^-)\rho^{a}(x_\perp ) 
\end{equation}
which reproduces Eq.(\ref{naiveint}) and is the form of the current
used in \cite{MV,AJMV}.  As explained in \cite{MV}, this form
is only valid in the gauge $A^-(x^-=0)=0$. In more general gauges the
current has to satisfy the covariant conservation condition
\begin{equation}
D^-J^+=0
\end{equation}
Our current (\ref{current}) evidently complies with this requirement.
This is a direct consequence of the residual gauge invariance.

All of the above considerations finally lead us to the following
effective action for the low $x$ DIS
\begin{eqnarray}
\label{action}
S&=&i\int d^2 x_\perp  F[\rho ^a(x_\perp )]\\ 
&&- \int\! d^4 x\, {1\over 4}G^2
 + {{i}\over{N_c}} \int\! d^2 x_\perp  dx^-
\delta (x^-)
 {\rm tr}\rho(x_\perp ) W_{-\infty,\infty} [A^-](x^-,x_\perp )
\nonumber 
\end{eqnarray}
which is the starting point of our approach.

We end this section with a comment about the nature of the action
Eq.(\ref{action}).  Although we use the term ``effective action'' when
referring to it, it should be understood that it is different in some
important aspects from the ``classic'' effective Lagrangians, like for
example the chiral effective Lagrangian of pion physics. The chiral
effective Lagrangian $L[\Pi^a]$ describes the dynamics of low momentum
pion fields, where the momentum cutoff is determined by the mass of
the $\sigma$ - particle, or alternatively the dimensionful pion
coupling $f_\pi$, $p<4\pi f_\pi$.  All the modes with momenta above
the cutoff, as well as all other heavy excitations of the fundamental
theory ($\rho$, $\phi$ mesons, etc.) have been integrated out to
arrive at this effective Lagrangian.  In this sense our effective
Lagrangian is similar. The gluon fields have longitudinal momenta
bounded by the cutoff $\Lambda^+$, while all higher momentum modes are
assumed to have been integrated out.

Importantly, the chiral physics has a sharp scale associated with it -
$f_\pi$. Consequently, {\it all} modes of the pion field with momentum
lower than the cutoff are described well by the chiral Lagrangian.  In
fact, the pions at low momenta interact very weakly, with the strength
proportional to $p^2/f_\pi$. Therefore the perturbation theory in the
chiral Lagrangian framework is well behaved and does not lead to large
corrections to the tree level results. Also, the description of the
low momentum pions is insensitive to the change of the cutoff
$\Lambda$.

In our case the situation is very different in this respect. There is
no sharp physical separation scale which would separate high from low
longitudinal momenta. The separation scale $\Lambda^+$ we impose is
arbitrary.  The interaction does not die away as we go far below
$\Lambda^+$.  Therefore there is no reason to expect that our
effective Lagrangian gives an adequate description for the modes with
momenta far below the cutoff.  In fact, quite to the contrary as we shall
see the perturbation theory in our effective theory gives larger
corrections the farther we go below the cutoff.  In this sense this
effective Lagrangian is inadequate for description of momenta $k^+\ll
\Lambda^+$. This is of course the manifestation of the absence of the
physical separation scale\footnote{In this respect our effective
  Lagrangian is more akin to ``fundamental'' Lagrangians of
  renormalizeable field theories than to effective Lagrangians of the
  chiral physics type.}.  This means, that if we want to describe low
momenta, $k^+\ll
\Lambda^+$, we have to correct the effective Lagrangian. The arguments
presented above however fix the form of all the terms in the
Lagrangian apart from $F[\rho]$.  If the form of the Lagrangian
remains the same under the $k^+$ evolution, the only thing that can
change is the statistical weight $e^{-F[\rho]}$.

Physically it is quite clear what should happen. As we move to smaller
values of the longitudinal momentum $k^+$, all the gluons with momenta
between $k^+$ and $\Lambda^+$ are transferred from the category of
``soft'' (or slow) into the category of ``fast''. They cease to be
dynamical degree of freedom of interest (hence the dynamical fields
have lower cutoff on the longitudinal momentum), but give extra
contribution to the static color charge density $\rho(x_\perp )$.
Effectively therefore as we go to lower $x$, the color charge density
as seen by the soft glue changes. Since the distribution of the charge
density is governed by the statistical weight $e^{-F[\rho]}$, this
means that $F$ should change as we lower the longitudinal cutoff
$\Lambda^+$. This is the physical origin of the renormalization group
flow we have referred to earlier.  In the following section we will
see explicitly how this happens.  First, however let us describe how
to set up the perturbation theory in the present framework.

\section{Perturbative calculation of gluonic observables.}
\label{sec:pert}

The perturbation theory for the effective Lagrangian Eq.(\ref{action})
was developed in \cite{MV}.  It is organized in the following way.
First one fixes the configuration of the color charge density, and
performs perturbative expansion in $\alpha_s$ at fixed $\rho(x_\perp
)$. The charge density is not considered to be small, thus this
perturbation theory is different from the standard one in that the
calculation is performed in a non vanishing background field.  In the
second step the averaging over $\rho(x_\perp )$ should be performed.
This part of the calculation is contingent on the knowledge of
$F[\rho]$ and is completely nonperturbative. In fact the counterpart
of this step in the standard perturbative analysis would be the
specification of various gluon operator averages in the hadronic
state. Conceptually therefore, the first, perturbative part of the
calculation can be thought of as the calculation of the generalized
``splitting functions'' (which includes however the mixings between
operators of different twist and is not intrinsically organized as
an expansion in powers of $1/Q^2$), while the second, nonperturbative
part is parallel to the calculation of operator averages in the
hadronic state. In the standard perturbation theory, of course one
does not have to know the operator averages in order to derive the
evolution equation. As we shall see, the exact same thing happens in
our calculation. It is only the perturbative part of the calculation
that has to be under control in order to derive the renormalization group
equation for $F$.  In this section therefore we will discuss the
perturbation theory in $\alpha_s$ at fixed $\rho(x_\perp )$.

\subsection{The tree level.}

As in every perturbative calculation the first step is to find the
classical solution to the equations of motion.  The equations of
motion that follow from the action Eq.(\ref{action}) are
\begin{equation} 
D_\mu G^{\mu\nu}={g\over{N_c}} 
\delta (x^-) 
\rho^{b}(x_\perp ) 
{\rm tr}\Bigg[T_b W_{-\infty,x^+} [A^-]T_a W_{x^+,\infty} [A^-]\Bigg]
\label{eom1}
\end{equation}

As explained in the previous section, these equations are invariant under
the residual $x^-$ independent gauge transformation
\begin{eqnarray}
A \to V\left[ A + \frac{i}{g}\partial\right] V^\dagger
\end{eqnarray} 
with
\begin{eqnarray}
V(x) = \exp\left[ i \Lambda(x_\perp ,x^+)\right], \ \ \Lambda
\rightarrow_{x^+\rightarrow\pm\infty}0
\label{gauge}
\end{eqnarray}

As a consequence, the equations of motion at fixed $\rho$ have an
infinite number of solutions. To properly set up perturbation theory,
we should choose one of these solutions. Technically this is achieved
by gauge fixing the residual gauge freedom. There are of course many
possible gauge fixings. From the calculational point of view it is
convenient to choose a gauge in which the classical solution is static
($x^+$ independent).  It is important to realize that the condition of
staticity of the classical solution is still insufficient. Even though
it completely eliminates the gauge freedom of Eq.(\ref{gauge}), there
are still many solutions to the equations of
motion. This is a consequence of the
remaining gauge symmetry of our problem, with gauge functions 
$\Lambda$ which do not vanish at $x^+\rightarrow \pm\infty$, but rather
are periodic in $x^+$. With the transformation Eq.(\ref{gauge}) moded out,
those are
\begin{eqnarray}
\label{gaugep}
&&A \to V\left[ A +\frac{i}{g}\partial\right] V^\dagger\\
&&\rho \to V^\dagger \rho V\nonumber\\
&& V(x) = \exp\left[ i \Lambda(x_\perp )\right]\nonumber
\end{eqnarray}

To see that this is indeed the case, consider the equations
Eq.(\ref{eom1}) for static fields (note that all static solutions have
vanishing $A^-$)
\begin{eqnarray}
&&\partial_i 
\partial^+ A^i +g[ A_i , \partial^+ A^i] = g \rho(x_\perp )\delta(x^-)\\
&&F^{ij}=0\nonumber
\end{eqnarray}
The general solution to this equation has the form
\begin{eqnarray}
A_i^V[\rho]={i\over g}
\theta (x^-)U(x_\perp )\partial_i U^\dagger(x_\perp )+{i\over g}\theta(-x^-)V(x_\perp )
\partial_iV^\dagger(x_\perp )
\label{gsolution}
\end{eqnarray}
where the $SU(N)$ matrices $U$ and $V$ satisfy
\begin{eqnarray}
\partial_i\left[V^\dagger U\partial_i(U^\dagger V)\right]=-g^2V\rho V^\dagger
\end{eqnarray}
This equation obviously has a solution for any $V(x_\perp )$.  The
matrix $V$, which labels these solutions is closely related to the
gauge transformation matrix of Eq.(\ref{gaugep}), although this
relation is rather subtle. Obviously any two solutions
Eq.(\ref{gsolution}), $A_i^V[\rho]$ and $A_i^{V'}[\rho]$ are not
related by a gauge transformation, since they solve the equation of
motion with {\it the same} $\rho$, while the gauge transformation
Eq.(\ref{gaugep}) acts nontrivially on $\rho$.  However it is easy to
see that {\it the set of solutions} $\{A_i^V[\rho]\}$ with fixed
$\rho$ and arbitrary $V$ is gauge equivalent to {\it the set of
  solutions} $\{A_i^{V'}[V^\dagger\rho V]\}$ with fixed $V'$
(which determines the asymptotics
at
$x^-\rightarrow - \infty$) and arbitrarily rotated
$\rho$.  Consequently, it would be redundant to take into account all
static classical solutions at fixed $\rho$ since we are subsequently
performing the unconstrained functional integral over $\rho$ with the
measure which is invariant under Eq.(\ref{gaugep}). We can therefore
gauge fix this extra gauge freedom by imposing, for example a fixed
boundary condition on $A_i$ at $x^-\rightarrow
-\infty$\footnote{Alternatively, one could impose a gauge condition on
  $\rho$, by requiring for example that $\rho$ be a diagonal matrix.
  Our choice here is dictated by calculational simplicity.}.

In this paper we will follow Ref.\cite{MV} and choose as the subsidiary
gauge condition
\begin{eqnarray}
\partial^iA_i(x^+,x_\perp ,x^-\rightarrow - \infty)=0.
\label{gaugecon}
\end{eqnarray}
This gauge has a nice feature that at $t\to -\infty$, at all finite
values of $z$ (i.e. $x^-\to-\infty$) the vector potential is required
to be the same as in the perturbative vacuum, $A_i=0$. This seems very
sensible, since at these times the hadron itself is still at $z\to
-\infty$ and could not have changed the quantum state at any finite
$z$.

Note that this gauge condition eliminates both, time dependent gauge
freedom Eq.(\ref{gauge}) and time independent gauge freedom
Eq.(\ref{gaugep}).  It is not ghost free, and therefore the measure in
the path integral in Eq.(\ref{action}) must be modified by the
appropriate Fadeev-Popov determinant
\begin{eqnarray}
\delta\left(\partial_i A_i (x^-\rightarrow -\infty)\right)
{\rm det}\left(\partial_i D_i \left[ A_i (x^-\rightarrow -\infty)\right]\right)
\label{FP}
\end{eqnarray}
This modification is harmless, since as we will see later 
the ghosts do not contribute to leading order in $\alpha_s$, which is the
order to which we calculate.

It is important to realize that the gauge fixing Eq.(\ref{gaugecon})
should be consistently used throughout the whole perturbative
calculation.  This means that not only does it determine the classical
solution we have to pick, but also the form of the propagator of the
fluctuations of $A_\mu$ around this solution to be used in the higher
order perturbative calculations. In this way all potential zero modes
in the propagator are eliminated and the calculation is unambiguous.
In the standard perturbation theory, although in principle the
situation is similar, in practice one can frequently get away without
specifying the gauge fixing condition for the residual gauge freedom.
The light cone gauge condition $A^+=0$ eliminates the major part of
the zero mode ambiguity and the rest of the zero modes start causing
problems only in higher orders.  It turns out that in our calculation
we have to be much more careful, and impose the residual gauge fixing
properly already in the lowest order.  This is related to a
nonstandard behavior of our fields at infinity.  On the classical
level this behavior is obvious from the form of the solution of the
equations of motion Eq.(\ref{gsolution}), which does not vanish at
$x^- \rightarrow \pm \infty$.  We will come back to this question in
Section~\ref{sec:sigmachi}.

Returning to Eq.(\ref{gsolution}) we see that in this gauge for a
generic fixed $\rho(x_\perp )$ there is a unique solution of the form
\begin{eqnarray}
\label{sol1}
&&A_i=\theta (x^-)\alpha_i(x_\perp )\\
&&\alpha_i(x_\perp )={i\over g}U(x_\perp )\partial_i U^\dagger(x_\perp )\nonumber
\end{eqnarray}
with the matrix $U(x_\perp )$ determined by
\begin{eqnarray}
\partial_i\alpha_i=-g\rho
\label{sol2}
\end{eqnarray}

Any gluonic observable in the tree level approximation is calculated as
\begin{eqnarray}
<O[A_\mu]>=\int D[\rho]e^{-F[\rho]}\ O\left[A^-=0, 
A_i=\theta(x^-)\alpha_i[\rho]\right]
\end{eqnarray}

For example, the unintegrated gluon density defined as
\begin{eqnarray}
g(x,k_\perp )=<\hat a^{\dagger a}_\lambda(x,k_\perp )\hat a^a_\lambda(x, k_\perp )>
\end{eqnarray}
where $\hat a$ and $\hat a^\dagger$ are the light cone gluon 
creation and annihilation operators,
is given by
\begin{eqnarray}
g(x,k_\perp )={1\over x}<\alpha_i^a(k_\perp )\alpha_i^a(-k_\perp )>
\label{tree}
\end{eqnarray}
Here the $<\ >$ denotes averaging over $\rho$ with the weight $\exp\{-F\}$.

This has a simple representation in terms of the standard Feynman
diagrams.  The classical field is given by the sum of the tree
diagrams for one point function in the background density.  Using
curly lines to represent gluons we have the following graphical
representation for the full classical solution $b$ and the first few
terms in its pertrubative expansion
\begin{eqnarray}
\label{eq:treeexp}
b &= & 
   \begin{minipage}[c]{2cm}
      \epsfig{file=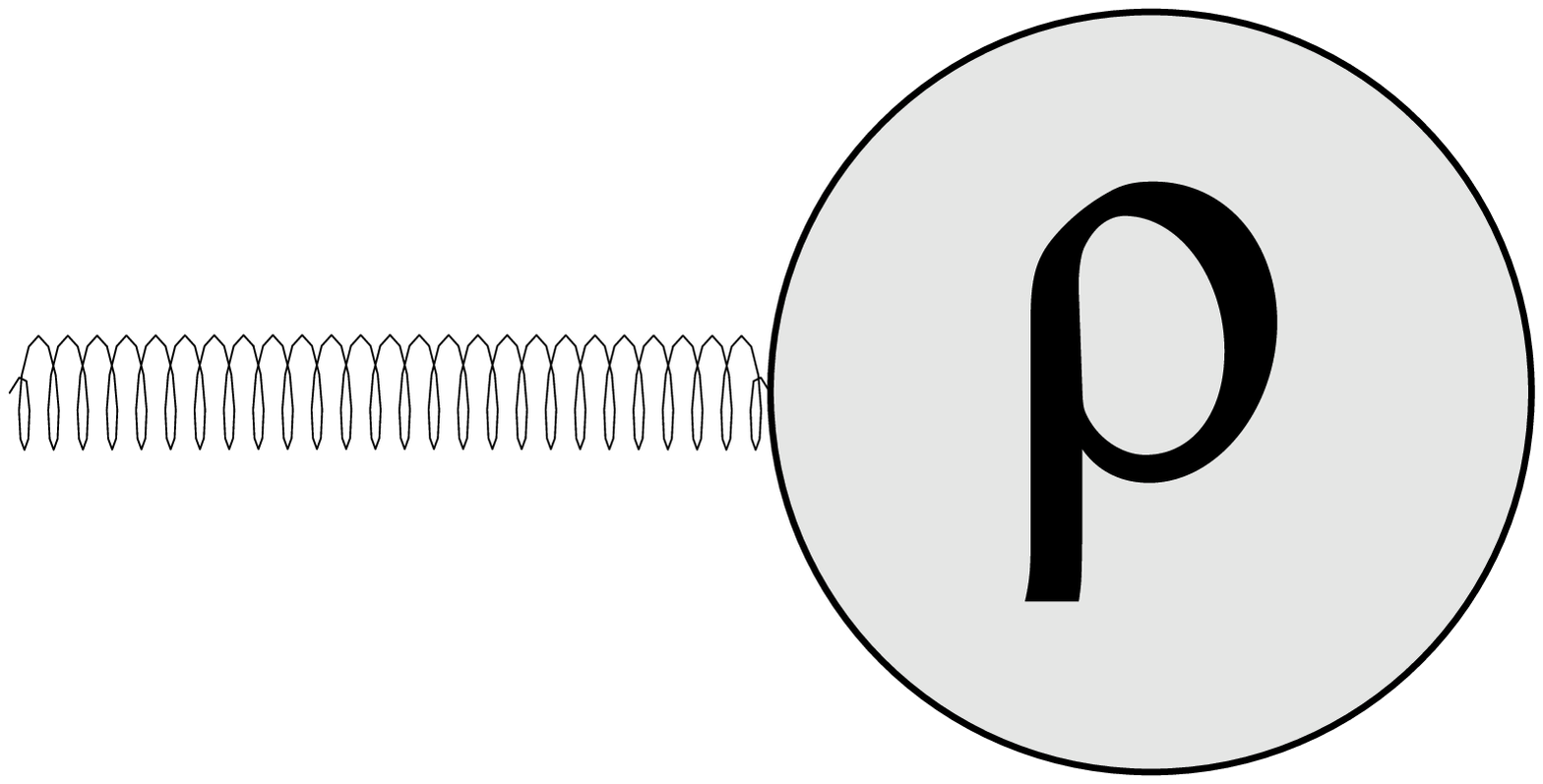,width=2cm} 
    \end{minipage}
=
    \begin{minipage}[c]{1cm}
      \epsfig{file=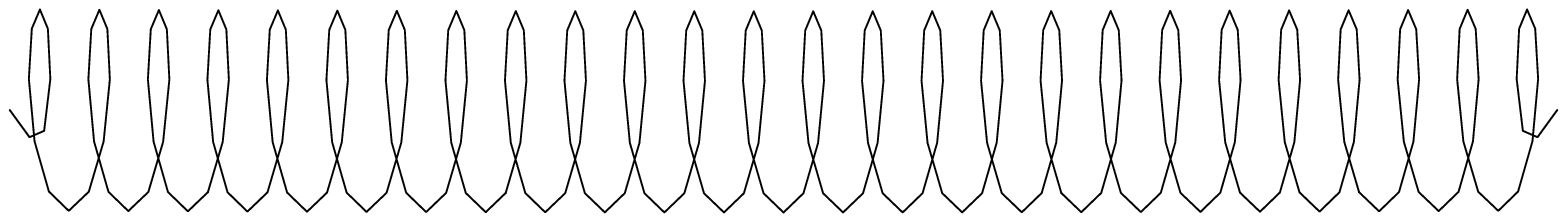,width=1cm} 
\end{minipage}\ { g \rho} 
+ g\ 
\begin{minipage}[c]{1.8cm}
\epsfig{file=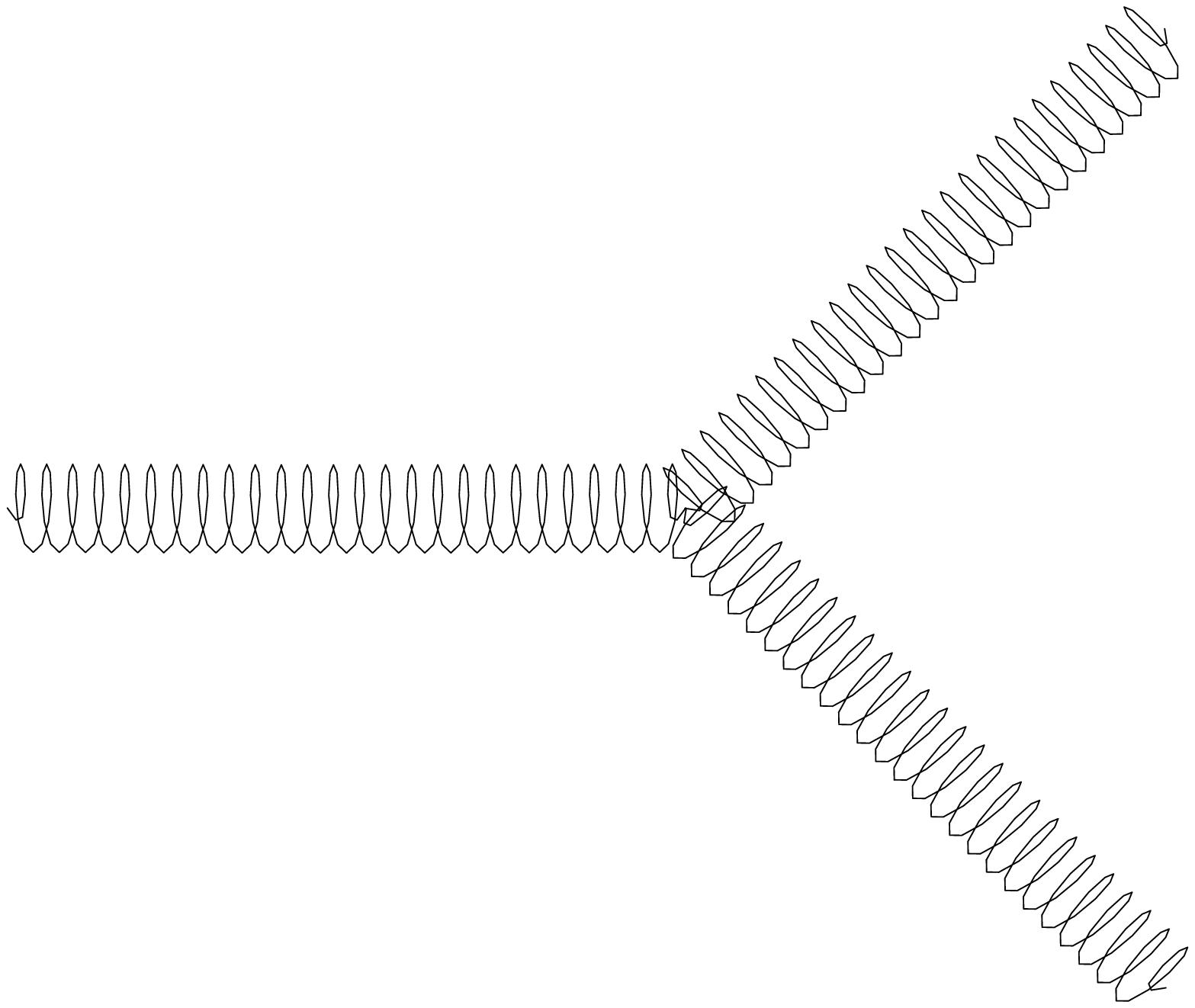,width= 1.8cm}
\end{minipage}
\begin{minipage}[c]{1cm}
${ g \rho}$\\ \vskip1.5cm ${ g \rho}$
\end{minipage}
\hspace{-0.2cm} 
+ {\cal O}(\rho^3)
\end{eqnarray}

The distribution function (up to some simple kinematical factors) is just
the square of the field averaged over $\rho$. To the order $\rho^2$
it is related in a simple way to the color charge density correlation function
\begin{eqnarray}
\label{relation}
g(x,k_\perp ) & = &
{1\over xk^2_\perp }<\rho(k_\perp )\rho(-k_\perp )>
\nonumber \\ & = &
<    \partial^i\  \begin{minipage}[c]{1.2cm}
      \epsfig{file=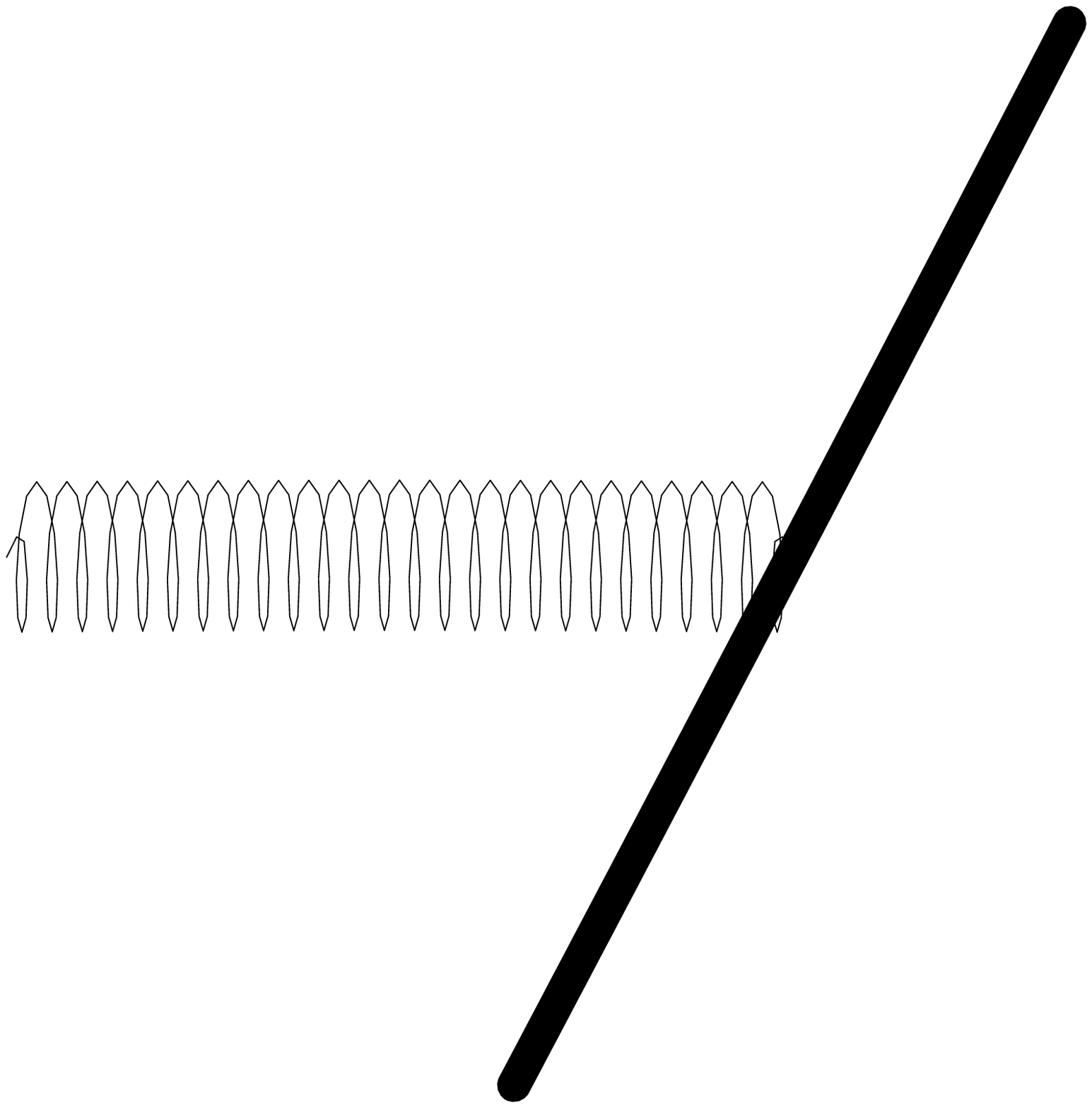,
        width=1.2cm} 
\end{minipage}
\hspace{0.2cm}\cdot\hspace{0.2cm}
      \partial^i \
      \begin{minipage}[c]{1.2cm}
        \epsfig{file=rhoghoriz4.eps,
        width=1.2cm} 
\end{minipage}
+ {\cal O}(\rho^3)
>
\end{eqnarray} 
where we have drawn the factors of $\rho$ as diagonal lines to
indicate that they are always associated with eikonal lines along the
$x^+$ direction that correspond to the worldlines of the fast
particles they represent. 

We stress that our goal in this paper is to perform the calculation to
all orders in $\rho$ in the first order in $\alpha_s$. Hence an
expansion in powers of $\rho$ as in Eqns.~(\ref{eq:treeexp}) and
(\ref{relation}) would not be sufficient for our purpose.  However,
the above representations are helpful in visualizing the physical
mechanism underlying the running of the charge density distribution
with $1/x$.  Also, even though our interest is in the phenomenon of
shadowing and saturation, which occur at large $\rho$, our
calculational procedure should be valid also at small color charge
density.  In this limit we should recover the known perturbative
results, which in the present context is the BFKL equation. Expansion
to leading order in $\rho$ of our result will therefore be an
important consistency check in the calculation.

\subsection{The first order perturbative corrections.}

\begin{figure}[htbp]
  \begin{center}
      \begin{tabular}{c @{\hspace{2cm}} c}
        \epsfig{file=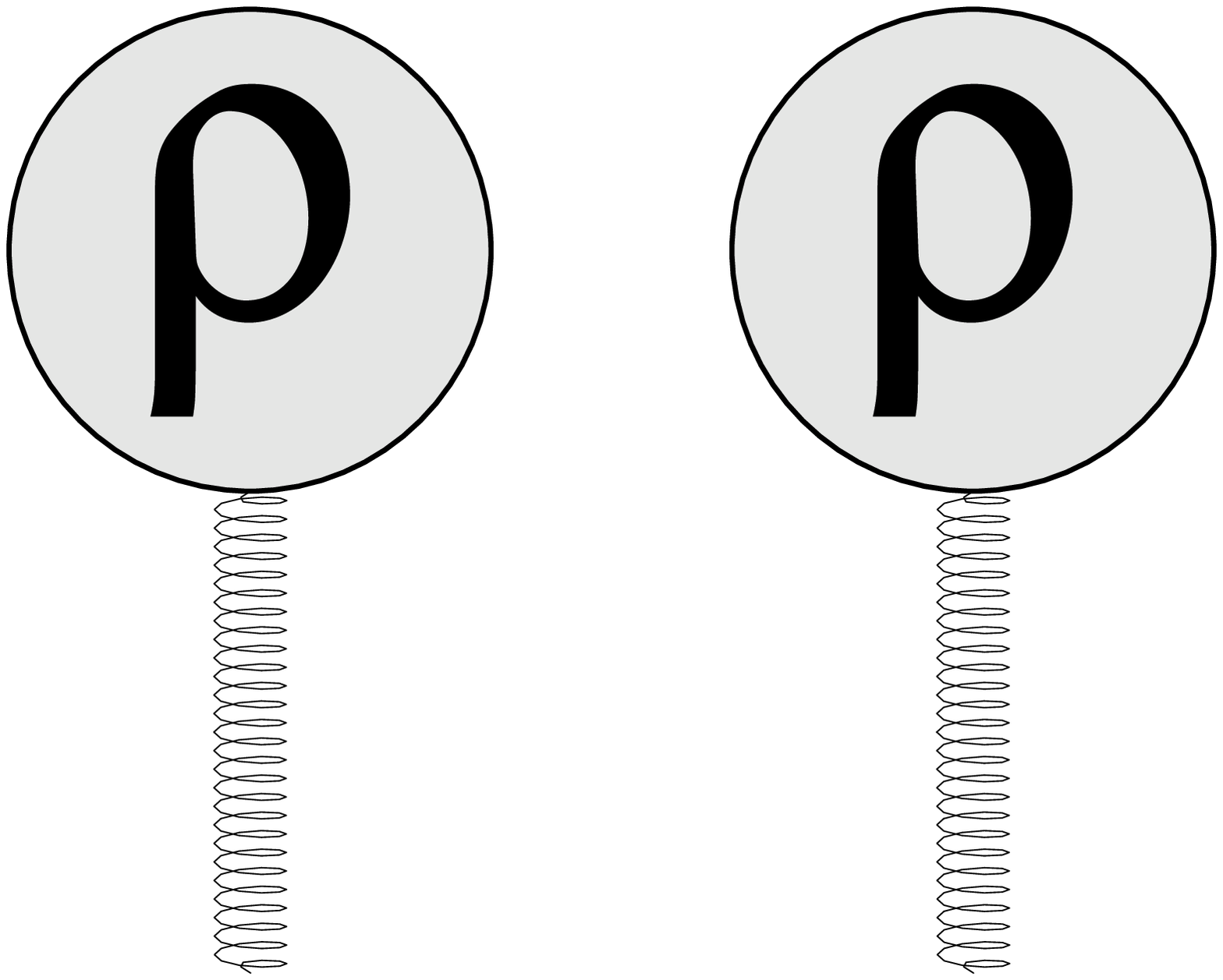, height=4.5cm} & 
        \epsfig{file=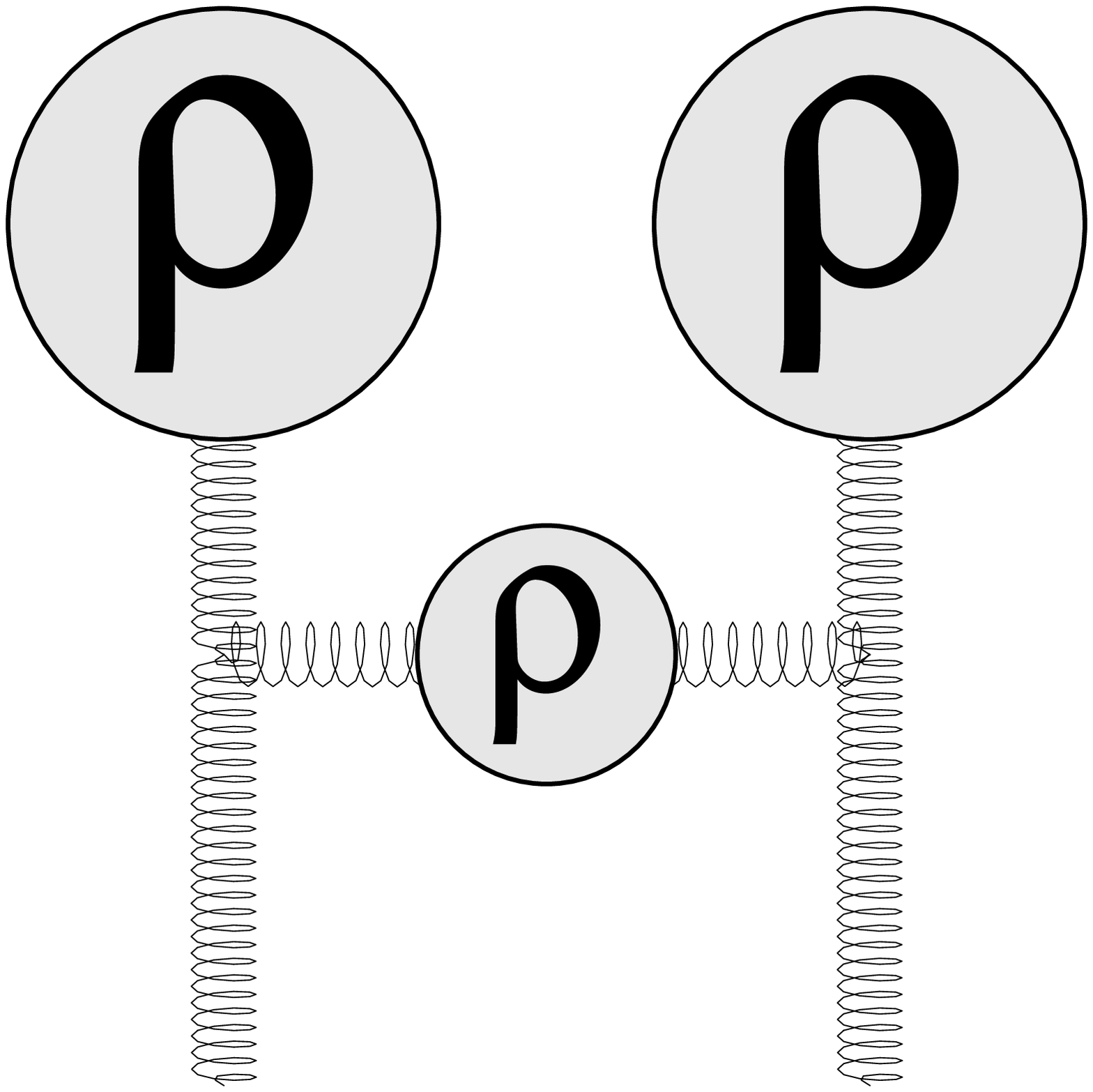, height=4.5cm} \\
    (a) & (b)
      \end{tabular}
    \caption{Diagram contributing to the gluon distribution at
      lowest order in $\alpha_s$ (but to all orders in $\rho$)
      (a) and a typical order $\alpha_s$ correction (b). The
      horizontal line represents a propagator in the presence of the
      full, $\rho$-induced background field
}
    \label{fig:distcorr}
  \end{center}
\end{figure}

One prominent feature of Eq.(\ref{tree}) is the {\em full} tree level
$x$ dependence of the gluon density. It is precisely the same as in
the leading order in the standard perturbation theory. We know,
that in the standard calculation this lowest order $x$
dependence feeds back through the higher order graphs and leads to
large perturbative corrections at small $x$. We expect therefore that
the same will happen in our perturbation theory.  Indeed, consider for
example the graph on Fig.\ref{fig:distcorr}b, which gives one of the contributions to
the gluon density at order $\alpha_s$.

\begin{figure}[htbp]
  \begin{center}
      \begin{tabular}{c @{\hspace{2cm}} c}
            \epsfig{file=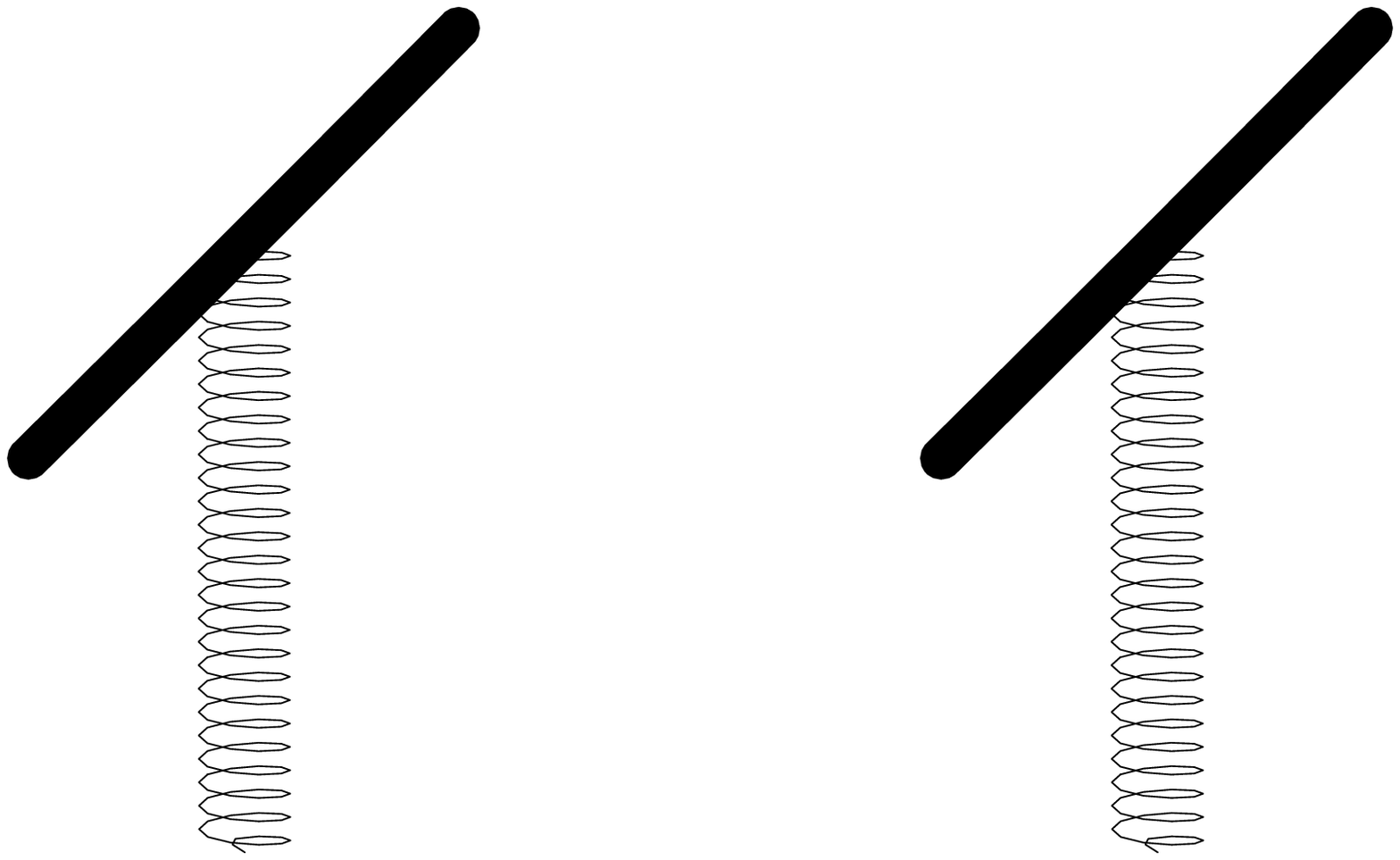, height=4.5cm} &
            \epsfig{file=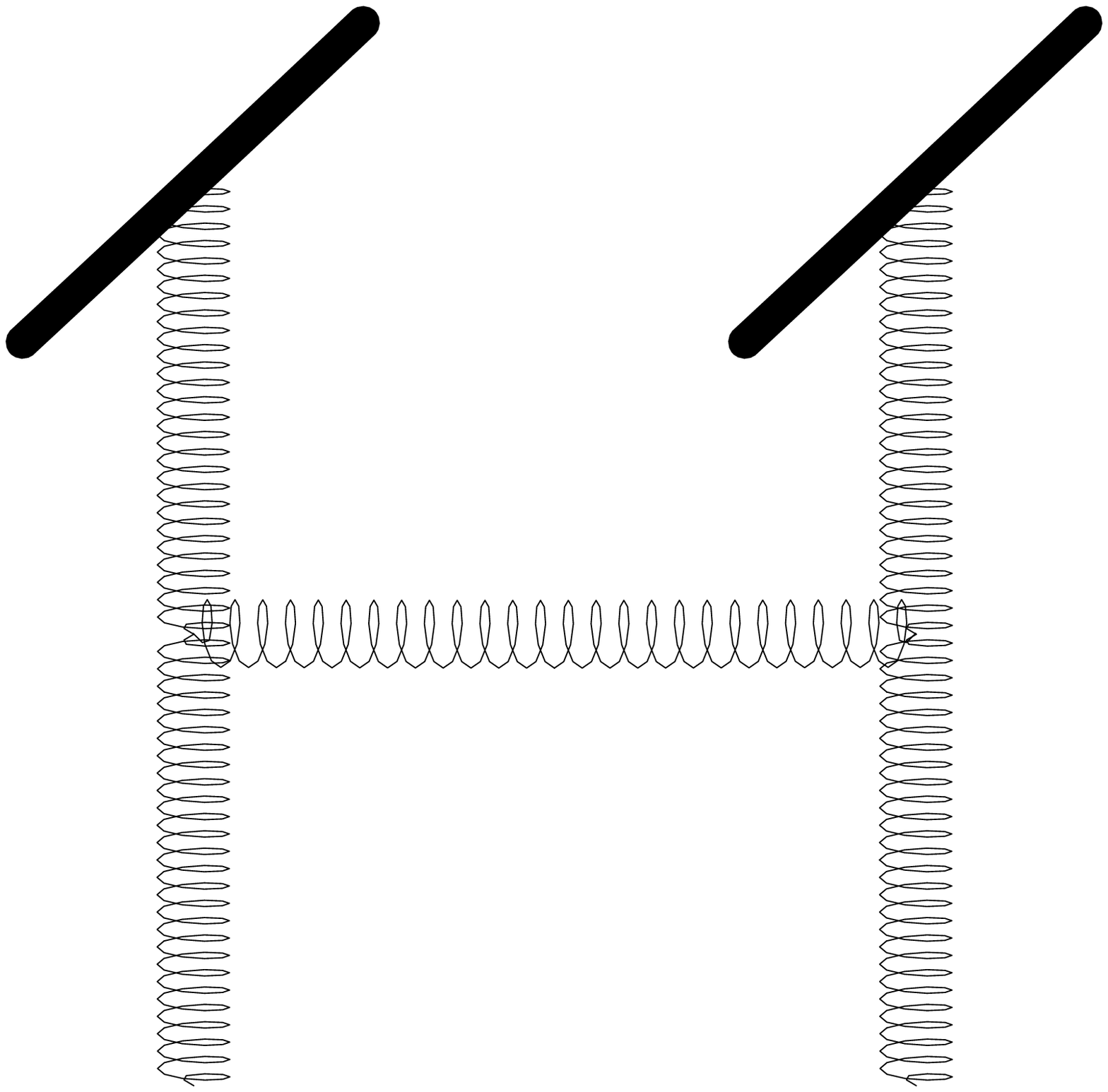, height=4.5cm} \\
    (a) & (b)
      \end{tabular}
    \caption{The small $\rho$ limit of the diagrams shown in Fig.
      \ref{fig:distcorr}: The lowest order terms shown in (a)
      correspond to those in Eq.(\ref{relation}). (b) is the small
      $\rho$ limit of Fig.\ref{fig:distcorr}b with all gluon
      propagators perturbative.}
    \label{fig:distcorrpert}
  \end{center}
\end{figure}
This contribution was discussed in \cite{AJMV}, and it was shown there
that it is indeed of order $\alpha_s\ln(1/x)$ relative to the leading
order result Eq.(\ref{tree}), or in the low density limit,
Eq.(\ref{relation}).  The reason for this enhancement is that when the
momentum on the external leg $l^+$ is much smaller than the maximal
longitudinal momentum allowed in the field, there is huge phase space
available to the emitted gluon $l^+<k^+<\Lambda^+$. The phase space
integral $\int {dk^+\over k^+}$ then gives the logarithmic enhancement
factor.

\begin{figure}[htbp]
  \begin{center}
    \begin{tabular}{c @{\hspace{2cm}} c}
      \epsfig{file=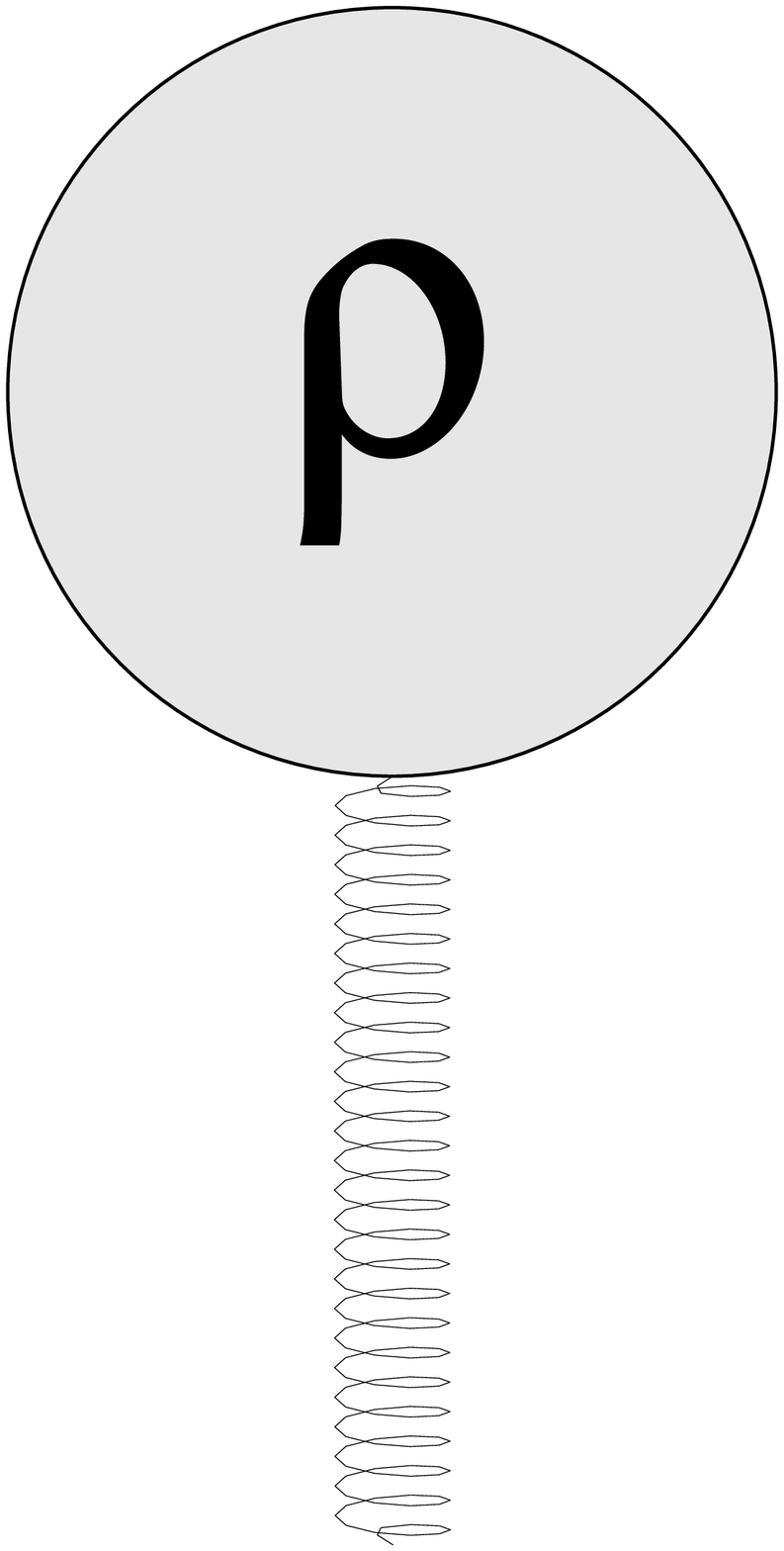,height=3cm} 
      \begin{minipage}[b]{2cm}
         \vskip2cm $l^+$
      \end{minipage}
& 
      \epsfig{file=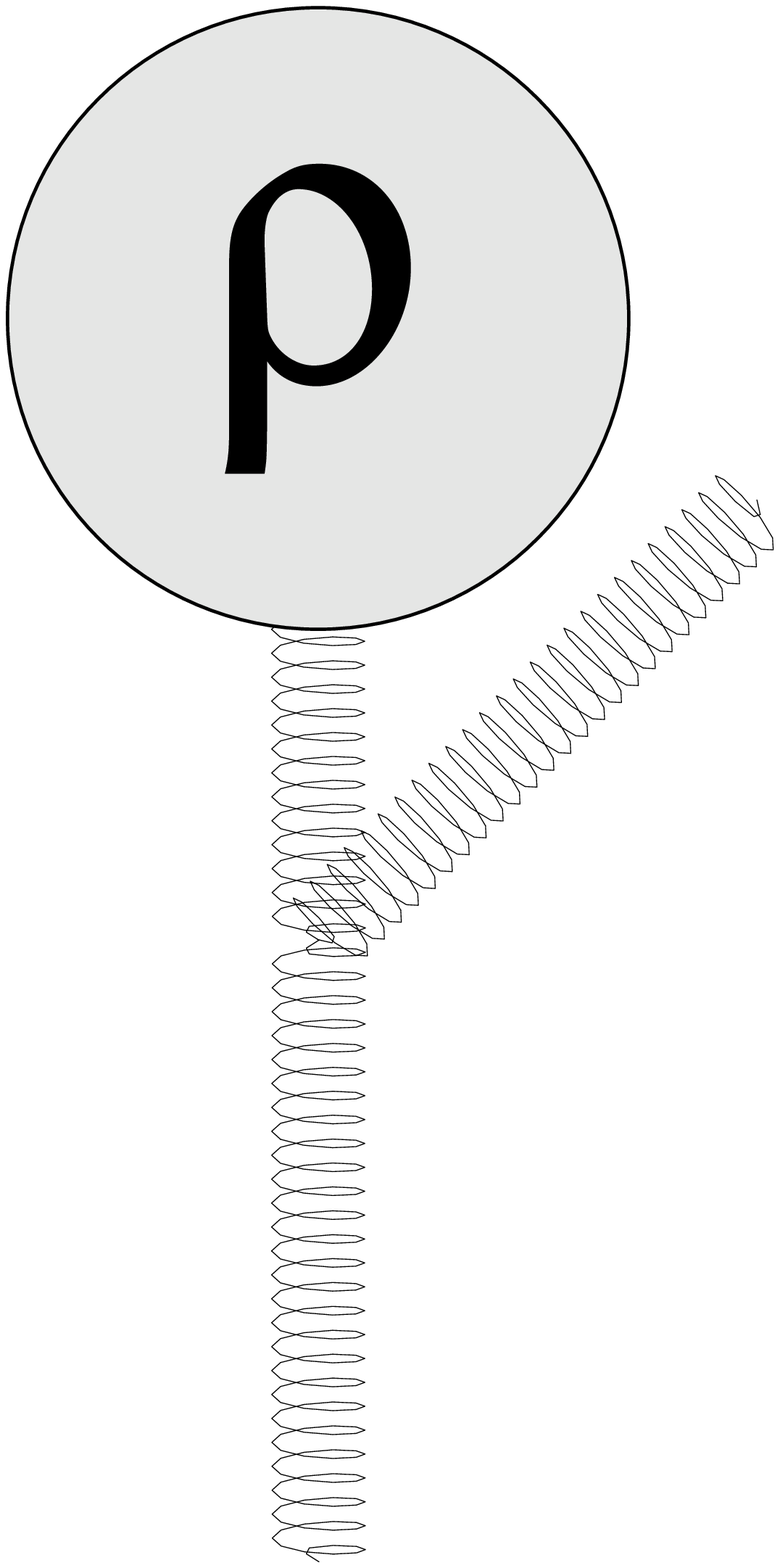,height=3.5cm} 
      \begin{minipage}[b]{2cm}
        \vskip1cm $k^+$\vskip1.5cm $l^+$
      \end{minipage}
\\ & \\
      (a) & (b)
    \end{tabular}
    \caption{Original vertex (a) and vertex modified by the
      emission of an additional fast gluon (b). ``$+$'' components of
      momenta are ordered from top to bottom}
    \label{fig:modvertex}
  \end{center}
\end{figure}

Carefully examining the $\alpha_s\ln 1/x$ corrections of
Figs.~\ref{fig:distcorr}b, \ref{fig:distcorrpert}b , we see that it
looks very similar to the tree level diagrams of
Figs.~\ref{fig:distcorr}a, \ref{fig:distcorrpert}a, except that the
soft gluon is emitted not from the original charge density $\rho$ as
depicted in Fig.\ref{fig:modvertex}a, but rather from a modified
charge density which in addition to $\rho$ contains one extra gluon.
One therefore can think of it as being emitted from the modified
vertex of Fig.\ref{fig:modvertex}b.  Since the large correction comes
from the region $l^+\ll k^+$, the emission from the modified vertex is
also eikonal.
So the source for emission of very soft gluons with $l^+\ll \Lambda^+$
effectively has been modified. This modification has to be taken into
account if we are to describe properly the soft glue distribution.
Fortunately, the change in the charge density is slow (logarithmic) so
that this is a perfect situation for the application of the Wilson
renormalization group ideas. We can integrate out the fluctuations
around the classical background perturbatively, gradually lowering the
longitudinal momentum cutoff $\Lambda^+$ on the remaining dynamical
degrees of freedom. This will generate the effective Lagrangian below
the new cutoff scale with modified $F[\rho]$.  So long as we keep the
change in the cutoff in every step of the RG small enough so that the
correction to $F$ is small relative to $F$ itself, the perturbative
procedure is justified. The condition for that is $\ln
{\Lambda^+-\delta\Lambda^+\over \Lambda^+}\sim 1$, $\alpha_s\ln
{\Lambda^+-\delta\Lambda^+\over \Lambda^+}\ll 1$.  In the next section
we describe in detail how to set up this renormalization group
procedure.

\section{The low x Wilson renormalization group.}
\label{sec:wilsonRG}

Let us introduce the following decomposition of the gauge field:
\begin{equation}
A^a_{\mu} (x) = b^a_{\mu} (x) + \delta A^a_{\mu} (x) + a^a_{\mu} (x)
\end{equation}
where $ b^a_{\mu} (x)$ is the solution of the classical equations of
motion Eq.(\ref{sol1}), $ \delta A^a_{\mu} (x) $ is the fluctuation
field containing longitudinal momentum modes $q^+$ such that
$\Lambda^+-\delta\Lambda^+\equiv\Lambda^{+'} < q^+ < \Lambda^+$ while
$a$ is a soft field with momenta $k^+<\Lambda^{+'}$.  Our aim is to
integrate out the fluctuation field $\delta A_\mu$ in the path
integral and compute the effective action for the soft field $a_\mu$
This integration is performed within the assumption that the
fluctuations are small as compared to the classical fields
$b^a_{\mu}$. More quantitatively, this requires that the coupling
constant is small $\alpha_s\ll 1$ and at each step of the
renormalization group procedure the ratio of the two cutoffs is not
too big, $\ln{\Lambda^+\over\Lambda^{+'}}\ll {1\over \alpha_s}$

To leading order in the coupling constant we should only keep the
terms up to second order in the fluctuation field $\delta A_\mu$ in
the expansion of the action around the classical solution $b^a_{\mu}
(x)$.
\begin{equation}
S= -{1\over 4}G(a)^2-{{1}\over{2}} \delta A_{\mu} [{\rm D}^{-1} 
(\rho)]^{\mu\nu} 
\delta A_{\nu} + 
ga^- J^{+\prime}+O((a^-)^2)
+ iF[\rho]
\label{effectiveaction}
\end{equation}
The inverse propagator of the fluctuation, $[{\rm D}^{-1}
(\rho)]^{\mu\nu}$ has a nontrivial dependence on the color charge
density.  Its explicit form is given in the next section,
Eq.(\ref{eq:action}).

\begin{figure}[htbp]
  \begin{center}
    $$\begin{minipage}{3cm}
      \epsfig{file=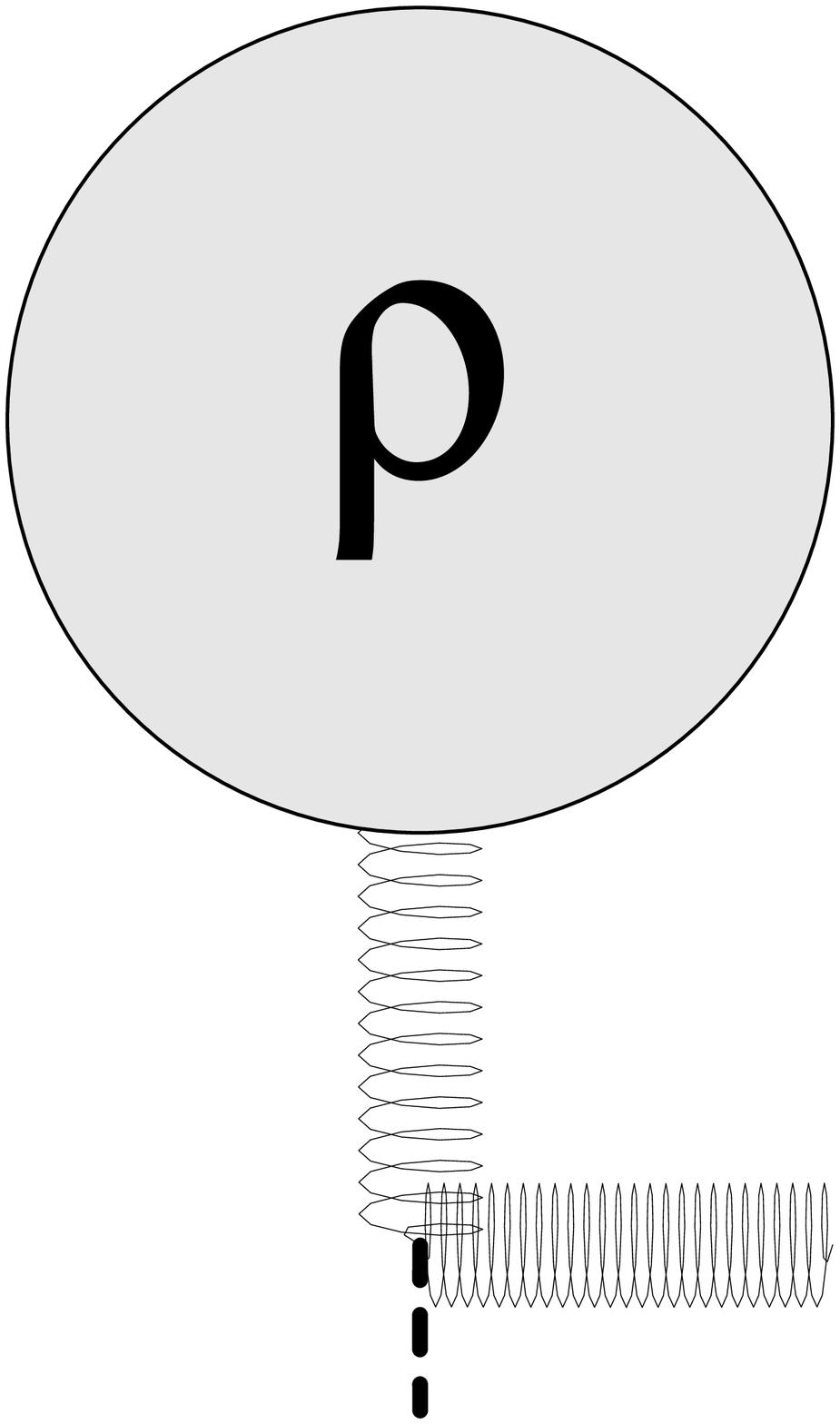,width=2.1cm}   \end{minipage}
    \circ \delta A + 
    \begin{minipage}{3cm}
      \epsfig{file=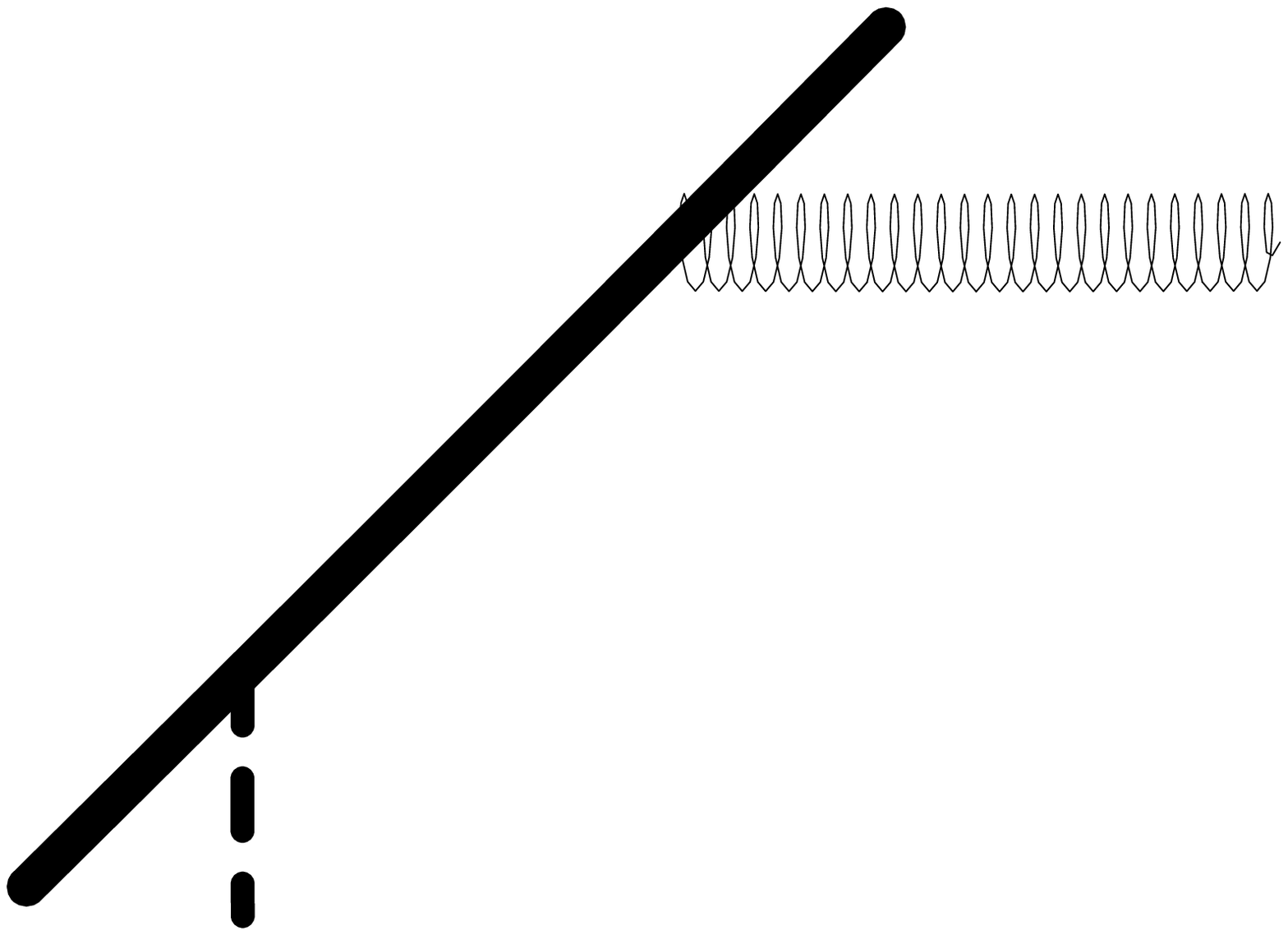,width=3cm}
    \end{minipage}
    \circ \delta A +     
    \begin{minipage}{2.2cm}
      \epsfig{file=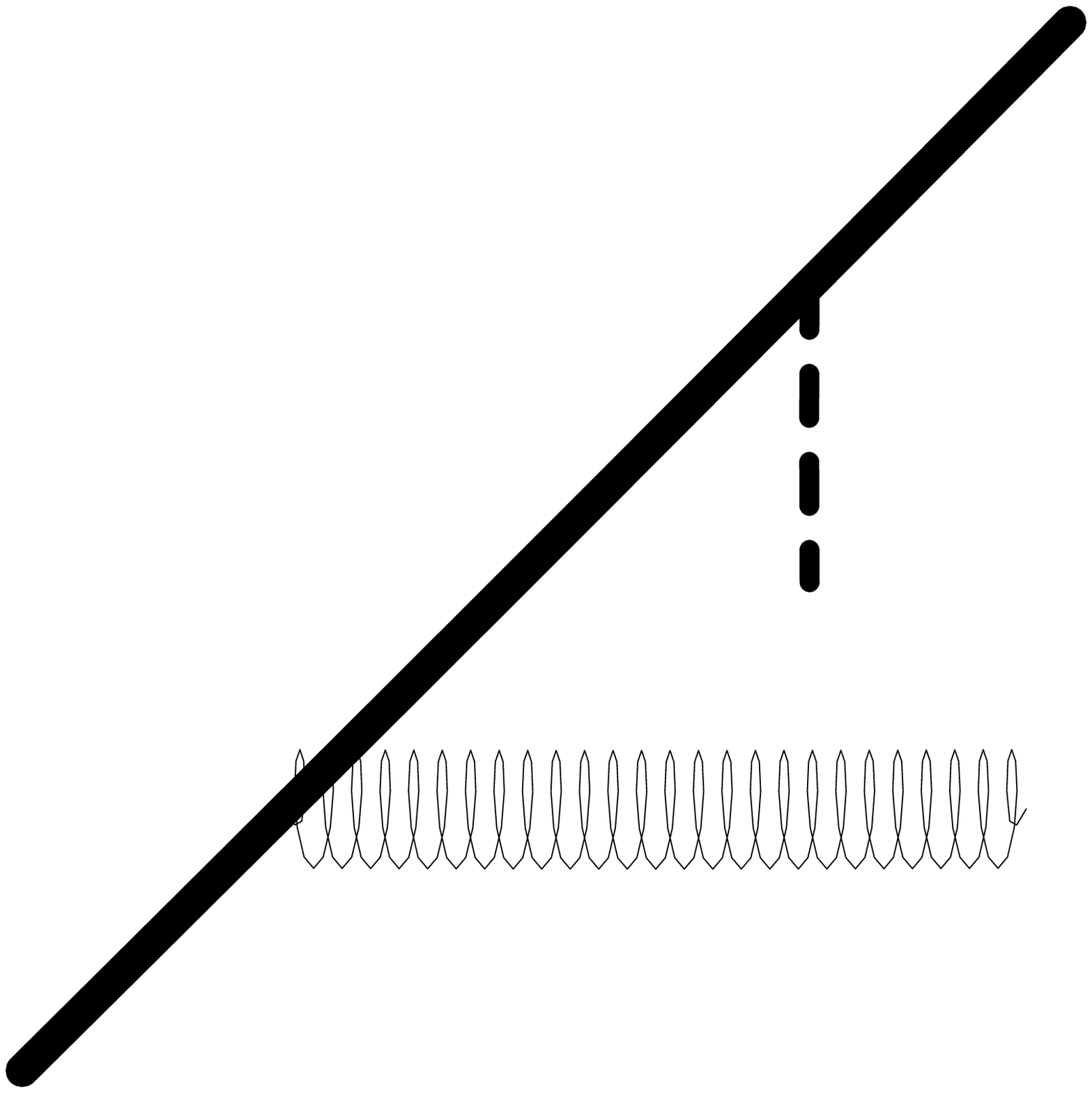,width=2.15cm}
    \end{minipage}\circ \delta A $$
    \caption{Diagrammatic representation of $\delta J_1$ in terms of 
      classical and fluctuation fields. The coupling to a $\delta A_\mu$
      field has been indicated by a curly line whereas
      slow modes $a_\mu$ have been symbolized by dashed ones.}
    \label{fig:deltaJ1}
  \end{center}
\end{figure}

\begin{figure}[htbp]
  \begin{center}
    \begin{minipage}{2cm}
      $ \delta A$ \hfill $\delta A$\vskip-0.3cm
    \epsfig{file=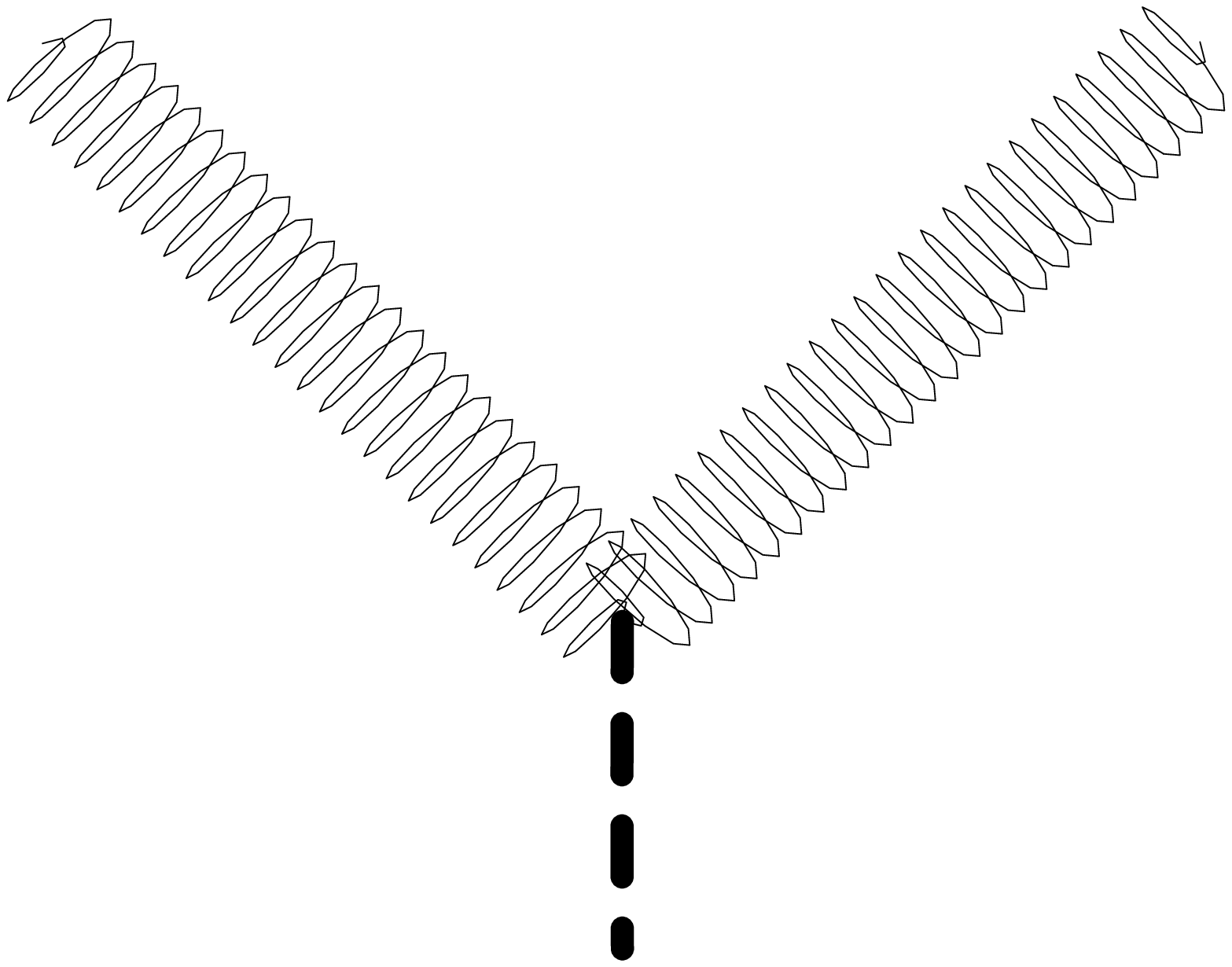,width=2cm}
    \end{minipage}
    +\ 
        \begin{minipage}{2.5cm}
    \epsfig{file=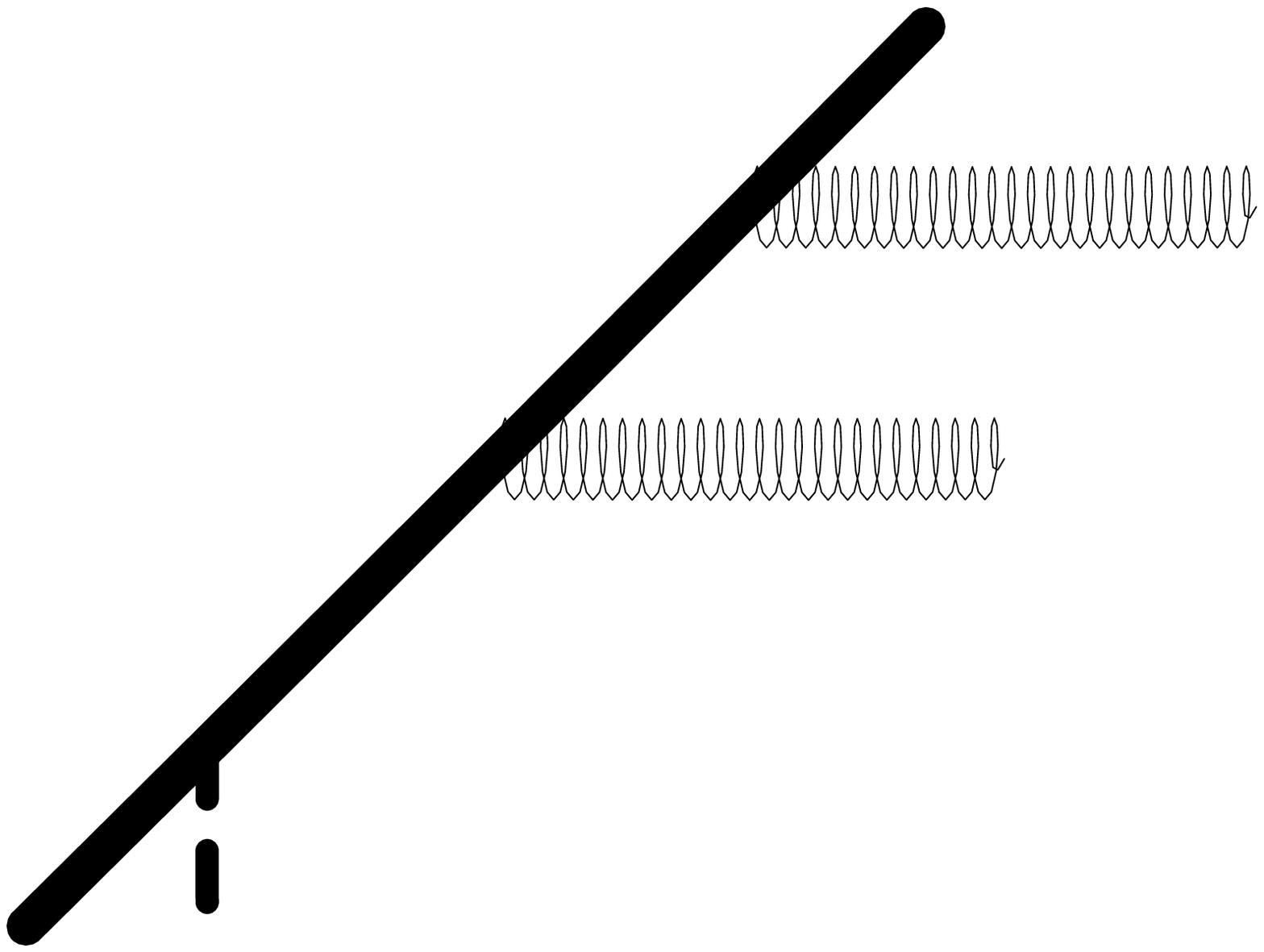,width=2.5cm}
    \end{minipage}
    \begin{minipage}{1cm}
      $\delta A$\vskip0cm$\delta A$\\\vskip1cm
    \end{minipage}
    + different time orderings
    \caption{Diagrammatic representation of $\delta J_2$.
      Symbols as in Fig.\ref{fig:deltaJ1}}
    \label{fig:deltaJ2}
  \end{center}
\end{figure}

We have introduced the modified color charge current $J^{+\prime}$,
whose explicit form in terms of the fluctuation fields is
\begin{equation}
J^{+\prime} =\rho(x_\perp )\delta(x^-)+ \delta J^+_1 + \delta J^+_2
\label{prime}
\end{equation}
with 
\begin{eqnarray}
\label{eq:deltaJ1}
  \lefteqn{\delta J^{+a}_1(x_\perp ,x^+) =  
   \delta (x^-) \Bigg[-2 f^{abc}
   \alpha^{b}_{i}\delta A^{c}_{i}(x^-=0)
   - {{g}\over{2}} f^{abc} \rho^{b} (x_\perp ) } 
\\ && \hspace{2cm}\times  \int\!dy^+ \Bigg[\theta
   (y^+ - x^+) - \theta (x^+ - y^+) \Bigg] \delta A^{-c}(y^+,
   x^-=0)\Bigg]
\nonumber
\end{eqnarray}
and
\begin{eqnarray}
\label{eq:deltaJ2}
\lefteqn{\delta J^{+a}_2(x) =
  -f^{abc} [\partial^+ \delta A^{b}_{i}(x)
  ]\delta A^{c}_{i}(x) - {{g^2}\over{N_c}} \rho^{b}(x_\perp )
  \delta(x^-) }
\\ && \nonumber \hspace{1cm} \times
  \int dy^+ \delta A^{-c}(y^+,x_\perp ,x^-=0) \int dz^+ \delta
  A^{-d}(z^+,x_\perp ,x^-=0) 
\\ && \nonumber \hspace{1cm} \times 
  \Bigg[\theta (z^+ -y^+)
  \theta (y^+ -x^+) {\rm tr} T^a T^c T^d T^b 
\\ &&\nonumber \hspace{1.2cm} 
  +\theta (x^+ -z^+) \theta (z^+ -y^+) {\rm tr} T^a T^b T^c T^d 
\\ && \nonumber \hspace{1.2cm} 
+\theta (z^+ -x^+) \theta (x^+ -y^+) {\rm tr} T^a T^d T^b T^c \Bigg]
\end{eqnarray}
The first term in both $\delta J_1^{+a}$ and $\delta J^{+a}_2$ 
arises from the expansion of $G^2$ in the action while the rest of the
terms proportional to $\rho(x_\perp )$ are coming from the expansion
of the Wilson line term.  The various terms with $\theta$ functions
correspond to different time orderings of the fields along the Wilson
lines.  Since the longitudinal momentum of $a^-$ is much lower than of
$\delta A$, we have only kept the eikonal coupling (the coupling to
$a^-$ only), which gives the leading contribution in this kinematics.
The contributions to $\delta J_1$ and $\delta J_2$ are depicted in
Fig.\ref{fig:deltaJ1} and \ref{fig:deltaJ2} respectively. Obviously
the first diagram in Fig.\ref{fig:deltaJ1} is nothing but our modified
vertex of Fig.\ref{fig:modvertex}b, now cast in a more precise
language. All other terms are nontrivial consequences
of the presence of a background of fast ``classical'' particles that
are encoded in the source term $\rho$ and a careful treatment of the
path integral over the modes in the interval $[\Lambda^+,\Lambda'^+]$.

We have not written out explicitly higher order in $a^-$ terms in the
effective action. There are of course such terms, which come from
expanding the Wilson line part of the action. Disregarding these terms
gives the effective action with the coupling of the field $a^-$ to the
charge density of the form $a^-J^+$.  However, imposing gauge
invariance on the final result together with the requirement that the
linear in $a^-$ term of the gauge invariant action should coincide
with the result of our calculation, the full gauge invariant form of
the effective action will be recovered. In the following therefore we
will concentrate on the linear term $a^-J^+$ only.
Note that the first term in Eq.(\ref{eq:deltaJ2}) does not have an
explicit factor of $\delta(x^-)$.  However we are only interested in
its low longitudinal momentum components since it couples directly to
$a^-$ in the effective action. In momentum space this contribution is
given by $f^{abc} \int dq^+[q^+ \delta A^{b}_{i}(q^+) ] \delta
A^{c}_{i}(-q^++k^+)$.  Since the leading logarithmic contributions
comes from the region $q^+\gg k^+$, to this accuracy this expression
does not depend on $k^+$ and can be therefore approximated by
$\delta(x^-)$ in coordinate space.  We then can define the modified
surface color charge density by
\begin{eqnarray} 
\label{eq:defrho}
&&
   \delta J^{+}(x_\perp ,x^-)=\delta\rho(x_\perp )\delta(x^-)
\nonumber \\ &&
   \delta \rho(x_\perp ) =\int\! dx^- \delta J^{+}(x_\perp , x^-)
\end{eqnarray} 
Formally $\delta\rho$ defined in this way is a function of $x^+$
as well as $x_\perp $. However, it is a function of $\delta A$'s which
only have longitudinal momenta much larger than the momenta in the
soft field $a$.  The (light cone) time variation scale of $\delta\rho$
is therefore ${1\over q^-}\sim{q^+\over q_\perp ^2}$ and is much
larger than the typical time variation scale of the on shell modes of
the field $a$. From this point of view $\delta \rho$ is therefore for
all practical purposes (light cone) time independent. Technically this
means that whenever we will need a correlation function of
$\delta\rho$'s, we will expand it to leading order in the time
derivatives
\begin{eqnarray} 
<\delta\rho(x_\perp , x^+)\delta\rho(y_\perp ,y^+)>=
<\delta\rho(x_\perp , x^+)\delta\rho(y_\perp ,x^+)>+...
\label{static}
\end{eqnarray} 
Corrections to this approximation are of order $q^-/k^-\sim x$.  We
will therefore not indicate the time dependence of $\rho$ explicitly.

The procedure now is the following. We first introduce the variable
$\rho^\prime$ in the path integral by 
\begin{equation} \int\!\!
  D[\rho,\delta A_\mu,a_\mu]e^{iS[a_\mu,\delta A_\mu,\rho]}= \int\!\!
  D[\rho^\prime,\rho,\delta A_\mu,a_\mu]
  \delta(\rho^\prime-\rho-\delta\rho[\delta A])e^{iS[a_\mu, \delta
    A_\mu, \rho]} 
\end{equation}
Here $\delta\rho[\delta A]$ is the functional of the fluctuation
fields defined by Eqs.(\ref{eq:deltaJ1},\ref{eq:deltaJ2},\ref{eq:defrho}).  
Now we
first have to integrate $\delta A_\mu$ at fixed $\rho$, and then integrate
over $\rho$.

This procedure generates the new effective action which symbolically
can be written as
\begin{equation}
\exp\{iS[\rho^\prime, a^\mu]\}=\exp\{-F^\prime[\rho^\prime]-{i\over
4}G^2(a) +iga\rho^\prime\}
\end{equation}
with
\begin{equation}
\exp\{-F^\prime[\rho^\prime]\}= \int D[\rho,\delta A]\ \delta
(\rho^\prime-\rho-\delta \rho[\delta A]) \exp\{-F[\rho] -{i\over 2}
\delta A D^{-1} [\rho] \delta A\}
\label{newf}
\end{equation}
Of course, to leading order in $\ln 1/x$ only terms linear in
$\alpha_s\ln 1/x$ should be kept in $F^\prime$.

The integration over the fluctuation field $\delta A_\mu$ is the most
technically involved part of this procedure. We will describe in
detail this part of the calculation in the next section.  The
structure of the result is however easy to understand from a simple
counting of powers of the coupling constant $\alpha_s$.  Consider
integration over the fluctuation field $\delta A_\mu$ at fixed $\rho$.
The counting of the powers of $\alpha_s$ is done most conveniently
after rescaling the fields and the charge density in the following
way\footnote{The reason for this rescaling can be traced back to
  Eq.(\ref{eq:treeexp}). Simple counting of powers of $g$ in the tree
  level graphs shows that for $\rho$ of order $g^{-2}$ all the tree
  level graphs are of the same order (see Appendix A). The classical
  field itself is then $O(g^{-1})$. This is also the magnitude of the
  field for which we expect to see the nontrivial shadowing and
  saturation effects. For parametrically smaller color charge
  densities an expansion in powers of the coupling constant
  automatically implies an expansion also in powers of $\rho$. Our
  primary interest is therefore in the charge densities of order
  $\alpha_s^{-1}$.}:
\begin{eqnarray} &&A_\mu\rightarrow {1\over g}A_\mu\\ 
&&\rho\rightarrow
{1\over g^2}\rho\nonumber\\ &&\delta\rho\rightarrow {1\over
g^2}\delta\rho\nonumber 
\end{eqnarray} 

Explicitly for the rescaled charge density we have
\begin{equation}
\label{prime1}
\rho^{+\prime} =\rho(x_\perp )+ \delta \rho^+_1 + \delta \rho^+_2
\end{equation}
with\footnote{We have used the identity
$\theta (z^+\!-\!y^+)\theta (y^+\!-\!x^+) \!+\! \theta (x^+\!-\!z^+)
\theta (z^+\!- \!y^+) \!+\!
\theta (z^+ \!- \!x^+)\theta (x^+ \!- \!y^+) \!=\! \theta (z^+-y^+)$
to simplify the expression for $\delta \rho^a_2$.}
                                       
\begin{eqnarray}
\label{rho11}
\delta \rho^{+a}_1(x_\perp ) &=& -2 f^{abc} \alpha^{b}_{i}\delta
A^{c}_{i}(x^-=0)
- {{1}\over{2}} f^{abc} \rho^{b} (x_\perp ) 
\\ && \nonumber \hspace{.5cm}\times
\int dy^+ \Bigg[\theta (y^+ - x^+) - \theta (x^+ - y^+) \Bigg] 
\delta A^{-c}(y^+,x_\perp ,x^-=0)
\end{eqnarray}
and
\begin{eqnarray}
\label{rho21}
  \delta \rho^{+a}_2(x) &=& 
  -f^{abc} \int d x^- [\partial^+ \delta A^{b}_{i}(x) ]\delta A^{c}_{i}(x) 
 \\ &&
  + {1\over{N_c}} \rho^{b}(x_\perp ) \int\! dy^+ 
  \delta A^{-c}(y^+,x_\perp ,x^-=0)
\nonumber \\ && \hspace{1cm} \times
  \int\! dz^+ \delta A^{-d}(z^+,x_\perp ,x^-=0) 
\nonumber \\ && \hspace{1cm} \times 
\Bigg[f^{ace}f^{bde}\theta (z^+ -x^+) \theta (x^+ -y^+)  \Bigg]
\end{eqnarray}
with
\begin{eqnarray}
\label{resa}
\alpha_i(x_\perp ) & = & i U(x_\perp )\partial_i U^\dagger(x_\perp)
\nonumber\\
\partial_i\alpha_i & = &-\rho
\end{eqnarray}
In terms of the rescaled fields the coupling constant $g$ disappears
from the expressions for $\delta \rho$, and appears only as the
overall factor $1/g^2$ in the action.  The propagator of the
fluctuation field is therefore of order $\alpha_s$. It immediately
follows from Eqs.(\ref{rho11}) and (\ref{rho21} that
\begin{eqnarray}
<\delta\rho>_{\delta A}=O(\alpha_s)\nonumber \\
<\delta\rho\delta\rho>_{\delta A}=O(\alpha_s)
\end{eqnarray}
while all other (connected) correlation functions of $\delta\rho$ are
higher order in $\alpha_s$. Since we are working to the lowest order
in $\alpha_s$ we can neglect all these other terms.  Therefore to
lowest order in $\alpha_s$, after integrating over $\delta A$ at fixed
$\rho$, we are left with the weight function for $\delta \rho$, which
generates only connected one- and two-point functions.  Such weight is
obviously a Gaussian.  Introducing the following notations
\begin{eqnarray}
&&<\delta\rho^a(x_\perp )>_{\delta A} =: \alpha_s\ln{1\over x}
\sigma^a(x_\perp )\nonumber \\
&&<\delta\rho^a(x_\perp ,x^+)\delta\rho^b(y_\perp ,x^+)>_{\delta A}=: 
\alpha_s\ln{1\over x}\chi^{ab}(x_\perp ,y_\perp )
\end{eqnarray}
we can write the result of the $\delta A_\mu$ integration in the form
\begin{eqnarray}
\lefteqn{ \int D[\rho,\rho^\prime] [ {\rm
Det}(\chi)]^{-1/2}\exp\left(-F[\rho]\right) }\\ && \times
\exp\left(-{1\over 2\,\alpha_s\,ln{1\over x}}
\left[\rho^\prime_x-\rho_x- \alpha_s\ln{1\over
x}\sigma_x\right][\chi^{-1}_{x y}]\left[
\rho^\prime_y-\rho_y-\alpha_s\ln{1\over x}\sigma_y\right]\right)
\nonumber
\label{next}
\end{eqnarray}
In the above equation we adopted condensed notations: the indices $x$
stand for the set of indices and coordinates $\{x_\perp ,a\}$, and repeated
indices are understood to be summed (integrated) over\footnote{We note
here that this result can be derived formally by introducing the
variable $\rho^\prime$ with the help of Lagrange multiplier
\begin{equation}
\delta (\rho'-\rho-\delta \rho[\delta A]) = \int D[\lambda]\
e^{i\lambda (\rho'-\rho-\delta \rho[\delta A])}
\end{equation}
and subsequently integrating out $\lambda$ in perturbation theory to
order $\alpha_s$.}.

The calculation of $\chi$ and $\sigma$ is the subject of the following
section.  However, the knowledge of the general structure of the
$\rho$ integral, Eq.(\ref{next}) is sufficient to perform the integral
over $\rho$ in Eq.(\ref{newf}) without the explicit knowledge of
$\chi$ and $\sigma$. The reason is that the integrand in
Eq.(\ref{next}) is a function very sharply peaked around
$\rho=\rho'+O(\alpha_s$), and the integral is calculable in the
steepest descent approximation.  This was done in \cite{nonlinear}.
The result is very simple

\begin{eqnarray}
\label{finals}
&&F^\prime=F+{\alpha_s\ln(1/x)\over 2}\left[\chi_{u v} {\delta^2
\over\delta\rho_u\delta\rho_v}F -{\delta^2\chi_{u v}
\over\delta\rho_u\delta\rho_v} -{\delta F\over \delta\rho_u} \chi_{u
v} {\delta F\over \delta\rho_v} \right. \nonumber \\ &&
\left. \hspace{1cm} +2{\delta F\over \delta\rho_u}{\delta\chi_{u
v}\over \delta\rho_v}+2{\delta \sigma_u\over\delta\rho_u} -2{\delta
F\over\delta\rho_u}\sigma_u\right]
\end{eqnarray}

Taking the derivative with respect to $\ln 1/x$ we obtain the Wilson
renormalization group equation for the functional $F$
\begin{eqnarray}
\label{finalrg}
{d\over d\ln(1/x)}F & = &{\alpha_s\over 2}\left[\chi_{u v} {\delta^2
\over\delta\rho_u\delta\rho_v}F -{\delta^2\chi_{u v}
\over\delta\rho_u\delta\rho_v} -{\delta F\over \delta\rho_u} \chi_{u
v} {\delta F\over \delta\rho_v} \right. \nonumber\\ &&
\left. \hspace{1cm} + 2{\delta F\over \delta\rho_u}{\delta\chi_{u
v}\over \delta\rho_v} +2{\delta \sigma_u\over\delta\rho_u} -2{\delta
F\over\delta\rho_u}\sigma_u\right]
\end{eqnarray}
This equation is extremely simple when written for the weight function
$Z\equiv \exp\{-F\}$
\begin{equation}
{d\over d\ln(1/x)}Z= \alpha_s \left\{{1\over 2}{\delta^2
\over\delta\rho_u\delta\rho_v}\left[Z\chi_{u v} \right] -{\delta
\over\delta\rho_u}\left[Z\sigma_u\right]\right\}
\label{final}
\end{equation}

Equations (\ref{finalrg}) and (\ref{final}) provide the closed form of
the renormalization group equation in terms of the functionals
$\sigma[\rho]$ and $\chi[\rho]$.  In the next section we will
calculate these two quantities.

Eq.(\ref{final}) can be written directly as evolution equation for the
correlators of the charge density. Multiplying Eq.(\ref{final}) by
$\rho_{x_1}...\rho_{x_n}$ and integrating over $\rho$ yields
\begin{eqnarray}
\lefteqn{
{d\over d\ln(1/x)}<\rho_{x_1}...\rho_{x_n}>
}
\\ &&=\alpha_s
\left[\sum_{0<m<k<n+1}<\rho_{x_1}...\rho_{x_{m-1}}\rho_{x_{m+1}}... 
\rho_{x_{k-1}}\rho_{x_{k+1}}...
\rho_{x_n}\chi_{x_m x_k}>
\right. \nonumber\\ && \left. \hspace{1cm}
+\sum_{0<l<n+1}<\rho_{x_1}...\rho_{x_{l-1}}\rho_{x_{l+1}}
...\rho_{x_n}\sigma_{x_l}>\right]
\nonumber 
\label{correl}
\end{eqnarray}

In particular, taking $n=2$ we obtain the evolution equation for the two
point function
\begin{equation}
{d\over d\ln(1/x)}<\rho_x\rho_y>=
\alpha_s\left\{<\chi_{x y}+\rho_x\sigma_y+\rho_y\sigma_x>\right\}
\label{prop}
\end{equation}
This equation is useful in making contact with standard evolution
equations, since the correlator of the color charge density at weak
fields is directly related to the unintegrated gluon density in a
hadron \cite{linear}. Eq. (\ref{prop}) can then be straightforwardly
rewritten as an evolution equation for the gluon density.

\section{Small fluctuations in the background field. The calculation of
$\sigma$ and $\chi$.}
\label{sec:sigmachi}

\begin{figure}[htbp]
  \begin{center}
    \epsfig{file=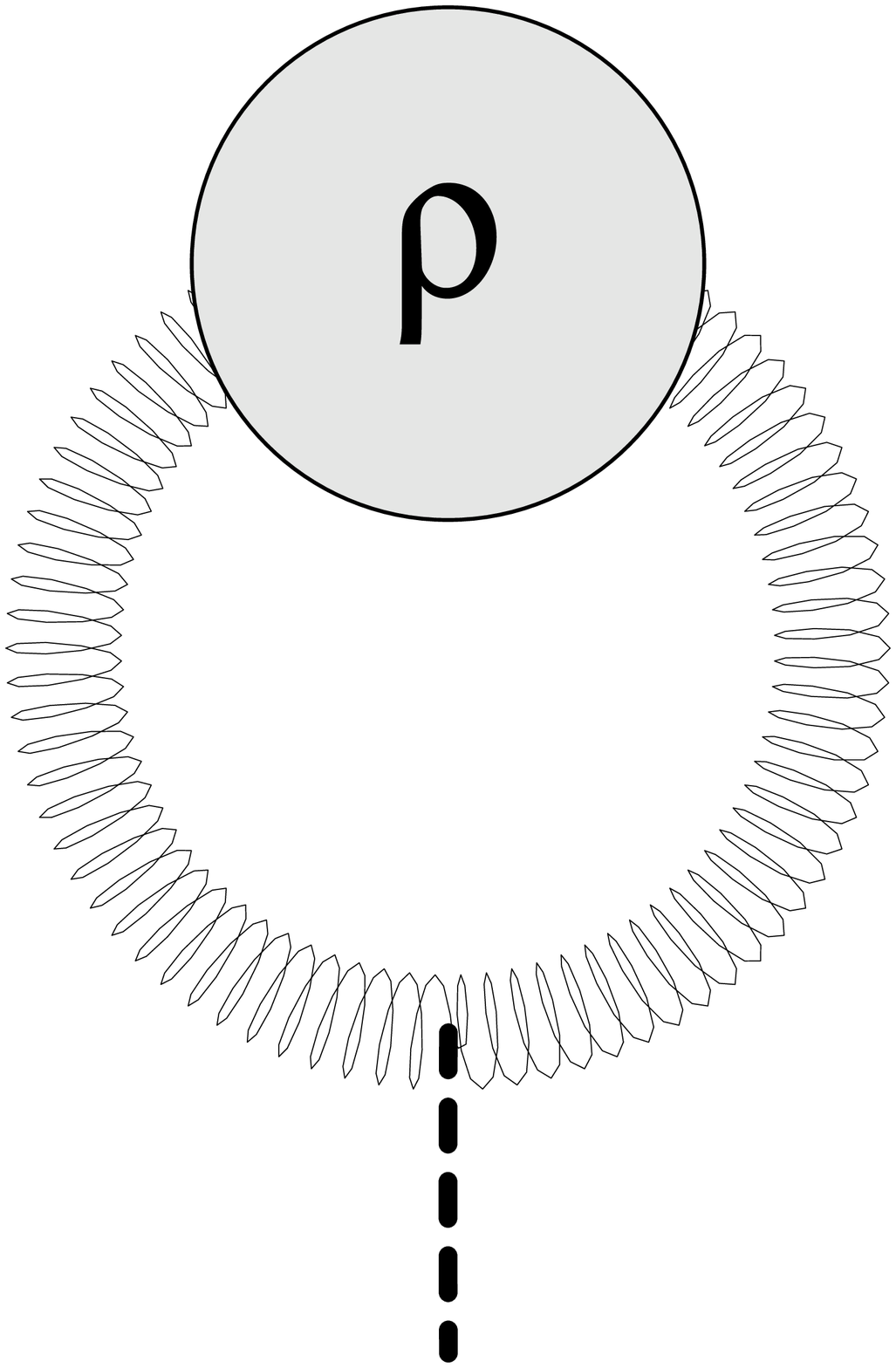, width=1.9cm}
    \hspace{.5cm}
    \epsfig{file=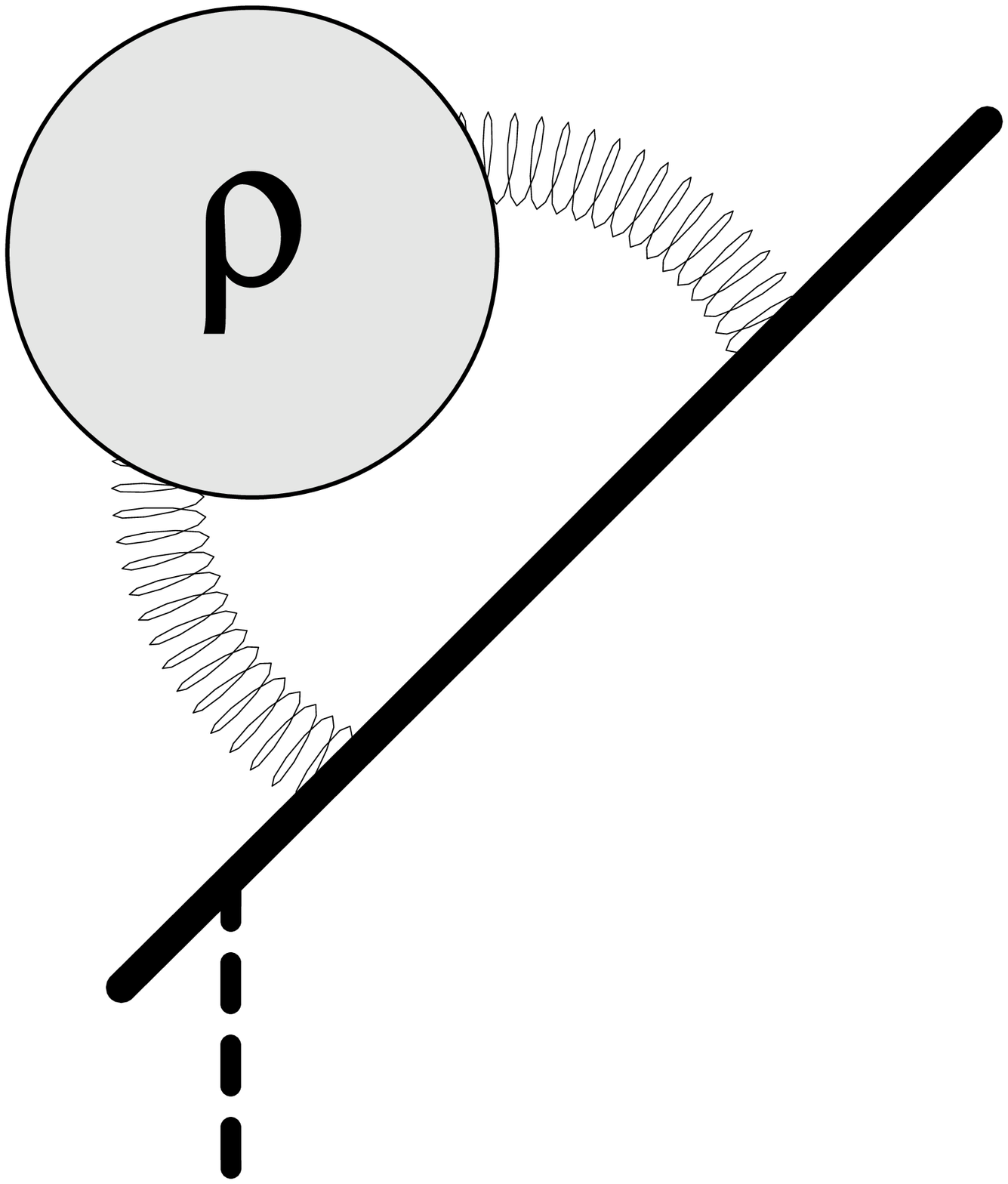, width=2cm}
    \hspace{.5cm}
    \epsfig{file=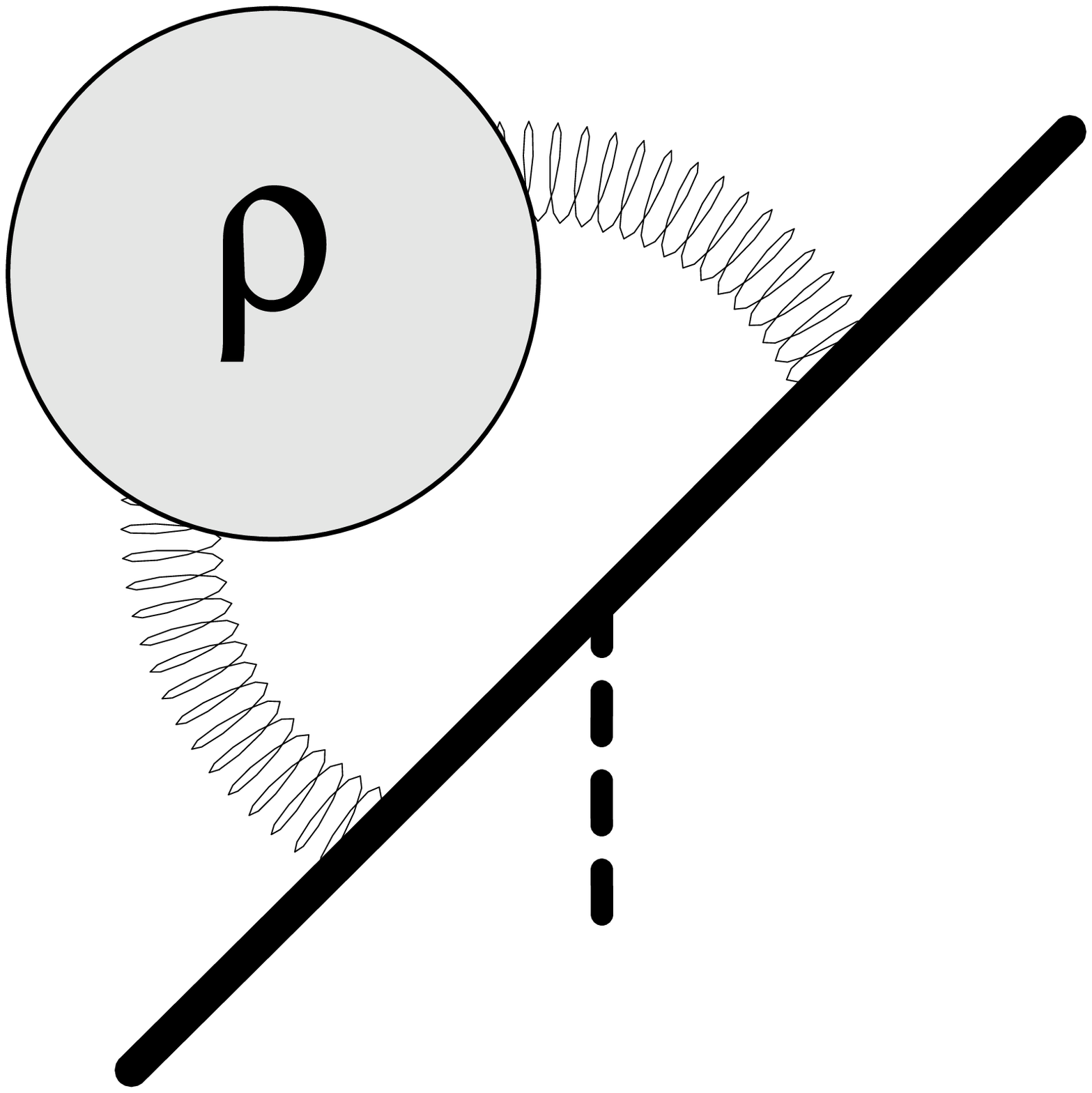, width=2cm}
    \hspace{.5cm}
    \epsfig{file=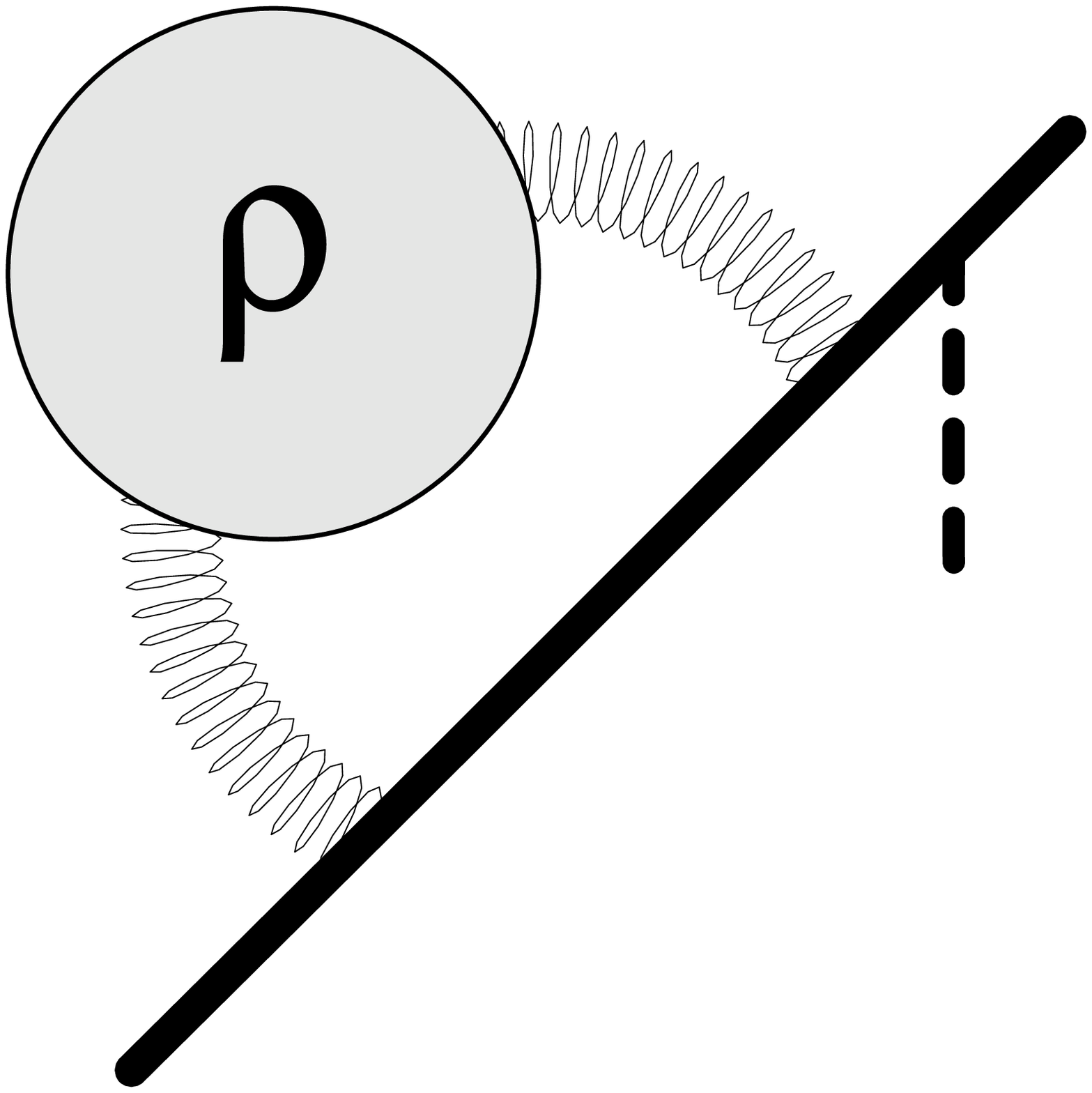, width=2cm}
    \caption{Diagrams contributing to $\sigma$. Note that these are all 
      virtual contributions resulting from contracting the $\delta A$
      lines of the diagrams shown in Fig.\ref{fig:deltaJ2}}
    \label{fig:sigma}
  \end{center}
\end{figure}

\begin{figure}[htbp]
  \begin{center}
    $$\left(
      \begin{minipage}{1.05cm}
      \epsfig{file=deltaJ1a.eps,width=1.05cm}   
    \end{minipage}
    + 
    \begin{minipage}{1.5cm}
      \epsfig{file=deltaJ1b.eps,width=1.5cm}
    \end{minipage}
    +
    \begin{minipage}{1.1cm}
      \epsfig{file=deltaJ1c.eps,width=1.1cm}
    \end{minipage}
    \right)\circ
    \begin{minipage}{1.5cm}
    \epsfig{file=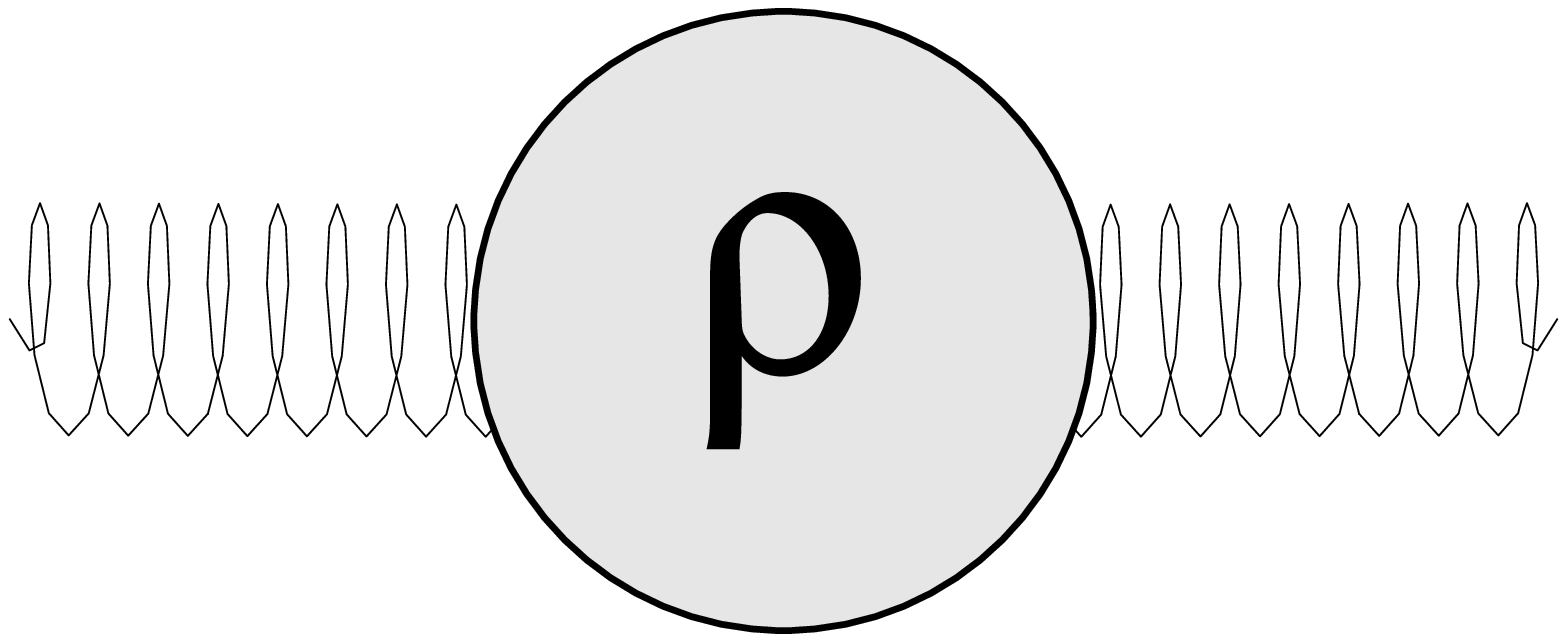,width=1.5cm}
    \end{minipage}
 \circ \left(
\begin{minipage}{1.05cm}
      \epsfig{file=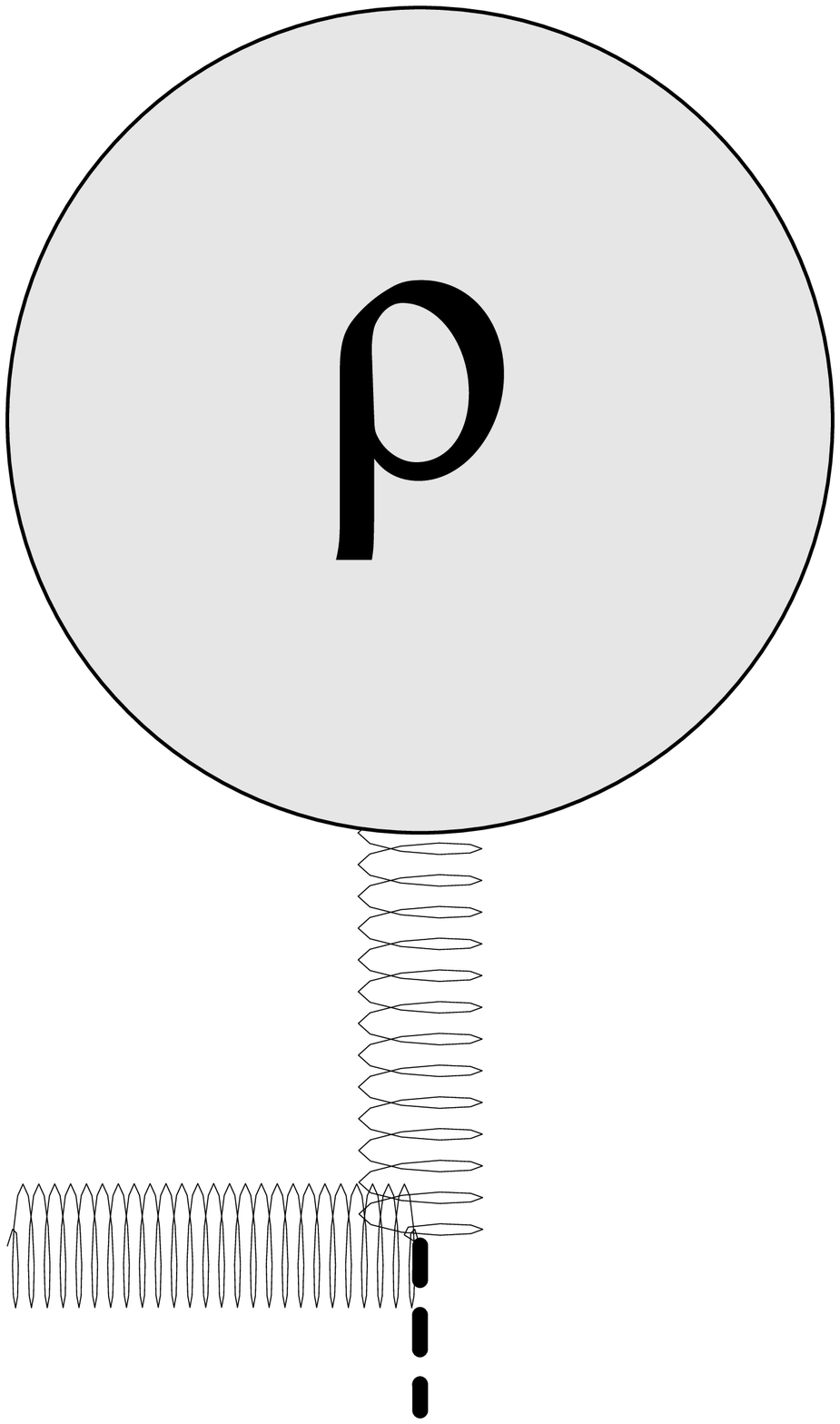,width=1.05cm}   
    \end{minipage}
    + 
    \begin{minipage}{1.1cm}
      \epsfig{file=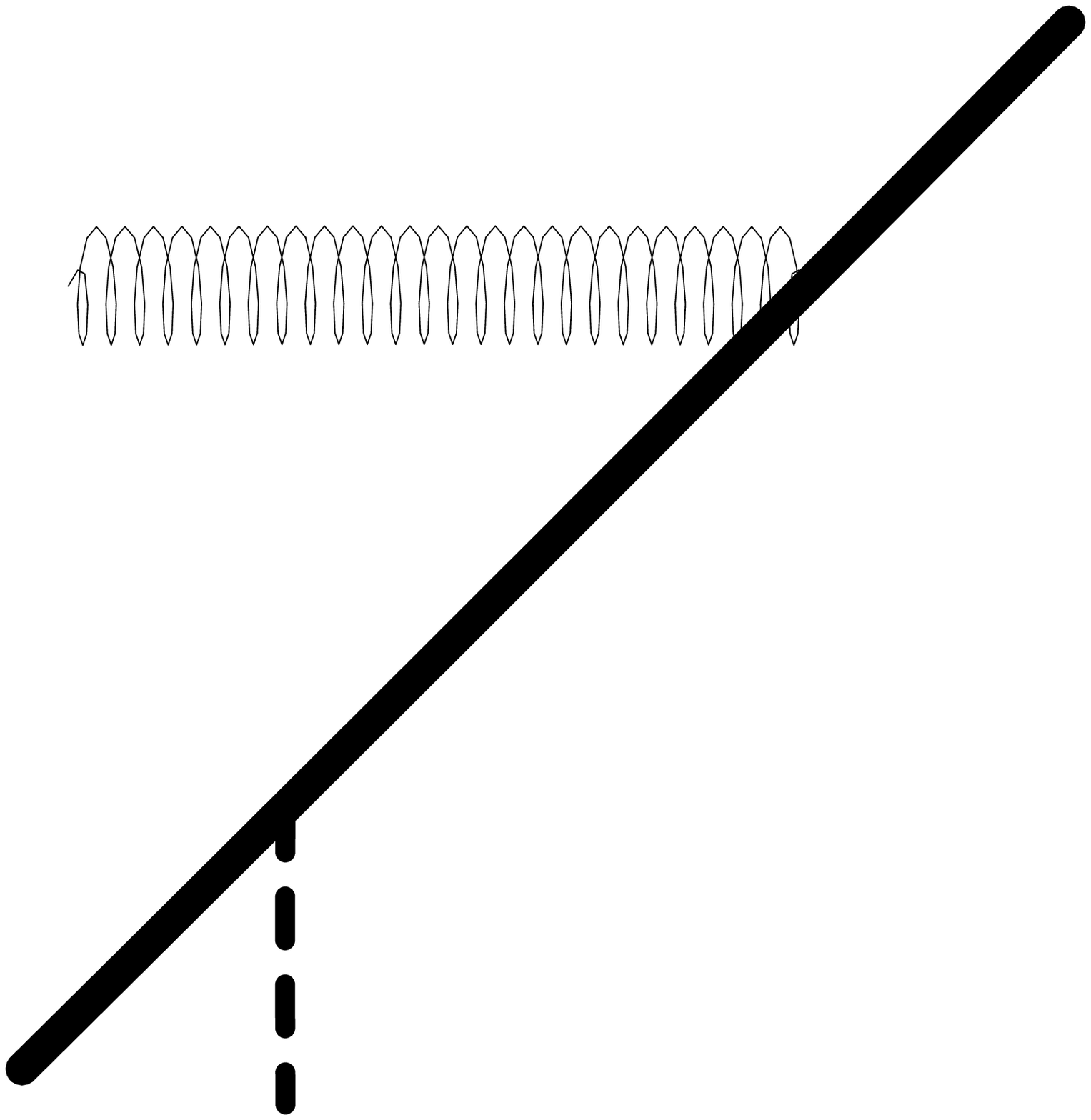,width=1.1cm}
    \end{minipage}
    +
    \begin{minipage}{1.5cm}
      \epsfig{file=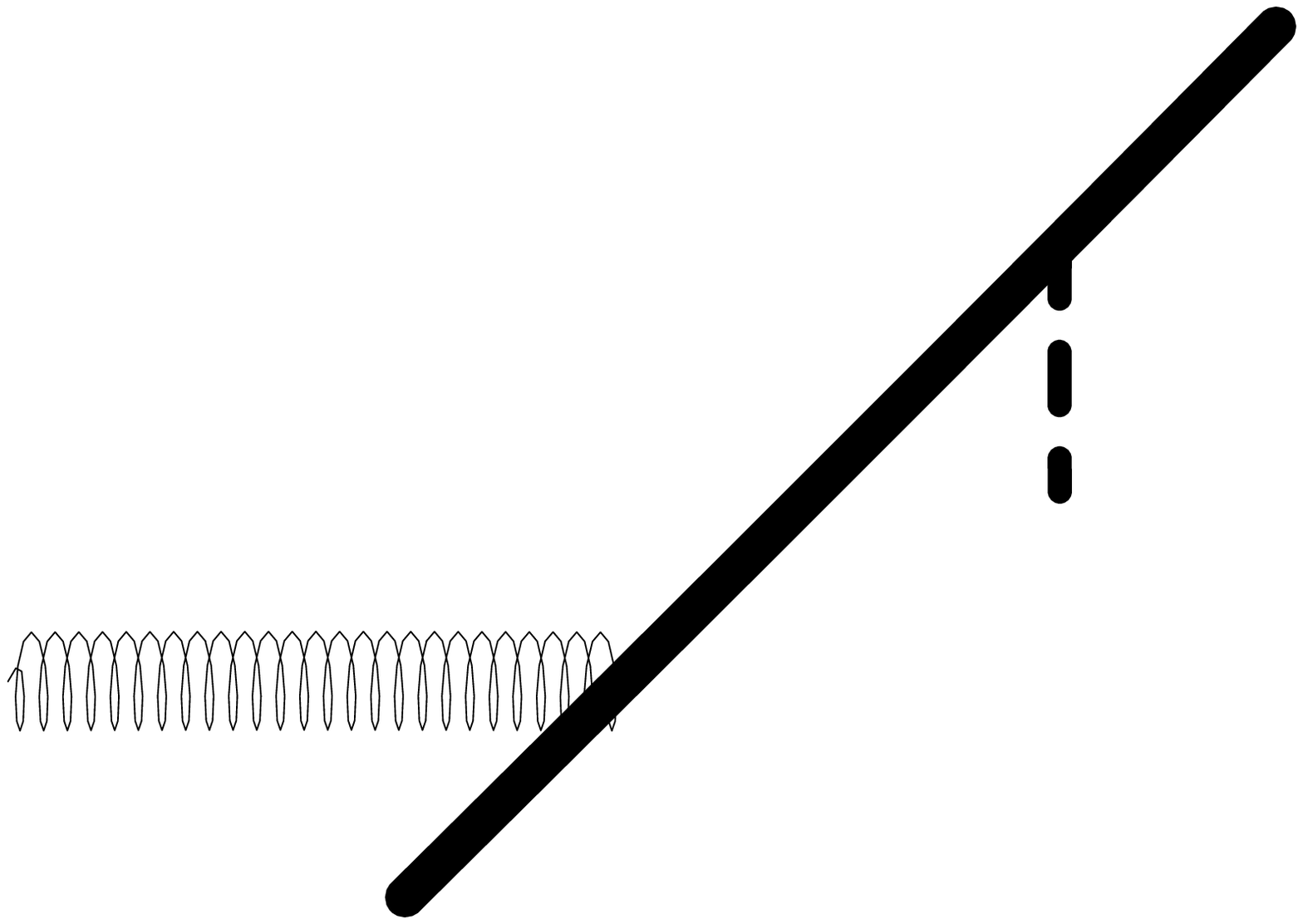,width=1.5cm}
    \end{minipage}
    \right)$$
    \caption{Diagrams contributing to $\alpha_s\ln1/x \chi$. Note
      that these arise from contracting to factors of the diagrams
      shown in Fig.\ref{fig:deltaJ1}. For a separate list of all 9
      contributions see Fig.\ref{fig:chifulllist}}
    \label{fig:chi}
  \end{center}
\end{figure}

In this section we calculate the one - point function $\sigma$ and two
- point correlation function $\chi$ of $\delta\rho(x_\perp )$,
Eq.(\ref{rho11}), (\ref{rho21}).  First, note that these quantities
are given by the Feynman diagrams of Fig.\ref{fig:sigma} and
\ref{fig:chi} respectively.

The propagator lines in these diagrams are the propagators of the
fluctuation fields $\delta A$ in the non vanishing background. This is
the inverse of the operator $D^{-1}_{\mu\nu}$ that appears in
Eq.(\ref{effectiveaction}).

At this point we see that the ghosts associated with our gauge fixing
do not contribute to order $\alpha_s$. The interaction of the ghost
fields with the rescaled fluctuation field is order one, see
Eq.(\ref{FP}).  However any insertion of a ghost vertex will lead to
an extra fluctuation propagator and this is proportional to
$\alpha_s$. We therefore forget about ghosts from now on.
\begin{figure}[htbp]
  \begin{center}
    \epsfig{file=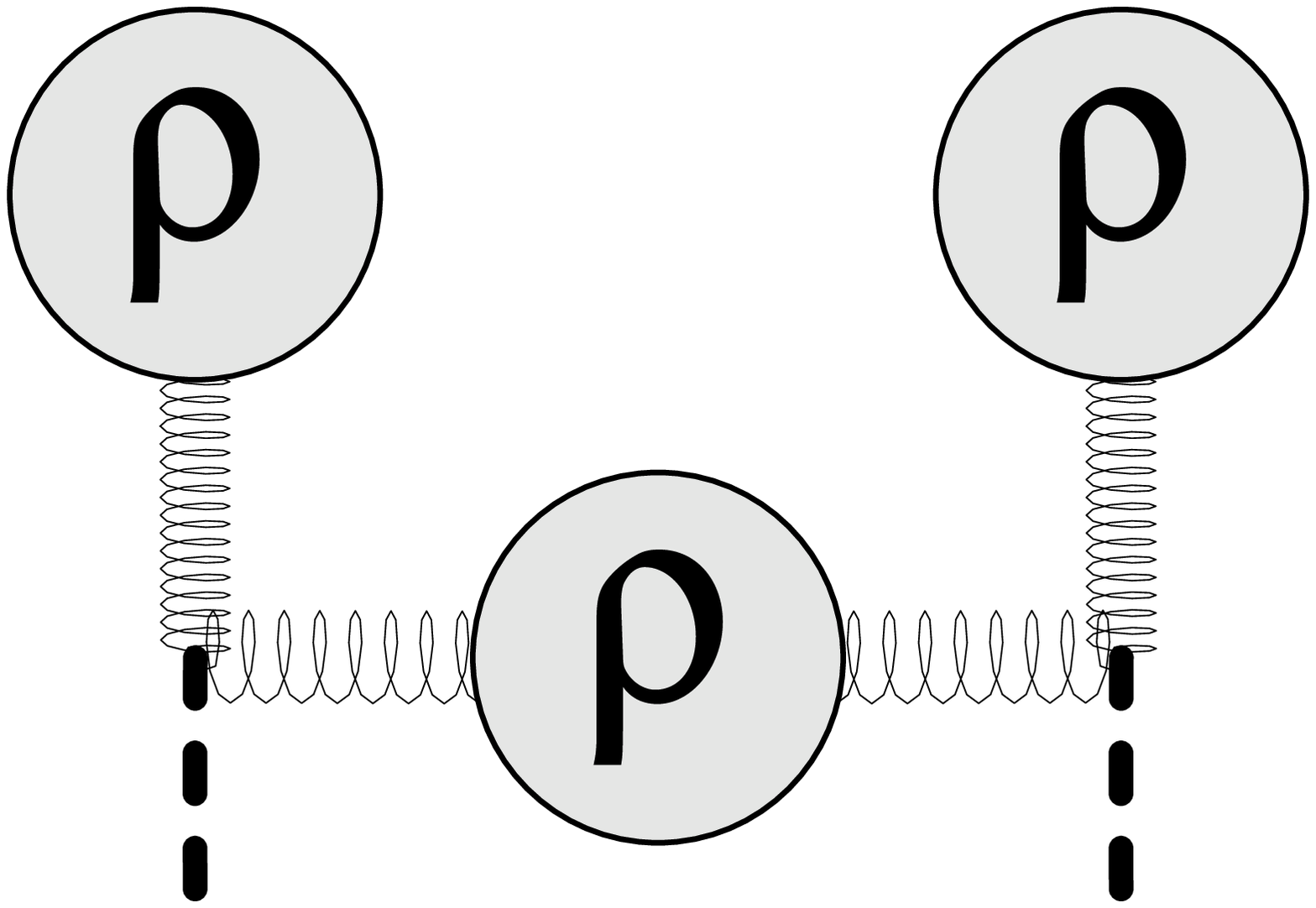,width=2.3cm}
    \epsfig{file=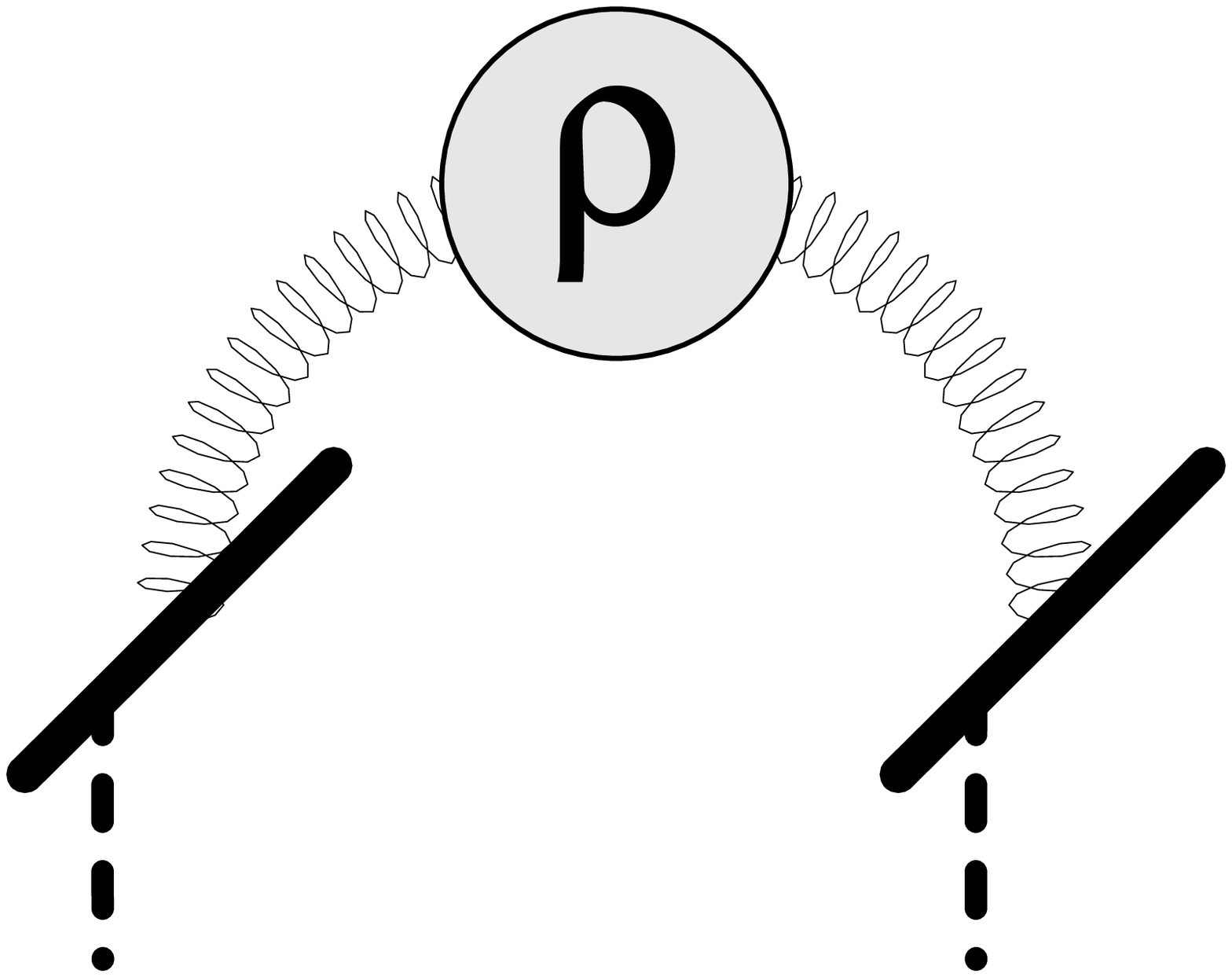,width=2.3cm}
    \epsfig{file=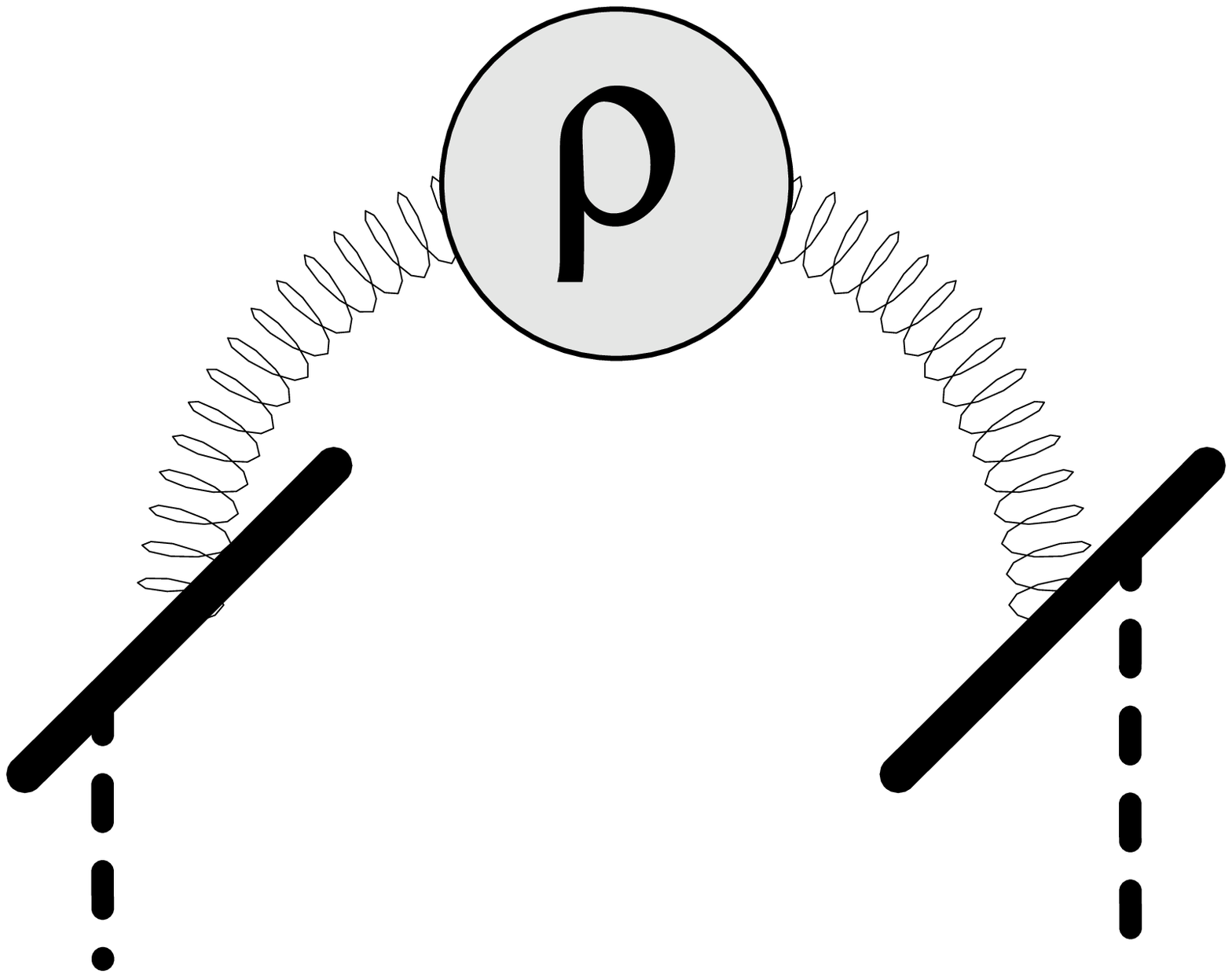,width=2.3cm}
    \epsfig{file=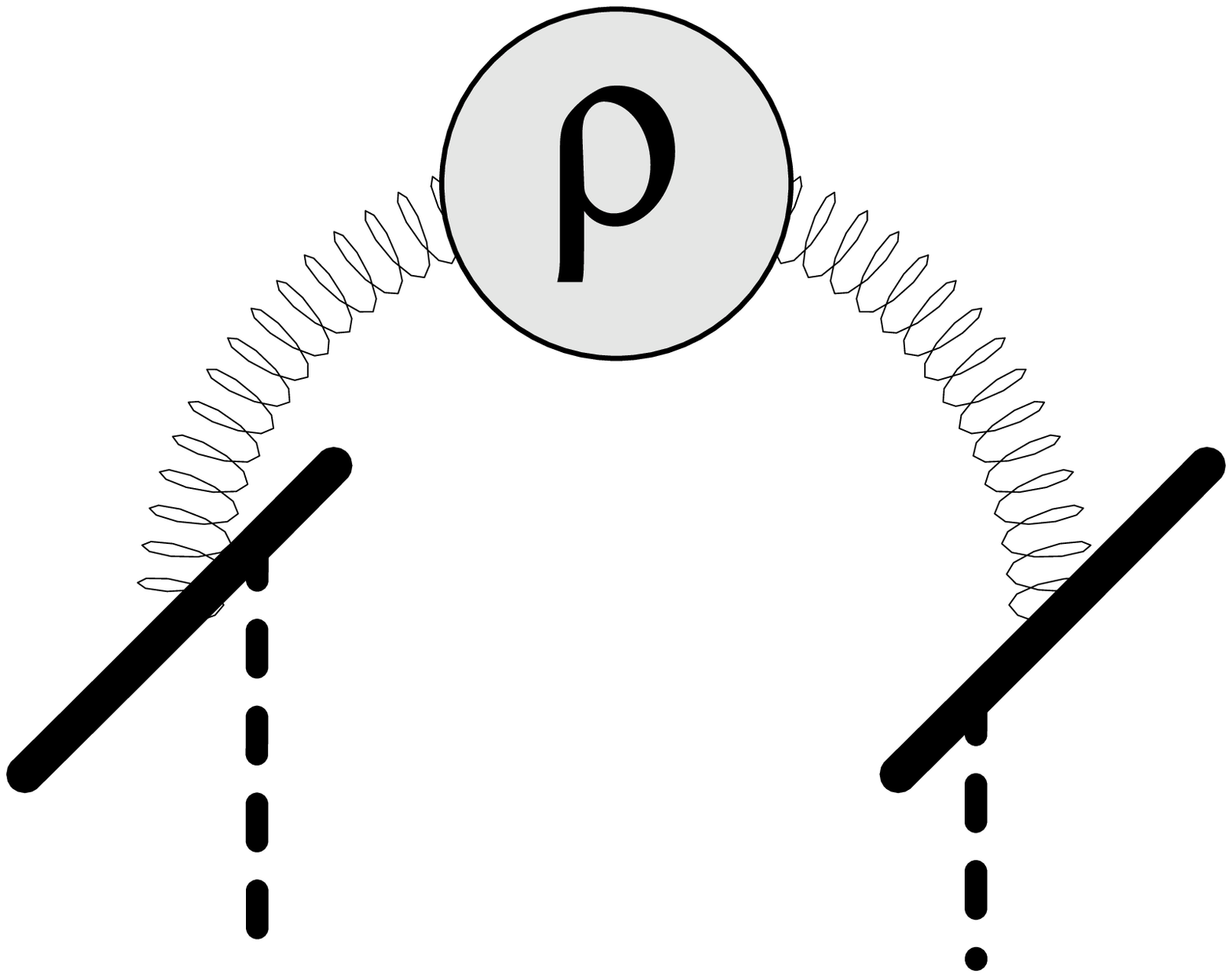,width=2.3cm}
    \epsfig{file=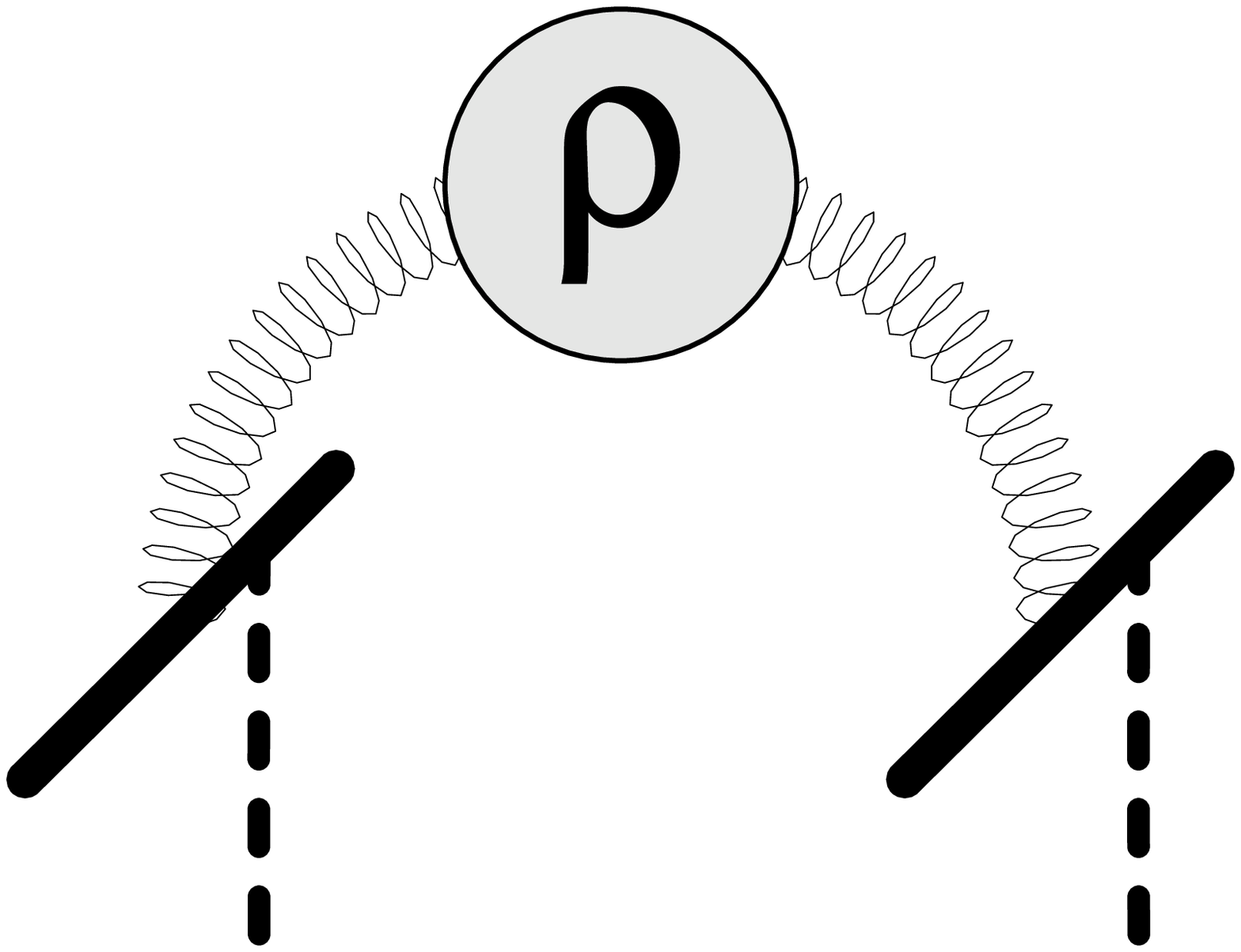,width=2.3cm}
      \epsfig{file=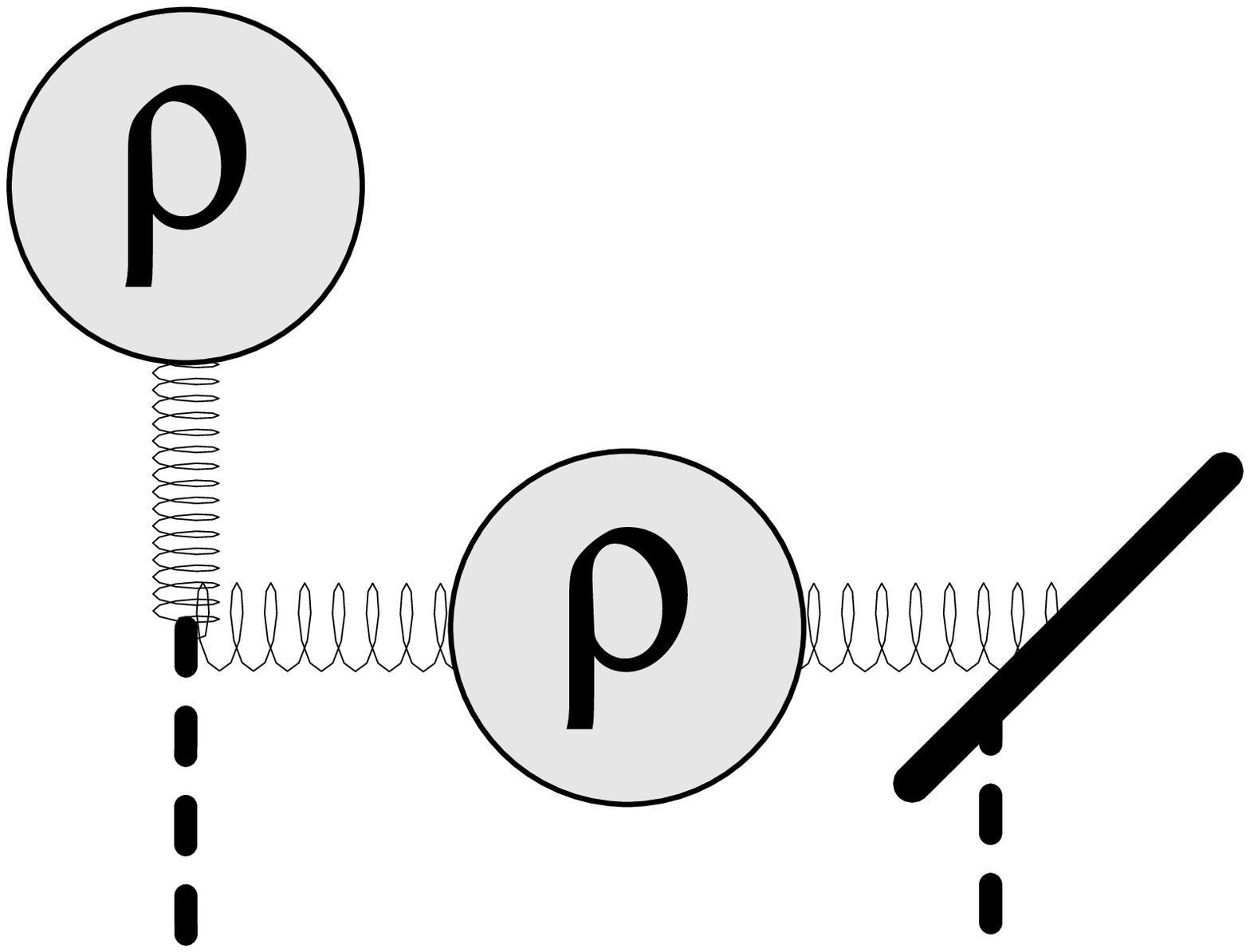,width=2.3cm}
    \epsfig{file=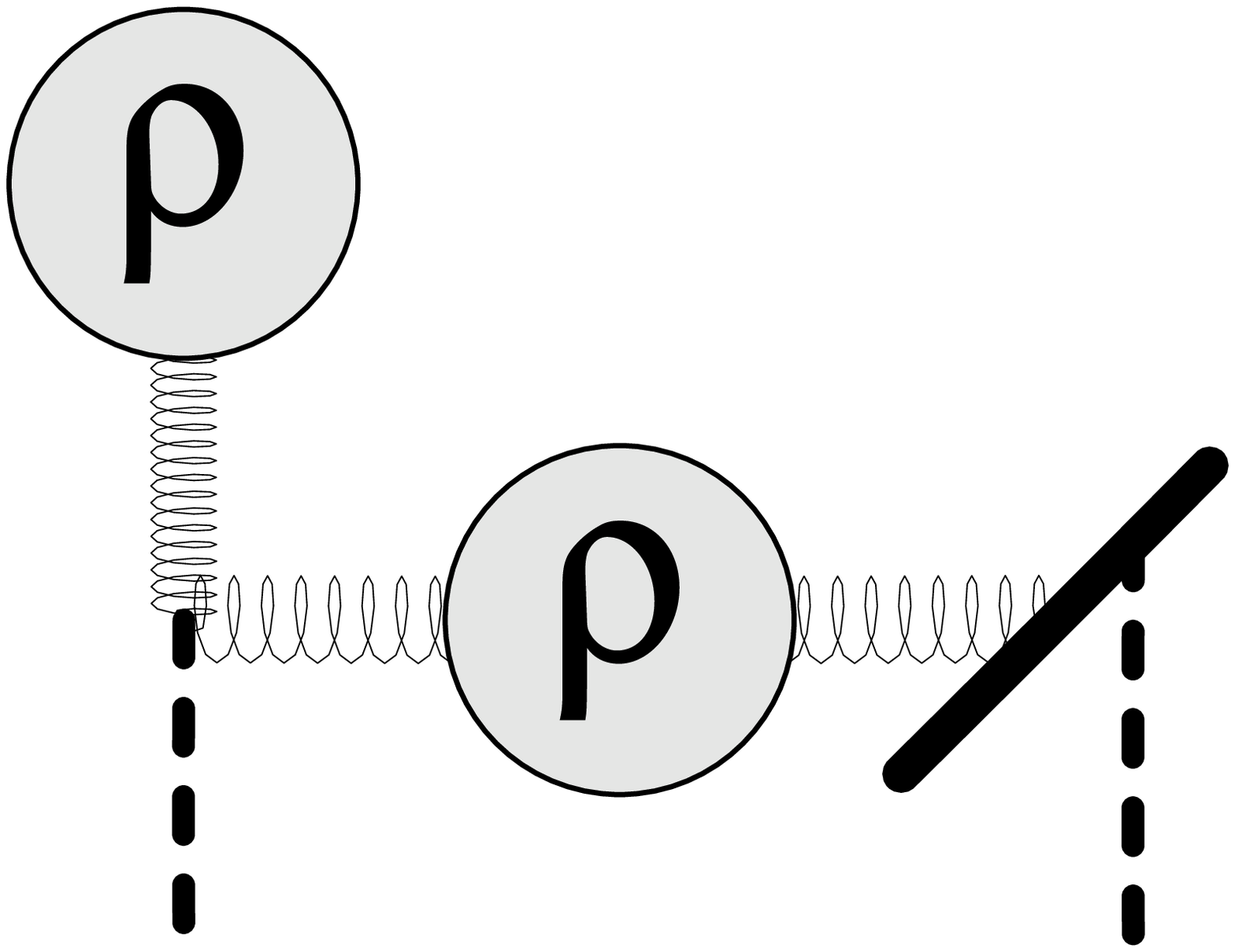,width=2.3cm}
    \epsfig{file=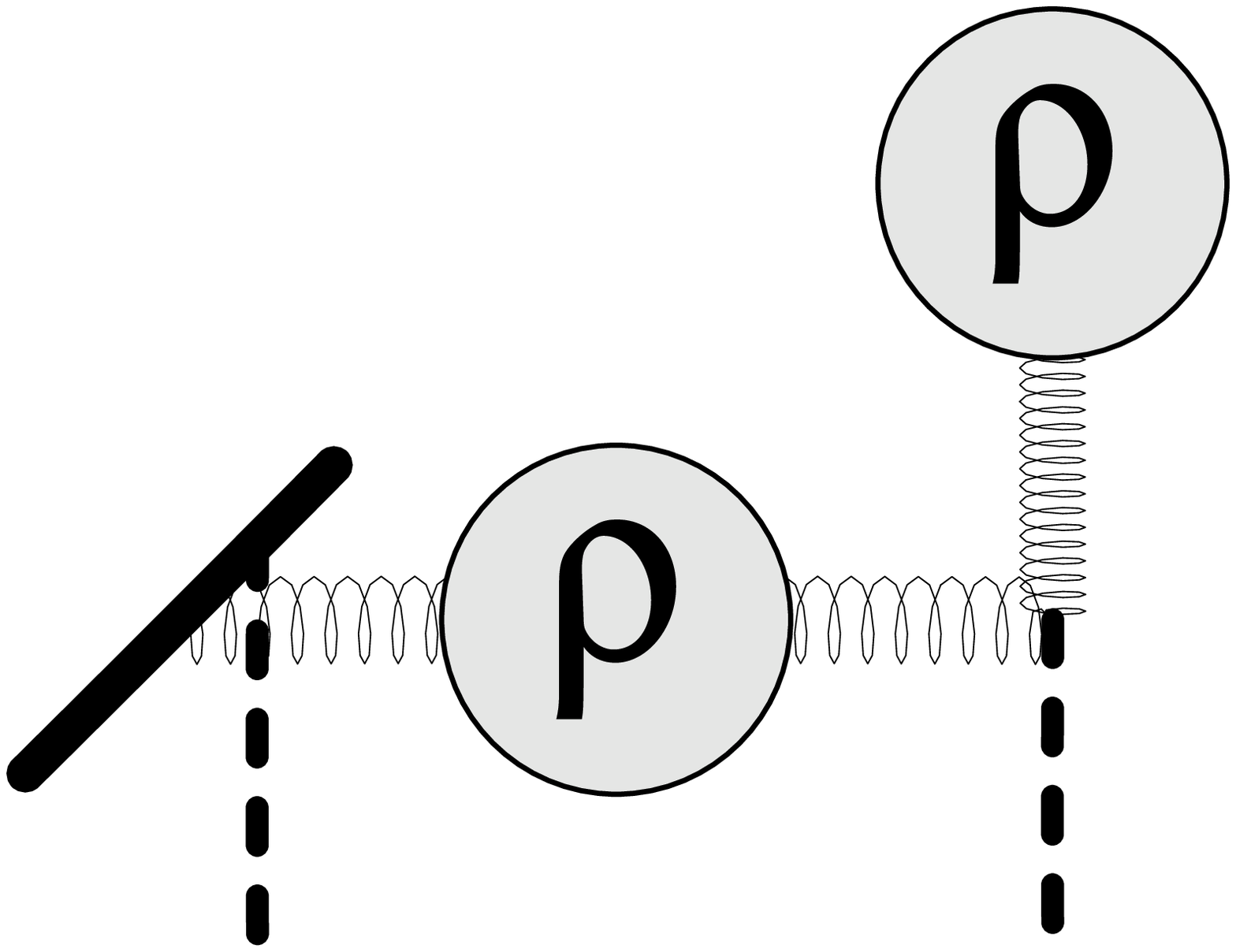,width=2.3cm}
    \epsfig{file=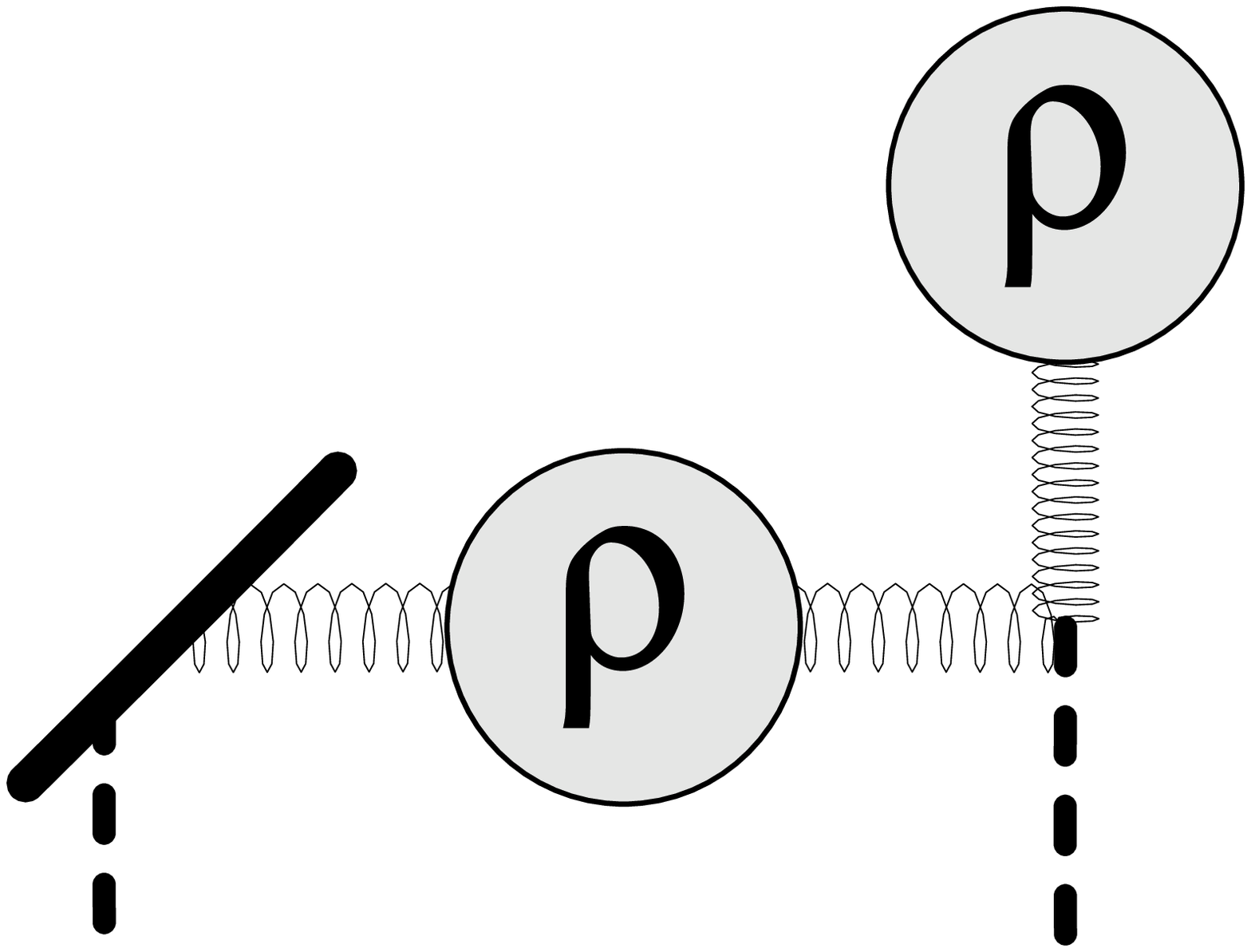,width=2.3cm}  
    \caption{Explicit list of the diagrams contributing to $\chi$}
    \label{fig:chifulllist}
  \end{center}
\end{figure}

Our goal therefore would seem to be the inversion of
$D^{-1}_{\mu\nu}$.  In fact our task is a little simpler than that,
since we only need to calculate the time independent average in
$\sigma$ and the equal time correlator in $\chi$. Those are determined
by the Wightman function of the fluctuation field $D_W$ rather than by
the Feynman propagator. The Wightman function satisfies the
homogeneous equation \begin{eqnarray} D^{-1}D_W=0 \end{eqnarray} and
is constructed from the eigenfunctions of $D^{-1}$ with the zero
eigenvalue.  Our first task is therefore to find the zero
eigenfunctions of $D^{-1}$.  For completeness we will present also the
eigenfunctions with nonzero eigenvalues, but will not construct
explicitly the Feynman propagator.

For clarity we have split our calculation into three main parts. We
determine the eigenfunctions in subsection~\ref{sec:sigmachieig}, find
their proper normalization in subsection~\ref{sec:sigmachinorm} and
use these results to express the main quantities of interest, $\chi$
and $\sigma$ in subsection~\ref{sec:sigmachiindcharge},
Eqs.(\ref{eq:finalchi}) and (\ref{eq:finalsigma})

\subsection{The eigenfunctions of $D^{-1}_{\mu\nu}$.}
\label{sec:sigmachieig}
The quadratic action for the fluctuation fields is
\begin{eqnarray} S=\frac{1}{2g^2}\Bigg\{ a^-_{x} K_{xy} a^-_{y} +2a^-(\partial^+ Da
+ 2 fa ) +2\partial^+a_i\partial^-a_i+-a_i \bigg[ D^2 \delta_{ij} +
D_{i}D_{j} \bigg] a_j\Bigg\}
\label{eq:action}
\end{eqnarray} 
Here we are using the following condensed notation 
\begin{eqnarray}
&&[fa]^{a}(x^+,x^-,x_\perp ) = f^{abc} \delta (x^-) \alpha^{ic}(x_\perp )
a^{ib}(x^+,x^-,x_\perp ) \nonumber \\ 
&&Da =
D_{i}[\alpha]a_{i}=(\partial_i\delta^{ab}+f^{abc}\theta(x^-)\alpha_i^c)a^b_i
\label{cov}
\end{eqnarray} 
and $x$ denotes the space time coordinates $x^{\pm} ,x^{i}$ as
well as color label. All repeated indices are summed (integrated)
over. The function $\alpha_i(x_\perp )$ is related to $\rho(x_\perp )$ through 
Eq.(\ref{resa}).
The operator $K$ is
\begin{eqnarray} K^{ab}_{xy} &=& - \bigg[
(\partial^{+})^2 \delta^{ab} + f^{abc}\rho^{c} \delta (x^-)
\frac{1}{\partial^-} \bigg]\nonumber\\ 
&=& - \bigg[ (\partial^{+})^2
\delta^{ab} -2 M^{ab} \delta (x^-)\bigg] 
\label{K}
\end{eqnarray} 
where we have defined \begin{eqnarray}
M^{ab}(p^-,x_\perp ) = {i \over 2p^-} f^{abc} \rho^{c}(x_\perp ) \equiv {i f\cdot
\rho(x_\perp ) \over 2 p^-}
\label{eq:matrixM}
\end{eqnarray}
Note, that $M$ is a color matrix locally defined in the transverse and 
frequency space, which does not depend on $x^-$.

For simplicity we will 
temporarily omit the factor $1/g^2$ in front of the action,
remembering to restore it in the expressions for the charge density by
appropriately scaling the small fluctuations propagator.

We have changed our notations from the previous section and denote the
fluctuation field by $a_\mu$ rather than $\delta A_\mu$.  We should
remember that the fluctuation fields contain longitudinal momenta only
above some scale $\Lambda^{+'}$. The question how exactly to impose
this cutoff is unimportant in the leading logarithmic approximation.
We find convenient to introduce it through the infrared cutoff in
coordinate space.  The longitudinal coordinate $x^-$ in our
expressions therefore varies between $-L$ and $L$. Whenever it is
harmless, we will take the limit $L\rightarrow\infty$, which
corresponds to the big cutoff ratio $\Lambda^+/\Lambda^{+'}\gg 1$.

Rather than writing down the eigenvalue equations for the quadratic
Lagrangian Eq.(\ref{eq:action}) it is more convenient first to
explicitly decouple the $a^-$ field. 
This is done by completing the square in Eq.(\ref{eq:action})
\begin{eqnarray}
S&=&\frac{1}{2}\bigg\{ \bigg[a^- + K^{-1} (\partial^+ Da + 2 fa )\bigg]_{x}
K_{xy}  \bigg[a^- + K^{-1} (\partial^+ Da + 2 fa )\bigg]_{y} \\
&-&  \bigg[\partial^+ Da + 2 fa \bigg]_{x} K^{-1}_{xy}  
\bigg[\partial^+ Da + 2 fa \bigg]_{y}
+2\partial^+a_i\partial^-a_i
- a_i \bigg[ D^2 \delta_{ij} + D_{i}D_{j} \bigg] a_j\bigg\}\nonumber
\label{eq:action1}
\end{eqnarray}
Defining
\begin{eqnarray}
\tilde a^-=a^-+ K^{-1} (\partial^+ Da + 2 fa )
\label{tildea}
\end{eqnarray}
we see that it decouples from $a_i$. Its correlator is given by
\begin{eqnarray}
<\tilde a^-_x\tilde a^-_y>=K^{-1}_{x,y}
\end{eqnarray}
The correlator of $a^-$ is then easily calculable once we know
$K^{-1}$ and the correlators of $a_i$.

The calculation of $K^{-1}$ is straightforward and is given in
Appendix A.  The result is
\begin{eqnarray}
K^{-1} =\Bigg\{ - {1 \over 2} |x^- - y^-| 
+ {1 \over 2} \eta (x_\perp )\bigg[ |x^-|
+ |y^-| \bigg] - {\eta(x_\perp )\over 2 M}\Bigg\}\delta(x_\perp ,y_\perp )
\label{eq:Kpropexpand}
\end{eqnarray}
The color matrix $\eta^{ab}(x_\perp )$ projects onto the nonzero
eigenvalue subspace of $M$. Together with the complementary projector
$\mu$ it satisfies the relations
\begin{eqnarray}
\mu M = 0,\;\;\; \eta M = M
\end{eqnarray}
and
\begin{eqnarray}
\mu + \eta = 1 ,\;\;\mu^{2} = \mu ,\;\; \eta^{2} = \eta
\label{eq:project}
\end{eqnarray}

We note, that the operator $K$ Eq.(\ref{K}) has zero modes 
of the form 
\begin{eqnarray}
f^a(x_\perp ,x^-,p^-)=\mu^{ab}f(x_\perp ,p^-)
\label{zero}
\end{eqnarray}
and is therefore strictly speaking non invertible. The result
Eq.(\ref{eq:Kpropexpand}) was obtained by excluding the zero modes and
inverting $K$ on the space of functions which does not include the
functions Eq.(\ref{zero}).  The normalizable zero modes of $K$ can not
be completely neglected in Eq.(\ref{eq:action}).  Expanding $a^-$ in
the basis of eigenfunctions of $K$
\begin{eqnarray}
a^{-} = \int d\lambda a^{-}_{\lambda} f_{\lambda}
\end{eqnarray}
we see immediately that $a^-_0$ drops out from the first term in
Eq.(\ref{eq:action}) but not from the second term.  As a result
$a^-_0$ does not decouple from $a_i$ and the Eq.(\ref{eq:action1})
should be slightly modified. In addition to the term quadratic in
$\tilde a^-$ we have
\begin{eqnarray}
S&=&
-{1 \over 2} \bigg[
\partial^+ Da + 2 fa \bigg]_{x} K^{-1}_{xy}  
\bigg[\partial^+ Da + 2 fa \bigg]_{y}
 +\partial^+a_i\partial^-a_i-{1 \over 2} a_i 
\bigg[D^2 \delta_{ij} + D_{i}D_{j} \bigg] a_j
\nonumber \\ && 
+ a^{-}_{0,x_\perp ,x^+}\mu_{x_\perp }
\int dx^-\bigg[\partial^+ Da + 2 fa \bigg]_x 
\label{eq:newaction}
\end{eqnarray}

Note that $a^-_0$ does not depend on $x^-$ since the zero mode of $K$
is constant in $x^-$.  The linear term in $a^-$ in
Eq.(\ref{eq:newaction}) is in fact nothing but the Gauss' law
constraint which remains after integrating out the $a^-$ component of
the vector potential. As we stressed before, our effective Lagrangian
is gauge invariant under the residual $x^-$ independent non-Abelian
gauge transformation. As a result, the Lagrangian expanded to second
order in the fluctuation field, Eq.(\ref{eq:action}) preserves the
linearized version of this transformation. It is in fact
straightforward to check that Eq.(\ref{eq:action}) is invariant under
\begin{eqnarray}
a_i\rightarrow a_i+D_i[\alpha]\lambda(x_\perp ,x^+), \ \ 
a^-\rightarrow a^-+\partial^-\lambda(x_\perp ,x^+)
\end{eqnarray}
with $D_i[\alpha]$ of Eq.(\ref{cov}), 
provided $\lambda(x_\perp ,x^+)\rightarrow_{x^+\rightarrow\pm\infty}0$.
The $x^-$ independent part of $a^-$ imposes
the Gauss' law constraint that corresponds to this transformation in the
Lagrangian Eq.(\ref{eq:action})
\begin{eqnarray}
\int dx^-K_{xy}a^-_y + (\partial^+ Da + 2 fa )_x=0
\label{gauss}
\end{eqnarray}
Decoupling $\tilde a^-$ is of course equivalent to integrating out $a^-$
from the path integral. This procedure solves Eq.(\ref{gauss}) for $a^-$ in
terms of $a_i$, except for the component of this equation which is
proportional to the zero mode of $K$, since this component does not 
contain $a^-$. This component of the equation is a constraint
that involves $a_i$ only and should be kept intact in the path integral
for $a_i$, Eq.(\ref{eq:newaction}). The field $a^-_0$ is just the
Lagrange multiplier that imposes this constraint.

Now that we have disposed of $a^-$ we have to find eigenfunctions and
eigenvalues of the operator $D^{-1}_{ij}[\rho]$ defined by the action
Eq.(\ref{eq:newaction}).  It is convenient to parameterize the fields
$a_{i}$ in the following way
\begin{equation}
\label{eq:newfield}
a^i  = \theta (x^-) \bigg[ a^{i}_{+}(x_\perp ,x^+,x^-) + 
\gamma^{i}_{+}(x_\perp ,x^+)\bigg]
+ \theta(-x^-) \bigg[ a^{i}_{-}(x_\perp ,,x^+,x^-) + \gamma^{i}_{-}(x_\perp ,x^+)\bigg]
\end{equation}
The reason we choose to use this parameterization is that the equations
of motion (the eigenvalue equations) as derived from Eq.(\ref{eq:newaction})
are first order in $\partial^+$ and contain coefficients of the form 
$\delta(x^-)$. We therefore expect the eigenfunctions to be discontinuous
at $x^-=0$. Also, since the classical background fields do not vanish
at $x^-\rightarrow\infty$, we should allow the same asymptotic
behavior in the fluctuations. We have
separated out for convenience the components of the field $\gamma^i_\pm$
which do not vanish as
$x^{-} \rightarrow \pm \infty$ so that by definition
\begin{eqnarray}
a^{i}_{\pm} \rightarrow_{x^{-} \rightarrow \pm \infty} 0
\end{eqnarray}
Substituting (\ref{eq:newfield}) into the 
action~(\ref{eq:newaction}) we obtain
\begin{eqnarray}
S&=& \int^{\infty}_{0}dx^- 
\bigg[ \partial^{-} a^{i}_{+} \partial^{+} a^{i}_{+} 
+ {1 \over 2} a^{i}_{+} D^{2}_{\perp} a^{i}_{+} \bigg] 
+ \int^{0}_{-\infty} d x^-\bigg[ \partial^{-} a^{i}_{-} \partial^{+} a^{i}_{-} 
+ {1 \over 2} a^{i}_{-} D^{2}_{\perp} a^{i}_{-} \bigg] 
\nonumber \\ &&
+ [ \gamma^{i}_{+} + v^{i}_{+}  ]\partial^{-} 
[ \gamma^{i}_{-} + v^{i}_{-}  ] 
+ [\partial^{-}\gamma^{i}_{-} v^{i}_{-}  - 
\partial^{-} \gamma^{i}_{+} v^{i}_{+} ]
\nonumber \\ && 
+ {1 \over 2} L \bigg[ \gamma^{i}_{+} [ D^{2}_{\perp} 
\delta^{ij} - D_{i}D_{j} ]
\gamma^{j}_{+} + \gamma^{i}_{-} [ \partial^{2}_{\perp} \delta^{ij} - 
\partial^{i}\partial^{j} ] \gamma^{j}_{-} \bigg] 
\nonumber \\ &&
+ {1 \over 2} \bigg[ \partial^{i} \gamma^{i}_{+} - D^{i} \gamma^{i}_{-} 
- \alpha^{i} [v^{i}_{+} + v^{i}_{+} \bigg] 
{\eta \over \partial \alpha } \partial^{-} 
 \bigg[ \partial^{j} \gamma^{j}_{+} - D^{j} \gamma^{j}_{-} 
- \alpha^{j} [v^{j}_{+} + v^{j}_{+} \bigg] 
\nonumber \\ && 
+ a^{-}_{0} \mu \bigg[ \partial^{i} \gamma^{i}_{+} - D^{i}\gamma^{i}_{-}
- \alpha^{i} [v^{i}_{+} + v^{i}_{-}]\bigg]
\end{eqnarray}
where 
\begin{eqnarray}
v^i_\pm=a^i_\pm(x^-=0)
\end{eqnarray}

The covariant derivative in this equation is 
\begin{eqnarray}
D_i^{ab}\equiv(\partial_i\delta^{ab}+f^{abc}\alpha_i^c)
\end{eqnarray}
We hope that the use of the same symbol as in Eq.(\ref{cov}) does not cause
confusion.

In this parameterization the linearized gauge transformation acts as 
\begin{eqnarray}
\delta \gamma^{i}_{+} = D^{i} \Lambda , \;\;\;
\delta \gamma^{i}_{-} = \partial^{i} \Lambda , \;\;\;
\delta a^{-}_{0} = \partial^{-} \Lambda
\end{eqnarray}

The equations for eigenfunctions are
\begin{eqnarray}
{\delta S \over \delta a_{i+}} &=& \theta (x^{-}) \bigg[
-2 \partial^{+}\partial^{-} + D^{2}_{\perp} \bigg]a^{i}_{+} 
+ \delta (x^{-}) \bigg[\partial^{-} [\gamma^{i}_{-} - \gamma^{i}_{+} 
- v^{i}_{+} + v^{i}_{-} ]
\nonumber \\ &&
+ \alpha^{i} {\eta \over \partial \alpha} \partial^{-} 
[ \partial \gamma_{+} - D\gamma_{-}
- \alpha (v_{+} + v_{-} )]
+ \alpha^{i} \mu a^{-}_{0}\bigg]\nonumber \\
&=& \lambda a^{i}_{+} 
\end{eqnarray}
\begin{eqnarray}
{\delta S \over \delta a_{i-}} &=& \theta (-x^{-}) 
\bigg[ -2 \partial^{+}\partial^{-} + \partial^{2}_{\perp} 
 \bigg] a^{i}_{-} + \delta (x^{-}) \bigg[\partial^{-} 
[\gamma^{i}_{-} - \gamma^{i}_{+} 
- v^{i}_{+} + v^{i}_{-}]
\nonumber \\ && 
+\ \alpha^{i} {\eta \over \partial \alpha} \partial^{-} 
[ \partial \gamma_{+} - D\gamma_{-}
- \alpha (v_{+} + v_{-})]
+\ \alpha^{i} \mu a^{-}_{0}\bigg]
\nonumber \\ &=& 
\lambda a^{i}_{-} 
\end{eqnarray}
\begin{eqnarray}
{\delta S \over \delta \gamma_{i+}} &=&
\partial^{-} [\gamma^{i}_{-} + v^{i}_{+} + v^{i}_{-} ] -
\partial^{i} \mu a^{-}_{0} 
\nonumber \\&&
-\ \partial^{i}{\eta \over \partial \alpha} \partial^{-} 
[ \partial \gamma_{+} - D\gamma_{-}
- \alpha (v_{+} + v_{-})] + 
L [D^{2}_{\perp} \delta^{ij} - D^{i}D^{j} ] \gamma^{i}_{+} \nonumber\\
&=& \lambda L \gamma^{i}_{+}
\end{eqnarray}
\begin{eqnarray}
{\delta S \over \delta \gamma_{i-}} &=&
-\partial^{-} [\gamma^{i}_{+} + v_{+} + v_{-}] +
D^{i} \mu a^{-}_{0} 
\nonumber \\ &&
+\ D^{i}{\eta \over \partial \alpha} \partial^{-} 
[ \partial \gamma_{+} - D\gamma_{-}
- \alpha (v_{+} + v_{-})] + 
L [\partial^{2}_{\perp} \delta^{ij} - \partial^{i}\partial^{j} ] 
\gamma^{j}_{-} \nonumber\\
&=& \lambda L \gamma^{i}_{-}
\end{eqnarray}
where all the derivatives are with respect to transverse coordinates
unless explicitly specified.
These equations are supplemented by the constraint
\begin{eqnarray}
\mu[ \partial \gamma_{+} - D\gamma_{-}
- \alpha (v_{+} + v_{-})]=0
\end{eqnarray}

First consider the zero eigenvalue $\lambda=0$.  Due to the gauge
symmetry, the equations for eigenfunctions have infinity of solutions
for $\lambda=0$. However, as stressed in Section~\ref{sec:pert} we
must work in a completely fixed gauge, which we have chosen as
$\partial_ia_i(x^-\rightarrow-\infty)=0$.  In the notations of this
section this means $\partial\gamma_-=0$. With this gauge fixing it is
straightforward to find the solution

\begin{eqnarray}
a^{i}_{p^-,r}& =& e^{ip^{-}x^{+}} \int d^{2}p_{t} \bigg[ \theta (-x^-)
\exp\left( i{p^{2}_{t} \over 2p^{-}}x^{-} - ip_{t}x_{t}\right)
v^{i}_{-, r}(p_{t}) 
\nonumber \\&&
+\ \theta (x^{-}) U^{\dagger} (x_\perp ) 
\exp\left( i{p^{2}_{t} \over 2p^{-}}x^{-} - ip_{t}x_{t}\right)
\left[Uv^{i}_{+,r}\right](p_\perp )
\nonumber \\ &&
+\ \theta (-x^{-}) \gamma^{i}_{-,r}(x_\perp ) + \theta (x^{-})
\gamma^{i}_{+,r}\bigg] 
\label{solut}
\end{eqnarray}
The frequency is a good quantum number since our background field
is static.
Here $r$
is the degeneracy label, which labels independent solutions with the
eigenvalue $\lambda=0$ and frequency $p^-$. In the free case it is
conventionally chosen as the transverse momentum, $\{r\}=\{p_i\}$.
The matrix $U(x_\perp )$ is the same $SU(N)$ matrix which defines
the classical field Eq.(\ref{sol1})
The auxiliary functions $\gamma^i_\pm, v^i_\pm$ are all determined in
terms of one vector function. We take this independent function as 
$v^i_-$\footnote{These expressions are valid up to terms of order $1/L$.
The omitted terms do not contribute to the leading order in $\ln1/x$.}.
Then
\begin{eqnarray}
v^{i}_{+,r}& =& \bigg[\delta^{ij} -2 D^{i}{1\over D^{2}}  D^{j} 
\bigg]
            \bigg[\delta^{jk} -2 \partial^{j} {1\over\partial^{2}}\partial^{k}
\bigg]
            v^{k}_{-,r} \\
\gamma^{i}_{+,r}& =& 2 D^{i} \bigg[{1\over\partial^{2}}\partial v_{-,r}  - 
 {1 \over D^{2}} Dv_{+,r}
\bigg] \nonumber\\
\gamma^i_{-,r}&=&0\nonumber\\
a^{-}_{0,r}& =& 2 {1 \over \partial^{2}} \partial v_{-,r} \nonumber
\end{eqnarray}

For the eigenfunctions corresponding to eigenvalues $\lambda\ne 0$ there
is no gauge invariance. Accordingly the functions vanish at 
infinity $\gamma_{\pm,\lambda\ne 0}=0$ and the solutions are 
\begin{eqnarray}
a^{i}_{\lambda ,p,r}& =& e^{ip^{-}x^{+}} \int d^{2}p_{t} \bigg[ \theta (-x^-)
\exp\left( i{\lambda + p^{2}_{t} \over 2p^{-}}x^{-} - ip_{t}x_{t}\right)
v^{i}_{-,\lambda , r}(p_{t}) 
\nonumber \\&&
+\ \theta (x^{-}) U^{\dagger} (x_\perp ) 
\exp\left( i{\lambda + p^{2}_{t} \over 2p^{-}}x^{-} - ip_{t}x_{t}\right)
\left(Uv^{i}_{+,r,\lambda}(p_\perp )\right)\bigg] 
\end{eqnarray}
with 
\begin{eqnarray}
v^{i}_{+}= [v^{i}_{-} - \alpha^{i}{\eta \over 
\partial\alpha}\alpha
(v_{+} + v_{-})]
\label{nonzero}
\end{eqnarray}
We now have to construct a complete set of solutions, which is
tantamount to picking $v^i_{\lambda,r}(x_\perp )$ for every eigenvalue
$\lambda$ as a complete basis of functions on the plane. This basis
should be chosen such that the solutions Eq.(\ref{solut}) are properly
normalized and are orthogonal for different values of $r$'s.

\subsection{The normalization of the eigenfunctions.}
\label{sec:sigmachinorm}
The orthonormality condition for the eigenfunctions
$a^i_{\lambda,p^-,r}$ is\footnote{One could ask whether the presence
  of the Lagrange multiplier $a^-_0$ in the Lagrangian can modify the
  normalization condition for the eigenfunctions.  It is shown in
  Appendix B that this is not the case, and the appropriate
  normalization condition is indeed Eq.(\ref{eq:ortho}).}.
\begin{eqnarray}
\int d^4x a^{ia}_{\lambda, p^{-}, r} (x) 
a^{ia}_{\lambda^{\prime}, p^{-\prime}, r^{\prime}} (x) = 
\delta (\lambda - \lambda^{\prime}) \delta (p^{-}-p^{-\prime}) 
\delta (r-r^{\prime})
\label{eq:ortho}
\end{eqnarray}
Although for the purpose of our calculation we only need 
eigenfunctions with the eigenvalue $\lambda=0$, it is convenient
to consider the orthonormality relation for arbitrary $\lambda$.
The reason is that if we take $\lambda=\lambda'$, the factor 
$\delta(\lambda-\lambda')$ gives a divergent constant, and it is difficult
to determine the numerical coefficient in front of it. Taking $\lambda$ and
$\lambda'$ generic, we can explicitly extract the $\delta$ - function factor
and determine the coefficient.

Let us consider the 
scalar product
\begin{eqnarray}
\int\! d^4x\  a^{ia}_{\lambda, p^{-}, r} (x) 
a^{ia}_{\lambda^{\prime}, p^{-\prime}, r^{\prime}} (x)& =& 
\delta (p^{-}-p^{-\prime}) \bigg[ i{2p^- \over \lambda - \lambda^{\prime} 
+ i\epsilon} M_{-} -  i{2p^- \over \lambda - \lambda^{\prime} 
- i\epsilon} M_{+}
\nonumber \\&& +\ L(N_{-} + N_{+})\bigg]
\end{eqnarray}
where  
\begin{eqnarray}
M_{-}&=&\int d^2 x_\perp  v^{ia}_{-,\lambda ,r}(x_\perp )
v^{ia\ast}_{-,r^{\prime},
\lambda^{\prime}}(x_\perp )\nonumber \\
M_{+}&=& \int d^2 x_\perp  v^{ia}_{+,\lambda ,r}(x_\perp )
v^{ia\ast}_{+,r^{\prime},
\lambda^{\prime}}(x_\perp )\nonumber \\
N_{-}&=&\int d^2 x_\perp  \gamma^{ia}_{-,\lambda ,r}(x_\perp ) 
\gamma^{ia\ast}_{-,r^{\prime},\lambda^{\prime}}(x_\perp )\nonumber \\
N_{+}&=&\int d^2 x_\perp  \gamma^{ia}_{+,\lambda ,r}(x_\perp ) 
\gamma^{ia\ast}_{+,r^{\prime},\lambda^{\prime}}(x_\perp )
\end{eqnarray}
here again we have kept the terms of order $L^{0}$ and 
dropped the cross terms since
\begin{eqnarray}
\int dx^- a^{i}\gamma^{i} \sim 1/L
\end{eqnarray}
and this can be ignored at large $L$.

Consider first the case when both eigenvalues $\lambda$ and 
$\lambda^{\prime}$ are non-zero.
It follows from Eq.(\ref{nonzero})
\begin{eqnarray}
Sv^{i}_{+} = S^{\dagger}v^{i}_{-}
\end{eqnarray}
where $S$ is the operator
\begin{eqnarray}
S = \delta^{ij} + \alpha^{i}{\eta \over \partial \alpha} \alpha^{j} 
\end{eqnarray}
This operator satisfies, 
\begin{eqnarray}
[S, S^\dagger]=0
\end{eqnarray}
and therefore $S^{-1}S^\dagger$ is unitary,
so that
\begin{eqnarray}
v_{+}^{\ast}v_{+} = v_{-}^{\ast}v_{-}
\end{eqnarray}
and $M_{+}=M_{-}$. 
The orthonormality 
condition~(\ref{eq:ortho}) then becomes
\begin{eqnarray}
\int\! d^4x\ a^{ia}_{\lambda, p^{-}, r_{j}} (x) 
a^{ia}_{\lambda^{\prime}, p^{-\prime}, r^{\prime}_{j}} (x) = 
2\pi\delta (\lambda - \lambda^{\prime}) \delta (p^{-}-p^{-\prime}) 
2|p^-| M_{-}(r,r^{\prime})
\label{normno}
\end{eqnarray}

It is clear now that for $\lambda \neq 0$, we can take our
orthonormalized basis to be
\begin{eqnarray}
[v^{bj}]_{-,r}^{ai}(x_\perp ) = {1 \over \sqrt{4\pi p^-} } \delta^{ab} \delta^{ij}
e^{ir_\perp x_\perp }\;\;\;\;\;\;\;\; \lambda \neq 0
\label{eq:nornonzero}
\end{eqnarray}

For $\lambda = 0$, there will also be a non-vanishing contribution
from the term which involves $\gamma_+$. 
The $x^-$ integral in the normalization condition Eq.(\ref{eq:ortho})
gives a factor of $L$. Comparing it with
Eq.(\ref{normno}) at $\lambda=\lambda'$ we identify this factor
as $4\pi |p^-| \delta(\lambda - \lambda^{\prime})$. 
We are then left
with 
\begin{eqnarray}
M_{-} + M_{+} + N_{-} + N_{+} ={1\over 4\pi|p^-|} \delta(r-r^{\prime})
\end{eqnarray}
It is easy to see that for the zero modes the
relation between $v_-$ and $v_+$ is also unitary.
Using $M_{-}=M_{+}$ and the explicit expressions for $\gamma_+$
we get
\begin{eqnarray}
{1\over 4\pi |p^-|}\delta(r-r^{\prime}) =
\int d^2x_\perp  d^2y_\perp  v^{\ast i}_{-r}(x_\perp ) 
O^{ij}(x_\perp,y_\perp)v^{j}_{-,r^{\prime}}(y_\perp) 
\end{eqnarray}
with
\begin{eqnarray}
O^{ij}(x_\perp,y_\perp)
&  = &  \left[\delta^{ik}-2\partial^i\frac{1}{\partial^2}\partial^k\right] 
 \nonumber \\ && \hskip .1cm \times
\left[\delta^{kl}+
2[(\delta^{kn}-\partial^k\frac{1}{\partial^2}D^n)
(\delta^{nl}-D^n\frac{1}{\partial^2}\partial^l)-
(\delta^{kl}-D^k\frac{1}{D^2}D^l)]\right]
 \nonumber \\ && \hskip .1cm \times
\left[\delta^{lj}-2\partial^l\frac{1}{\partial^2}\partial^j\right](x_\perp,y_\perp)
\label{eq:opo}
\end{eqnarray}

This relation means that for proper normalization
one should choose the functions $v^i_{-,r}$ as
eigenfunctions of the operator $O$ 
The normalization of $v^{i}_{-,r}$ should not be one, but rather
$1/\sqrt{k}$ where $k$ is the appropriate eigenvalue of the operator
$O$. The degeneracy label $r$ therefore numbers the vectors of this
particular basis. Since $O$ is a Hermitian operator, its
eigenfunctions form a complete basis, and therefore the basis of our
eigenfunctions is also complete.  Therefore we have
\begin{eqnarray}
\int d^2 r_\perp  v^{ia}_{-r}(x_\perp ) v^{\ast jb}_{-r}(y_\perp ) = 
{1\over 4\pi |p^-|}[O^{-1}]^{ab}_{ij} (x_\perp ,y_\perp )
\label{ort} 
\end{eqnarray}
All of our results then will be expressed in terms of $O^{-1}$ where
\begin{eqnarray}
\lefteqn{
[O^{-1}]^{ij}
} \\ \nonumber & = &\left[\delta^{ik} 
- 2{\partial^{i}\partial^{k} \over \partial^{2}}\right]
\left[1+
2[(1-\partial\frac{1}{\partial^2}D)
(1-D\frac{1}{\partial^2}\partial)-
(1-D\frac{1}{D^2}D)]\right]^{-1kl}
\left[\delta^{lj} - 2{\partial^l
\partial^j \over \partial^2 }\right]
\label{eq:inverseo}
\end{eqnarray}

\subsection{The induced charge density.}
\label{sec:sigmachiindcharge}
We are now ready to calculate the induced charge density $\delta
\rho$.  As was mentioned in the beginning of this section, since we
are interested in the equal time correlations of the fluctuation
fields $a_i$ we will only need the on shell propagators, and therefore
only eigenfunctions for the eigenvalue $\lambda=0$.  To see this
explicitly let us consider a typical expression we have to evaluate in
order to calculate the charge density correlator
\begin{eqnarray}
\lefteqn{<a^i(x_\perp ,x^-=0)a^j(y_\perp , x^-=0)>} 
\\ & = & 
\int \! d x^-{d\lambda\over\lambda+i\epsilon}
dp^-d^2k_\perp d^2p_\perp  
\ e^{i{2\lambda+ p_\perp ^2+ k_\perp ^2\over 2p^-} x^-}
F(k_\perp ,p_\perp )
v^i_{-,\lambda,r}(p_\perp )v^{j*}_{-,\lambda,r}(k_\perp )\delta(x^-)
\nonumber
\end{eqnarray}
We have regulated the $\lambda$ dependence of the integrand by moving
slightly away from $x^-=0$. At nonzero $x^-$ we can close the
integration contour in the $\lambda$ plane. At every fixed value of
$p^-$ the contour can be closed either
in the upper or in the lower halfplane, depending on the sign of
$x^-$. The only contribution to the integral comes from the pole at
$\lambda=0$.  If the contour is closed upstairs the integral vanishes,
while if it is closed downstairs there is a contribution $2\pi i$.
Therefore for every $p^-$ the $\lambda$
integral gives a factor $2\pi i\theta(x^-)$ or $2\pi i\theta(-x^-)$,
depending on the sign of $p^-$.  Multiplying by
$\delta(x^-)$ and integrating over $x^-$ gives the factor $1/2$ in
either case.  The result of the $\lambda$ integral is therefore that
it puts the propagator on shell ($\lambda=0$) and gives the numerical
factor $\pi i$.

In the following formulae, $\lambda=0$ is therefore assumed.

First, we calculate $a^-$. In fact as is obvious from the explicit 
expressions for the charge density, Eqs.(\ref{rho11}) and (\ref{rho21}) 
we need only $a^-(x^-=0)$.
Also, as can be easily checked $\tilde a^-$ does not contribute to the
order $\ln 1/x$, and we omit it in the following. Then, using
Eqs.(\ref{solut}), 
(\ref{tildea}) we find
\begin{eqnarray}
a^{-}(x^-=0) &=& - K^{-1}(0,y^-) \bigg[\partial^+ Da + 2 fa \bigg](y^-)\\
&=&ip^- e^{ip^- x^+} \bigg[ {Dv_{+,p^-,r} (x_\perp ) \over D^{2}} + 
{\partial v_{-,p^-,r}(x_\perp ) \over \partial^{2}}\bigg]f_{p^-,r}\nonumber
\end{eqnarray}
the integration over $p^-$ and $r$ is implied in this equation.
The objects $f_{p^-,r}$ are the coefficients in the expansion of the
fields $a_i$ in the basis of the eigenfunctions of the operator $D^{-1}_{ij}$
\begin{eqnarray}
a_i(x)=\int\! d\lambda dp^- dr\ a_{i, \lambda, p^-,r}(x)f_{\lambda,p^-,r}
\end{eqnarray}
Since our eigenfunctions are properly normalized, $f$'s
have standard correlator
\begin{eqnarray}
<f_{\lambda,p^-,r}f^*_{\lambda',p^{-'},r'}>={i\over \lambda+i\epsilon}\delta(\lambda-\lambda')
\delta(p^--p^{-\prime})\delta(r-r')
\end{eqnarray}

Now, using the Eqs.(\ref{rho11}),(\ref{rho21}) and (\ref{solut})
we find
\begin{equation}
\label{eq:totalreal}
\delta \rho^{a}_{1} = -ge^{ip^-x^+}{1 \over 2} f^{abc}\bigg\{2\alpha^{b} 
[v_{+,p^-,r} + v_{-,p^-,r}]^{c} - 2\rho^{b}
\bigg[{Dv_{+,p^-,r} \over D^{2}_{t}} + {\partial v_{-,p^-,r} \over 
\partial^{2}_{t}}
\bigg]^{c}\bigg\}f_{p^-,r}
\end{equation}
and
\begin{eqnarray}
\label{eq:totalvir}
\lefteqn{\delta \rho^{a}_{2} =
-g^2\bigg\{f^{abc}\bigg[v^{b}_{+,p^-,r} 
v^{*c}_{-,p^{-\prime},r'} +
{1 \over 2} \gamma^{b}_{+,p^-,r} [v^*_{+,p^{-\prime},r'} + 
v^*_{-,p^{-\prime},r'} ]^{c}
}
\\ && \hspace{6cm}
 - {1 \over 2}
[v_{+,p^-,r} + v_{-,p^-,r} ]^{b}\gamma^{*c}_{+,p^{-\prime},r'} \bigg]
\nonumber \\&&
+\ {1\over N_c} f^{ace}f^{bde} \rho^{b}(x_\perp )
\bigg[{Dv_{+,p^-,r} \over D^{2}_{t}} + {\partial v_{-,p^-,r} \over \partial^{2}_{t}}
\bigg]^{c}{(x_\perp )} \bigg[{Dv^*_{+,p^{-\prime},r'} \over D^{2}_{t}} + 
{\partial v^*_{-,p^{-\prime},r'} 
\over \partial^{2}_{t}}\bigg]^{d}{(x_\perp )}
\nonumber \\  &&
-\ f^{abc}\bigg[
\int^{0}_{-\infty} dx^- 
[\partial^{+} a^{b}_{-,p^-,r}]a^{*c}_{-,p^{-\prime},r'} + \int^{\infty}_{0} dx^- 
[\partial^{+}a^{b}_{+,p^-,r}]a^{*c}_{+,p^{-\prime},r'} \bigg] 
\bigg\}
\nonumber \\ && \times e^{i(p^--p^{-\prime})x^+}f_{p^-,r}f^*_{p^{-\prime},r'}\nonumber
\end{eqnarray}

Now it is straightforward to evaluate $\sigma$ and $\chi$. Since $\delta\rho_1$
is linear in $a_i$, clearly $<\delta\rho_1>=0$. Also, $\delta\rho_1$ is
order $g$ while $\delta\rho_2$ is order $g^2$. Therefore 
to order $\alpha_s$ only $\delta\rho_1$
contributes to $\chi$, and only $\delta\rho_2$ contributes to $\sigma$.
The frequency integral is trivial. The
only $p^-$ dependence is in the normalization factor  $1/|p^-|$, 
Eq.(\ref{ort}). The integral over $p^-$ then gives the logarithmic factor
which we identify with $\ln 1/x$.

Using the explicit expressions for $v_{+}$ and $\gamma_{+}$ in terms of
$v_{-}$ and the normalization condition Eq.(\ref{ort}) we obtain
\begin{eqnarray}
\label{eq:finalchi}
\lefteqn{\chi^{ad}(x_\perp ,y_\perp ) = f^{abc}f^{def}}
\\&\times &
\bigg\{ \alpha^{i}_{b}\bigg[(T^{ij}-L^{ij})
(t^{jk}-l^{jk}) + \delta^{ik}\bigg]^{cg}+ 
\rho^{b}\bigg[
{D^i \over D^2_{t}}
(t^{ik} - l^{ik}) - {\partial^{k} \over \partial^2_\perp }\bigg]^{cg}
\bigg\}_{(x_\perp ,z_\perp )}
\nonumber\\
&\times &
\bigg[O^{-1}(z_\perp ,\bar z_\perp )\bigg]^{kn}_{gh}
\nonumber\\
&\times&
\bigg\{ \alpha^{l}_{e} \bigg[ (T^{ls}-L^{ls})
(t^{sn}-l^{sn}) + \delta^{ln}\bigg]^{fh}+
\rho^{e}\bigg[{D^l \over D^2_{t}}
(t^{ln} - l^{ln}) - {\partial^{n} \over \partial^2_\perp }\bigg]^{fh}
\bigg\}_{(y_\perp,\bar z_\perp )}\nonumber
\end{eqnarray}
and
\begin{eqnarray}
\label{eq:finalsigma}
\lefteqn{\sigma^{a}(x_\perp ) = -f^{abc}\bigg[(T^{ij} - L^{ij})(t^{jk} - l^{jk})
\bigg]^{bd}_{(x_\perp ,y_\perp )}
\bigg[O^{-1}(y_\perp ,x_\perp )\bigg]^{ki}_{dc}}
\nonumber\\ &&
+2 f^{abc}\bigg[(T^{ij} - L^{ij})(t^{jk} - l^{jk})+ \delta^{ik}
\bigg]^{bd}_{(x_\perp ,y_\perp )}
\nonumber \\ && \hspace{1cm}\times
\bigg[O^{-1}(y_\perp ,\bar y_\perp )\bigg]^{kn}_{de}
D^{i}_{cf}\bigg[{\partial^n 
\over \partial^2_\perp } + {D^m \over D^2_{t}}(t^{mn} - l^{mn})
\bigg]^{fe}_{(x_\perp,\bar y_\perp )}\nonumber\\
&&-{ 1\over N_c} f^{ace}f^{bde}\rho^b(x_\perp) \bigg[ {D^i \over D^2_{t}}
(t^{ij} - l^{ij}) - {\partial^{j} \over \partial^2_\perp }\bigg]^{cg}_{(x_\perp ,y_\perp )}
\nonumber \\ && \hspace{1cm}\times
\bigg[O^{-1}(y_\perp ,\bar y_\perp)\bigg]^{jl}_{gh}
\bigg[{D^k \over D^2_{t}}
(t^{kl} - l^{kl}) - {\partial^{l} \over \partial^2_\perp }
\bigg]^{dh}_{(x_\perp ,\bar y_\perp )}\nonumber\\
&&- f^{abc}R^{bc}(x_\perp )
\end{eqnarray}
where $R^{bc}(x_\perp )$ is 
\begin{eqnarray}
R^{bc}(x_\perp )= \int {d^2p_\perp  \over (2\pi)^2}{d^2q_\perp  \over (2\pi)^2}
p^{2}_{t} e^{-i(p_\perp  -q_\perp )x_\perp } \bigg[
{v^{ib}_{-}(p_\perp )v^{\ast ic}_{-}(q_\perp ) \over p^2_\perp  - q^2_\perp  -i\epsilon } - 
{v^{ib}_{+}(p_\perp )v^{\ast ic}_{+}(q_\perp ) \over p^2_\perp  - q^2_\perp  +i\epsilon }\bigg]
\end{eqnarray}
and we have defined the projection operators
\begin{eqnarray}
&&T^{ij}\equiv \delta^{ij} - {D^i D^j \over D_\perp ^2},\ \ \ \ \ \ \
L^{ij}\equiv  {D^i D^j \over D_\perp ^2}\nonumber\\
&&t^{ij}\equiv  \delta^{ij} - {\partial^i \partial^j \over \partial_\perp ^2},\
 \ \ \ \ \ \ 
l^{ij}\equiv  {\partial^i \partial^j \over \partial_\perp ^2}
\label{eq:tproj}
\end{eqnarray}
These expressions can be somewhat simplified. The inversion of $O$ can
be performed explicitly as far as the transverse index structure is
concerned.
\begin{eqnarray}
&&[(t-l)O^{-1}(t-l)]^{ab}_{ij}(x_\perp,y_\perp)=\\
&&<x_\perp|\delta^{ab}_{ij}-2\bigg[[\partial^i{1\over\partial^2}-D^i{1\over
D^2}]
S^{-1}[{1\over\partial^2}\partial^j-{1\over
D^2}D^j]\bigg]^{ab}|y_\perp>
\nonumber
\end{eqnarray}
Here $<x_\perp|...|y_\perp>$ denotes the configuration space matrix
element in a standard way.
We also find it simpler to use the matrix notation
\begin{equation}
\alpha_i^{ab}=f^{abc}\alpha_i^c,\ \ \ \ \rho^{ab}=f^{abc}\rho^c
\end{equation}
The rotational scalar operator $S$ is defined as 
\begin{eqnarray}
S&=&{1\over D^2}+2({D_i\over D^2}-{\partial_i\over\partial^2})
({D_i\over D^2}-{\partial_i\over\partial^2})\\
&=&{1\over D^2}-2{1\over \partial^2}\partial\alpha{1\over D^2}
+2{1\over D^2}D\alpha{1\over\partial^2}\nonumber
\end{eqnarray}

In terms of this operator we have
\begin{eqnarray}
\chi^{ab}(x_\perp,y_\perp)&=&-<x_\perp|\bigg
[2\alpha_i-\{\alpha,\partial\}{\partial_i\over\partial^2}-
\{\alpha,D\}{D_i\over D^2}\bigg]\nonumber \\
&\times&
\bigg[\delta_{ij}-2({\partial_i\over\partial^2}-{D_i\over D^2})
S^{-1}({\partial_j\over\partial^2}-{D_j\over D^2})\bigg]\nonumber \\
&\times&
\bigg[2\alpha_j-{\partial_j\over \partial^2}
\{\alpha,\partial\}-{D_j\over
D^2}\{\alpha,D\}\bigg]^{ab}|y_\perp>\nonumber\\
\sigma^a(x_\perp)&=&f^{abc}\bigg\{<x_\perp|-4{D_i\over
D^2}(D\partial){\partial_i\over \partial^2}\\
&-&2\bigg[{\partial_i\over \partial^2}+{D_i\over D^2}
-2{D_i\over D^2}(D\partial){1\over \partial^2}\bigg]
S^{-1}
\bigg[{\partial_i\over \partial^2}+{D_i\over D^2}
-2{1\over D^2}(D\partial){\partial_i\over \partial^2}\bigg]\nonumber \\
&+&4\bigg[{D_i\over D^2}(D\partial){1\over \partial^2}-{\partial_i\over
\partial^2}(\partial D){1\over D^2}\bigg]S^{-1}{D_i\over
D^2}|x_\perp>\nonumber \\
&+&{1\over N_c}<x_\perp|\bigg({D_i\over D^2}+
{\partial_i\over \partial^2}\bigg)
({D_i\over D^2}+{\partial_i\over \partial^2})\nonumber \\
&-&2\bigg({D_i\over D^2}+{\partial_i\over \partial^2}\bigg)
\bigg({D_i\over D^2}-{\partial_i\over \partial^2}\bigg)
S^{-1}\bigg({D_j\over D^2}-{\partial_j\over \partial^2}\bigg)
\bigg({D_j\over D^2}+{\partial_j\over \partial^2}\bigg)|x_\perp>
\rho(x_\perp) \nonumber\\
&-&R(x_\perp)\bigg\}^{bc}\nonumber
\label{eq:final}
\end{eqnarray}
Here $\{X,Y\}$ denotes anticommutator.

Equations (\ref{eq:finalsigma}), (\ref{eq:finalchi}) and
(\ref{eq:final})
are the
central result of this paper.  Those are expressions for the
coefficient functions of the renormalization group equations as
functions of the charge density $\rho$. These expressions do not look
very simple and clearly one will need to develop some intuition and
deeper understanding to be able to use them in either analytic or
numeric calculations.  This is the matter of future work. For now we
are content with being able to derive these explicit expressions.

As an important cross check on our results, we have checked that in
the weak field limit, expanding the renormalization group equation
Eq.(\ref{finalrg}) to leading order in the charge density $\rho$ we
recover the BFKL equation. The details of this calculation are given
in Appendix C.

\section{Discussion}\label{sec:disc}

The main result of this paper is the calculation of the one - and two
- point functions of the charge density induced by the gluonic
fluctuations.  This completes in the formal sense the derivation of
the renormalization group equation that describes the flow of gluonic
observables at low $x$ according to the ideology of
\cite{JKMW,linear,nonlinear}.  We want to point out that in fact, the
flow is described not by one RG equation, as in a theory with one
running coupling constant and not even by a finite set of equations,
as in a theory with finite number of relevant operators, but rather by
a functional equation Eq.(\ref{finalrg}). The functional equation is
equivalent, of course to an infinite number of ordinary equations.
This can be interpreted as indicating that the low $x$ RG flow has an
infinite number of relevant operators.  This is a rare example of the
renormalization group flow in an infinite dimensional space of
relevant couplings with all "$\beta$ - functions" calculable
explicitly.

Much work remains to be done to understand the physics of the full
nonlinear evolution equation. It is probably wise first to see whether
one recovers the simpler known equations as its particular limits. As
we have mentioned, we have checked explicitly (see \cite{linear} and
Appendix C) that in the leading order expansion in powers of the
charge density our equation reduces to BFKL equation.  The DLA limit
of the DGLAP evolution is also obtained if we expand to leading order
in $\rho$, impose the transverse momentum ordering in the rungs of the
ladder in the real diagrams of Fig.\ref{fig:chifulllist} and neglect
the virtual contributions of Fig.\ref{fig:sigma}.  With a little more
work one should be able to recover the GLR equation \cite{GLR},
\cite{MQ}.  To this end one has to expand our result to the next to
leading order in the charge density $\rho$ and impose the DLA
kinematics. Without imposing the DLA transverse momentum ordering the
next to leading order expansion should reproduce the triple pomeron
vertex.

These are important consistency checks on our calculation and they
should certainly be performed. In the framework of the full nonlinear
problem, there are two very interesting questions which can be asked
immediately. First, does the flow described by Eqs.(\ref{finalrg}),
(\ref{eq:finalsigma}) and (\ref{eq:finalchi}) have a fixed point. The
presence of such a fixed point would mean in our framework the
saturation of gluonic observables at low $x$.  Needless to say, if
such a fixed point exists and the fixed point value of $F[\rho]$ can
be determined, it would be extremely interesting.  It would describe a
universal behavior of DIS observables at low $x$, independent of the
hadron that is being considered.  The statistical weight
$e^{-F[\rho]}$ defines a two dimensional Euclidean field theory.  It
is interesting to note, that the evolution equation itself does not
contain any scale. Therefore, if the fixed point exist it could
be scale invariant, and in that case very likely also conformally
invariant\footnote{We thank M. W\"usthoff for this observation.}. It
may be therefore possible to study it with the methods of two
dimensional conformal field theory.

Second, even if the fixed point does not exist it would be interesting
to investigate what is the impact of the low $x$ evolution on
observables at small transverse momentum.  Starting from some
reasonable initial condition at $x_0$ and evolving $F$ to low enough
values of $x$, one could study the low $k_\perp $ behavior of the
resulting two dimensional theory.  Again, it could be that at low
$k_\perp $ the two dimensional theory becomes conformal and can be
analyzed analytically.

Another outstanding question, is how to generalize this approach to
include not just DIS but also hadron - hadron collisions.  An
effective Lagrangian for the hadron collision in multi Regge
kinematics was derived by Lipatov \cite{Li}. It would be worthwhile to
extend the renormalization group approach to this case.  We note that
the leading order of the perturbative calculation for two hadron
collision was considered in \cite{twohad} to first order in $\rho$ and
numerical work is in progress to understand the nonlinear effects to
leading order in $\alpha_s$ \cite{KV}.

We want to conclude with a discussion of the physical picture of the
difference between the BFKL limit and the low $x$ DLA DGLAP limit as
it emerges from our approach. Although this does not have direct
relation to the nonlinear problem considered in this paper, the
observation can hopefully help to put our approach
in a more general perspective.

It is interesting to interpret our calculational procedure from the
perspective of the Born - Oppenheimer approximation. The Born -
Oppenheimer approximation is standard in systems that have two
distinct time scales. One considers the slow degree of freedom $X$ as
a static background and solves the dynamics of the fast degree of
freedom $Y$ with given background $X$. Integrating out $Y$ generates a
change in the Lagrangian for $X$. This is of course the standard
procedure for deriving effective Lagrangians in theories which contain
well separated fast and slow degrees of freedom. An example is the
chiral Lagrangian, where light pions are slow, and heavy $\rho$,
$\phi$ etc.  mesons are fast.  From this point of view, in our case we
would like to think of the partons with large longitudinal momentum
$p^+>\Lambda^+$ as the slow modes, since their frequency
$p^-\sim{1\over p^+}$ is small\footnote{Note that those are the modes
  that we called ``fast'' in Section~\ref{sec:effact}.  We hope this
  does not cause confusion. Indeed these modes are ``fast'' if we
  consider their variation in time $t$, since their energy is large.
  However, in light cone time $x^+=z+t$ these same modes are almost
  static, since their light cone time dependence is given by
  $\exp\{ip^-x^+\}$ and $p^-$ is small. In this section we are
  interested in the light cone time variation.}. The charge density
$\rho$ due to these partons is considered to be static while
integrating out the fluctuation fields $\delta A$ with momenta
$\Lambda^+>k^+>\Lambda^{'+}$.  However as stressed before, our system
does not have two sharply separated time scales, but rather a
continuum of time scales. We are integrating over $\delta A$ in order
to derive the Lagrangian for the soft fields $a$.  Those contain light
cone frequencies $l^-\sim{1\over l^+}\gg k^-$. The fluctuation fields
$\delta A$ would appear as fast modes relative to $\rho$ but as slow
relative to soft fields $a$.  The situation in fact is slightly more
complicated, since in principle the field modes $\delta A$ contain
also vastly different transverse momenta. For $k_\perp ^2\ge p_\perp
^2$ they are in fact faster than $\rho$, however for very small
transverse momenta $k_\perp ^2\ll p^2_\perp $ the frequency of the
$\delta A$ field can be as small or even smaller than $p^-$.  The
splitting in terms of longitudinal momenta $k^+$ therefore does not
exactly correspond to the Born - Oppenheimer type splitting in terms
of th?e frequency.

\begin{figure}[htbp]
  \begin{center}
       \epsfig{file=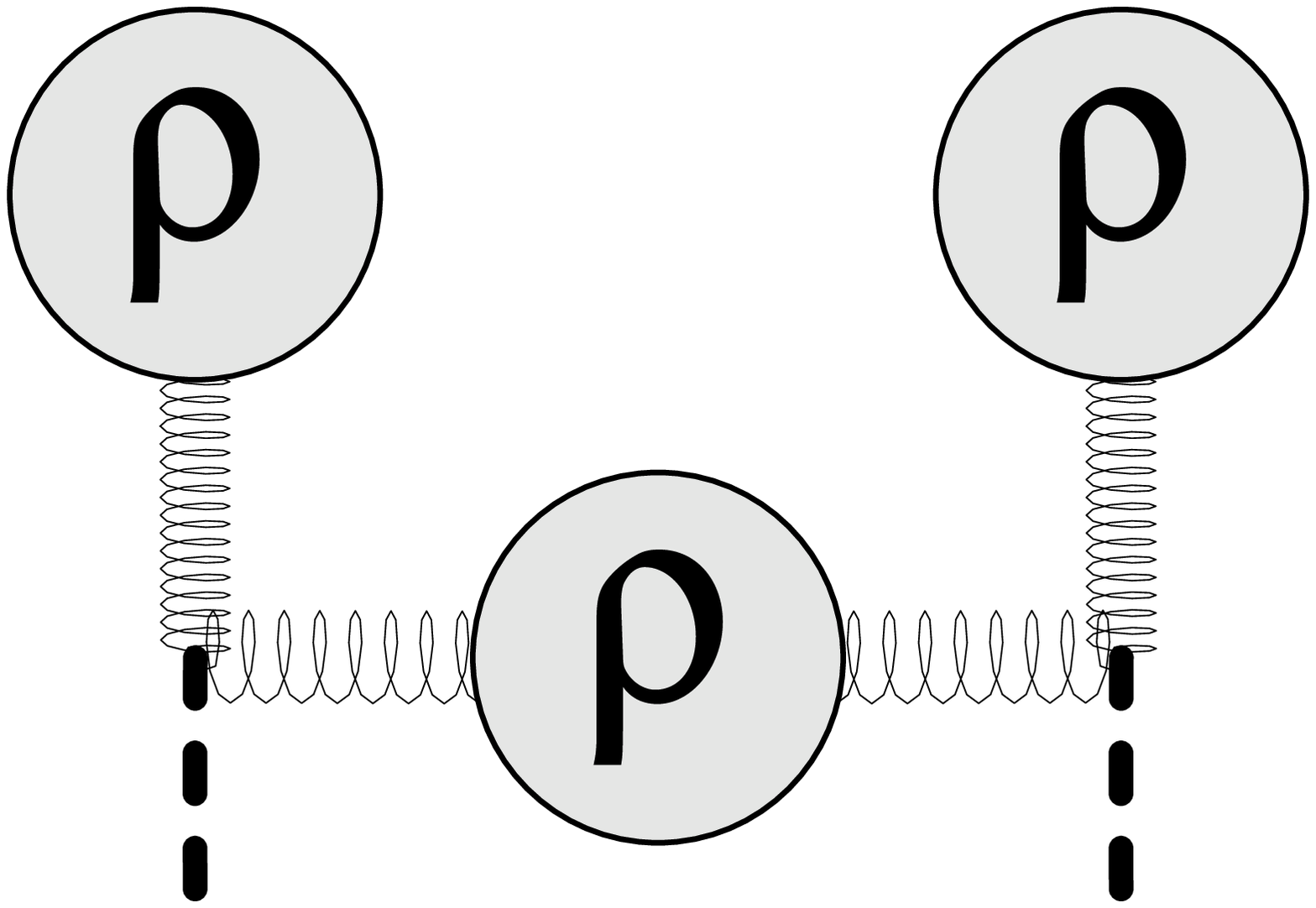,width=2.8cm}
    \caption{Part of the real contribution to the induced color charge}
    \label{fig:chiladder}
  \end{center}
\end{figure}

The charge density $\delta \rho$ that is induced by the fluctuations
$\delta A$ accordingly contains two types of contributions.  First,
there are contributions with frequencies of order $k^-\gg p^-$. Those
come from the real diagrams of Fig.\ref{fig:chifulllist} with ordered
transverse momentum, $k_\perp ^2\ge p^2_\perp $.  For convenience of
reference we redraw one diagram from this set in Fig.\ref{fig:chiladder}
The frequency of this component of the induced charge density
$\delta \rho_{faster}$ is ${k^2_\perp \over k^+}\gg {p^2_\perp \over
  p^+}$.

There is another component $\delta\rho_{slower}$ which comes from the
unordered region of the transverse momenta $k^2_\perp \sim p^2_\perp
$. This component is in the same frequency range as the original
$\rho$. Since it is not contributed by faster modes it is not a Born -
Oppenheimer type contribution.
  In principle there is also a contribution from the region with
  opposite ordering $k_\perp ^2\ll p_\perp ^2$. This component of the
  induced charge density would contain frequencies which are even
  smaller than those in $\rho$.  However it is exactly canceled by the
  virtual corrections of Fig.\ref{fig:sigma} as can be checked
  explicitly from the expression for the BFKL kernel\footnote{One can
    be tempted to think of these virtual corrections as the Born -
    Oppenheimer type backreaction, which ``renormalizes'' the
    effective distribution of the slow component of the charge
    density. This is not the case since the contribution in the
    virtual diagrams comes mostly from the low transverse momentum
    region in the integral, and is therefore due to slow modes in the
    loop.}.

So, to reiterate, as we move to lower $x$
we get two distinct contributions. 
First, there is a Born - Oppenheimer type contribution.
It appears because we ``renormalize'' the notion of staticity by 
including new (faster) modes in the category of static.
This is the
$\delta\rho_{faster}$. Physically, this relates to 
the processes at smaller $x$ that happen at ever faster time
scales. Second, there is a new contribution to the charge density which
has a long wavelength but also low frequency. This is $\delta\rho_{slower}$.
This is not a contribution of the Born - Oppenheimer type and appears 
since the splitting in terms of the longitudinal momentum 
does not strictly coincide with the splitting in terms of frequency.

Now, the DLA approximation to DGLAP assumes transverse momentum ordering and 
therefore includes only $\delta\rho_{faster}$ in the induced charge density.
The BFKL evolution on the contrary includes both contributions.
Therefore the DLA DGLAP includes in the induced density only contributions
which are faster than previously present, and is evolution
simultaneously in the longitudinal momentum and frequency. The BFKL evolution
includes contributions which at every step in the evolution modify
also the slow component of the charge density, and is therefore evolution
only in the longitudinal momentum.

The $\delta\rho_{slower}$ contribution to the BFKL kernel is known to
be problematic, since it creates a channel through which the
nonperturbative small transverse momentum modes couple to the
evolution. The result is the infamous random walk in the transverse
momentum space \cite{BFKL,askew} in the asymptotic solution of the
BFKL equation. Our full nonlinear procedure is similar to the BFKL
approach in the sense that it does not discriminate between the field
modes on the basis of their frequency.  Whether it still suffers from
the same low transverse momentum problem is not clear to us at this
point. It is possible that the nonlinearities in the equation suppress
the low transverse momentum contributions by generating a dynamical
scale somewhat like what happens in the finite temperature and finite
density equilibrium systems.  It would be very interesting to develop
a renormalization group procedure in which the evolution parameter is
not the longitudinal momentum (like in our approach) and not the
transverse momentum (like in the DGLAP evolution) but rather directly
the time resolution scale.  This kind of approach would consider the
contributions of low $k_\perp $ modes at low $x$ as part of the
initial condition rather than part of the evolution and would thereby
provide a cleaner separation of perturbative and nonperturbative
effects.  A nonlinear evolution obtained in this type of approach
should be closely related to the nonlinear evolution equation of
Laenen and Levin\cite{LL}.

\appendix

\section{High and low density situations: the parametric size of $\rho$}
\label{app:size}

The situation that interests us in this paper is the one when the
color charge density is high enough so that the nonlinearities are
important.
Already when looking at the tree graph expansion of a classical field
generated by a color charge density $\rho$, we are in a position to
judge how big $\rho$ has to be parametrically for {\em all} of the
terms to contribute equally to the result.  

To do this we observe that one may build all diagrams that make up the
classical field starting from the linear term
\begin{eqnarray}
  \label{eq:der.start}
   \begin{minipage}[c]{1cm}
      \epsfig{file=tree1.eps,width=1cm} 
\end{minipage}\ { g \rho} 
\end{eqnarray}
by stripping off a factor $g \rho$ and replacing it successively by either
\begin{eqnarray}
\nonumber
g \ \begin{minipage}[c]{1.8cm}
\epsfig{file=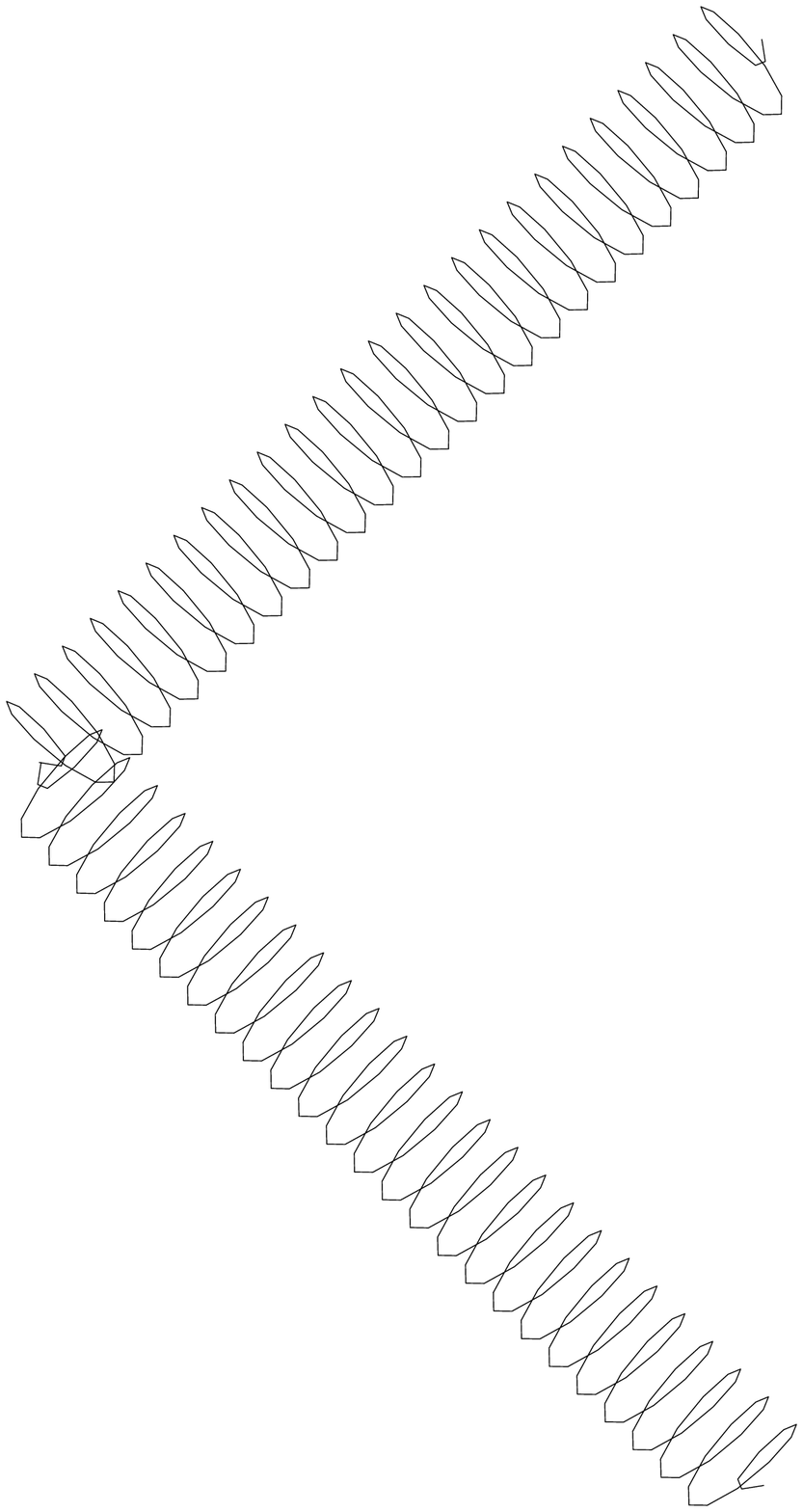,height= 1.8cm}
\end{minipage}
\hspace{-0.8cm}
\begin{minipage}[c]{1cm}
${ g \rho}$\\ \vskip1.5cm ${ g \rho}$
\end{minipage}  
\qquad \mbox{or} \qquad 
g^2 \ \begin{minipage}[c]{1.8cm}
\epsfig{file=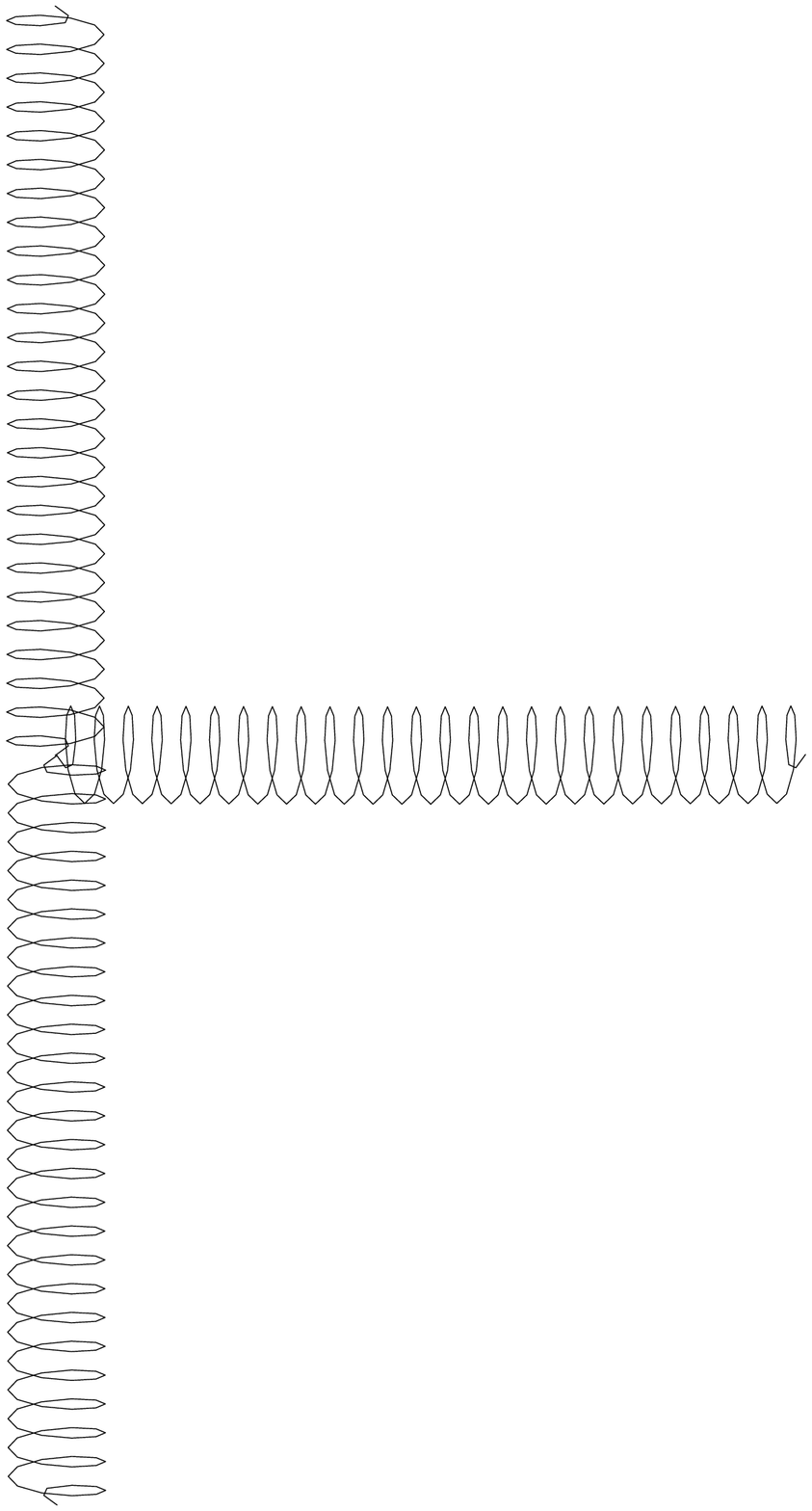,height= 1.8cm}
\end{minipage}
\hspace{-0.5cm}{ g \rho}
\hspace{-1.4cm}
\begin{minipage}[c]{1cm}
${ g \rho}$\\ \vskip1.5cm ${ g \rho}$
\end{minipage}  
\end{eqnarray}
To generate diagrams which are all of the same order we need to
simultaneously satisfy the equations
\begin{eqnarray}
  \label{eq:der.3gadd}
  g \cdot {\rm order}(\rho) & = & g \left(g \cdot {\rm order}(\rho)\right)^2
\\
  \label{eq:der.4gadd}
  g \cdot {\rm order}(\rho) & = & g^2 \left(g \cdot {\rm order}(\rho)\right)^3 
\end{eqnarray}
This fixes $ 1 = g^2 \cdot {\rm order}(\rho) $ i.e. $\rho$ to be of
order $g^{-2}$.

It is therefore for $\rho\sim O(g^{-2})$
that the nonlinear effects described by the RG evolution derived in this
paper are physically important.

\section{Inverting $K$.} 
In this appendix we invert the operator $K$ which appears in the small
fluctuation action
\begin{eqnarray} 
K^{ab}_{xy} &=& - \bigg[
(\partial^{+})^2 \delta^{ab} + f^{abc}\rho^{c} \delta (x^-)
\frac{1}{\partial^-} \bigg]\nonumber\\ 
&=& - \bigg[ (\partial^{+})^2
\delta^{ab} -2 M^{ab} \delta (x^-)\bigg] 
\label{K1}
\end{eqnarray} 

First, note that the frequency and  the transverse coordinate dependence
of $K$ is trivial, and therefore $p^-$ and $x_\perp $ are conserved quantum numbers
in this inversion. For the purpose of this calculation we can imagine that
the color matrix $M$ has been diagonalized at every point in $x_\perp $ and 
we will therefore treat it formally as a number.

Let $f^{a}_{\lambda} (x) $ be the set of eigenfunctions of $K$
\begin{eqnarray}
\label{eq:eigenvalue}
\int dy^- K^{ab}(x,y) f^{b} (p^-, y^- ,x_\perp ) = 
\lambda f^{a}(p^-, x^- ,x_\perp )
\end{eqnarray}
Assuming for the moment that the eigenvalues are positive, the
eigenfunctions have the form
\begin{equation}
f^{a}_{p^-,x_\perp }(x) = \theta (-x^-) e^{ip^+x^-} U^a(x_\perp ,p^-) + 
\theta (x^-) \bigg[e^{ip^+x^-} 
F^a (x_\perp ,p^-) + 
e^{-ip^+x^-} G^a (x_\perp ,p^-) \bigg]
\end{equation}
where $F$ and $G$ are to be determined. Requiring that $f^{a}$ is continuous
at $x^- =0 $ gives $ U^a = F^a + G^a $. The first derivative of $f^a$ 
should be discontinuous such that $\partial^+ $ when acting 
on it will cancel the 
$\delta (x^-)$ in the equation (\ref{K1}). 
In other words, the discontinuity  in the first
derivative must be equal to the integral of the delta function across 
the discontinuity. This gives
\begin{eqnarray}
f^{a}_{cont} = U^{a}_{\alpha} e^{ipx} -i \theta (x^-) \frac{\sin p^+ x^-}
{p^+ p^- } f^{abc} U^{b}_{\alpha} \rho^{c} (x_\perp ).
\label{eq:conteigen}
\end{eqnarray}
These eigenfunctions correspond to a positive eigenvalue
\begin{eqnarray}
\lambda=p^{+2}\nonumber
\end{eqnarray}
To determine the proper normalization of the ``polarization vector''
$U^{a}_{\alpha}$ it is easier to work with the symmetric and
anti-symmetric combinations
\begin{eqnarray}
f^{a}_{\alpha sym} = U^{a}_{\alpha} \cos p^+x^- 
-i \frac{1}{2p^+p^-} f^{abc} 
U^{a}_{\alpha} \rho^{c} \sin p^+ x^- \bigg[\theta (x^-) - 
\theta (-x^- )\bigg]
\nonumber
\end{eqnarray}
and
\begin{eqnarray}
f^{a}_{\alpha asym} = i U^{a}_{\alpha} \sin p^+ x^- 
-i \frac{1}{2p^+ p^-} f^{abc}U^b_\alpha
\rho^{c} \sin p^+ x^-
\nonumber
\end{eqnarray}
The antisymmetric functions vanish at $x^-=0$ and therefore are the same 
as for $M=0$. Their normalized form is just
\begin{eqnarray}
f^a_{\alpha asym}=\sin(p^+x^-)\delta^a_\alpha
\nonumber
\end{eqnarray}

The overlap matrix for the symmetric eigenfunctions is
\begin{eqnarray}
\int dx^- \bigg[f^{a}_{(\alpha ,p^+)} (x)\bigg]^{\dagger} 
\bigg[f^{a}_{(\beta ,p^{+'})}(x)\bigg] = 
\delta ( p^+ - p^{+ \prime } ) \frac{U^{\dagger} U}{2} 
\bigg[1- \frac{(f\cdot \rho)^2}{(2p^+ p^- )^2}\bigg]
\nonumber
\end{eqnarray}
and we get
\begin{eqnarray}
U^{\dagger} U = \frac{2 (p^+)^2}{(p^+)^2 + M^2}
\nonumber
\end{eqnarray}
These are the eigenfunctions for ``continuum'' states, the ones
corresponding to positive eigenvalues.

Note that since $tr f\cdot \rho =0$, the matrix $M$ has both negative
and positive eigenvalues. Therefore the eigenvalue
equation~(\ref{eq:eigenvalue}) must also have negative eigenvalues and
corresponding ``bound state'' solutions.  The bound state wave
function must be symmetric under $x^- \rightarrow -x^-$ and has the
same form as the continuum solution except that $p^+$ is imaginary.
Requiring that it decay exponentially at large $x^-$, we have
\begin{eqnarray}
p^+ = i M \theta (M)\nonumber
\end{eqnarray}
The eigenfunction then is
\begin{eqnarray}
f^{a}_{bs} = N V^{a}_{\alpha} \bigg[ \theta (x^-) e^{- M x^-} + 
\theta (-x^-) e^{M x^-} \bigg]
\label{eq:bs}
\end{eqnarray}
where $N$ is a normalization factor and $V^{a}_{\alpha}$ are the set
of eigenfunctions of $M$ with positive eigenvalue.  The normalization
factor $N$ is easily calculated and is given by $N^2 = M$.

It is easily checked that the set of our eigenfunctions is complete.
The completeness relation is
\begin{eqnarray}
I \equiv \int \frac{d^4 p}{(2\pi )^4} \bigg[a^{a}(x,p)\bigg]^{\dagger} 
\bigg[a^{b}(y,p)\bigg] =
\delta (x-y) \delta^{ab}
\label{eq:completeness}
\end{eqnarray}
This can be written as
\begin{eqnarray}
I = \delta (x_\perp  -y_\perp ) \int \frac{dp^- }{2\pi} I_{1}(x^-,y^-,p^-)
\nonumber
\end{eqnarray}
where for the continuum solutions we have
\begin{eqnarray}
I_{1}&=& \theta (x^-)\theta(y^-) \bigg[ \frac{-1}{2} (\mid M \mid + M)
         \theta (x^- + y^-) e^{-\mid M \mid (x^- +y^-)} \nonumber \\
     &-& 
         \frac{1}{2} (\mid M \mid - M)
         \theta (-x^- - y^-) e^{\mid M \mid (x^- +y^-)}\nonumber \\
     &+& \delta (x^- +y^-) +
         \int \frac{dp^+}{2\pi} 2 \sin p^+ x^- \sin p^+ y^- \bigg]\nonumber \\
     &+&
          \theta (x^-)\theta(-y^-)\bigg[  \frac{-1}{2} (\mid M \mid + M)
          \theta (x^- - y^-) e^{-\mid M \mid (x^- -y^-)}\nonumber \\
     &-&     \frac{1}{2} (\mid M \mid - M)
          \theta (-x^- + y^-) e^{\mid M \mid (x^- -y^-)} + 
          \delta (x^- -y^-)\bigg]\nonumber \\
     &+&  ( x^-,y^- \rightarrow -x^-, -y^-)
\end{eqnarray}
while for the bound state 
\begin{eqnarray}
I_{1}&=& \frac{1}{2} (\mid M \mid + M) \bigg[ \theta (x^-)\theta(y^-) 
          e^{-\mid M \mid (x^- +y^-)}  +  \theta (x^-)\theta(-y^-)
           e^{-\mid M \mid (x^- -y^-)}\nonumber \\
     &+& ( x^-,y^- \rightarrow -x^-, -y^-)\bigg]
\end{eqnarray}
All $M$ dependent terms cancel out between the two contributions. 
Performing the $p^+$ 
integration and adding all the terms establishes the completeness 
relation~(\ref{eq:completeness}).

Now that we have the complete set of eigenfunctions, we are ready to
invert the operator $K$. But first we should understand the zero modes
$p^+ =0$.  These eigenfunctions do not vanish at
$x^-\rightarrow\pm\infty$ and we should be careful with them, for
example when integrating by parts.  In fact when calculating the
propagator, these eigenfunctions should be excluded entirely from the
sum as explained in Section~\ref{sec:sigmachi}.  The convenient way to
do that is to regulate the factor $1/\lambda$ which enters in the
calculation of the propagator as
\begin{eqnarray}
\frac{1}{\lambda} \equiv \frac{\lambda}{\lambda^2 + \epsilon^4}
\nonumber
\end{eqnarray}
taking the limit $\epsilon\rightarrow 0$ at the end of the
calculation.  We use the same regulator to regulate a possible
singularity $M\rightarrow 0$ in the bound state eigenvalue.
\begin{eqnarray}
\frac{1}{M^2} \rightarrow \frac{M^2}{M^4 + \epsilon^4}
\nonumber
\end{eqnarray}
The propagator $K^{-1}$ is calculated as
\begin{equation}
K^{-1}(x^-,x_\perp ,p^-,y^-,y_\perp ,p^{-'})
=\int d\lambda {1\over\lambda}f_{\lambda, p^-,x_\perp }(x^-)
f^*_{\lambda, p^-,x_\perp }(y^-)
\delta(p^--p^{-'})\delta(x_\perp -y_\perp )
\end{equation}
The result is
\begin{eqnarray}
K^{-1}&=& \frac{-1}{4i\sqrt{i}\,\epsilon}
                                       \Bigg\{ \theta (x^- -y^-) 
           \bigg[ e^{i\sqrt{i}\,\epsilon (x^- -y^-)}
            -i e^{-\sqrt{i}\,\epsilon (x^- -y^-)}\bigg]\nonumber \\
                                       &+& M\Bigg[ \theta (x^-)\theta (y^-)
            \bigg[\frac{ e^{i\sqrt{i}\,\epsilon (x^- +y^-)}}
                       {M + i\sqrt{i}\,\epsilon}-i
                   \frac{ e^{-\sqrt{i}\,\epsilon (x^- +y^-)}}
                    {M -\sqrt{i}\,\epsilon} \bigg]\nonumber \\
                                        &+&  \theta (x^-)\theta (-y^-)
                    \bigg[ \frac{ e^{i\sqrt{i}\,\epsilon (x^- -y^-)}}
                       {M + i\sqrt{i}\,\epsilon}-i
                   \frac{ e^{-\sqrt{i}\,\epsilon (x^- -y^-)}}
                    {M -\sqrt{i}\,\epsilon} \bigg]\nonumber \\
                                       &+& ( x^-,y^- \rightarrow -x^-, -y^-)
                                           \Bigg\}\delta(p^--p^{-'})\delta(x_\perp -y_\perp )
\label{eq:Kprop}
\end{eqnarray}
Expanding the above expression in powers of $\epsilon$ to order one, we get
\begin{equation}
\label{eq:Kpropexpand1}
K^{-1} = \bigg\{- {1 \over 2} |x^- - y^-| + {1 \over 2} \eta \bigg[ |x^-|
+ |y^-| \bigg] - {\eta \over 2 M} + {\mu \over 2\sqrt{2} \epsilon}\bigg\}
\delta(p^--p^{-'})\delta(x_\perp -y_\perp )
\end{equation}
where we have defined the projection operators $\eta$ and $\mu$ that
project on nonzero and zero eigenvalue subspaces of $M$ respectively
\begin{eqnarray}
&&\mu M = 0,\;\;\; \eta M = M\\
&&\mu + \eta = 1 ,\;\;\mu^{2} = \mu ,\;\; \eta^{2} = \eta
\label{eq:project1}
\end{eqnarray}

Note that the last term in Eq.(\ref{eq:Kpropexpand1}) diverges in the
limit $\epsilon\rightarrow 0$. However examining carefully equations
in Section~\ref{sec:sigmachi}, we see that $K^{-1}$ always acts on a
particular combination of fields $B$ which satisfies the constraint
$\mu B=0$ by virtue of the Gauss' law.  We can therefore omit this
term from the expression for $K^{-1}$ altogether, which is what we did
in the text Eq.(\ref{eq:Kpropexpand}).

\section{Proper normalization of eigenfunctions.}

In this appendix we show how to properly normalize eigenfunctions in a
theory with a Lagrange multiplier field.  Consider a quadratic form
\begin{eqnarray}
S=a_iM_{ij}a_j,\ \ \ \ \ \ \ \ \ i=1,...,N
\end{eqnarray}
where the variables $a_i$ are constrained by $m$ linear conditions
\begin{eqnarray}
c^\alpha_ia_i=0 ,\ \ \ \ \ \ \ \alpha=1,...,m
\label{acon}
\end{eqnarray}
Our problem is to invert $M$ on the subspace whose vectors satisfy
Eq.(\ref{acon}).  This is the precise analog of the system we deal
with in Section~\ref{sec:sigmachi}.  Let us assume that all $m$
vectors $c^\alpha_i$ are linearly independent.  In that case they span
an $m$ - dimensional subspace $C$ of the original $N$ dimensional
vector space $V$. Let $l^\alpha$ be an orthonormal basis on this
subspace.  We can then construct the projection operator
\begin{eqnarray}
P_{ij}=l^\alpha_i l^\alpha_j
\end{eqnarray}
which projects on $C$.

Then instead of considering the matrix $M$ we should consider
\begin{eqnarray}
\tilde M=(1-P)M(1-P)
\end{eqnarray}
and invert it on $V-C$.
The eigenvalue and eigenfunction equations for this problem are
\begin{eqnarray}
\tilde Ma=\lambda(1-P)a
\end{eqnarray}
or alternatively
\begin{eqnarray}
&&(1-P)Ma=\lambda a\nonumber\\
&&c^\alpha a=0\nonumber
\label{ah}
\end{eqnarray}
Clearly the eigenfunctions have to be normalized in the standard way
\begin{eqnarray}
a_i^\lambda a_i^{*\lambda'}=\delta_{\lambda,\lambda'}
\label{norma}
\end{eqnarray}
The inverse of $M$ on $V-C$ is constructed as
\begin{eqnarray}
\tilde M^{-1}=\sum_{\lambda}{a_i^\lambda a_j^{*\lambda}\over\lambda}
\end{eqnarray}
since
\begin{eqnarray}
\tilde M\tilde M^{-1}=(1-P)M\tilde M^{-1}=
\sum_{\lambda}a_i^\lambda a_j^{*\lambda}=P_{ij}
\end{eqnarray}

An alternative to explicit construction of the projection operator $P$
is introduction of the Lagrange multiplier field, as we did in
Section~\ref{sec:sigmachi}.  We add to the Lagrangian the term
\begin{eqnarray} 
b^\alpha c^\alpha a
\end{eqnarray}
The eigenvalue equations we have to solve now are
\begin{eqnarray}
&&Ma+b^\alpha c^\alpha=\lambda a\nonumber \\
&&c^\alpha a=o\nonumber
\label{oh}
\end{eqnarray}
Now since, $Pc^\alpha=c^\alpha$ and $Pa=0$,
this equation gives
\begin{eqnarray}
PMa+b^\alpha c^\alpha=0
\end{eqnarray}
Solving this for $b^\alpha$ and substituting back into Eq.(\ref{oh})
we obtain again Eq.(\ref{ah}). This proves that the eigenfunctions
obtained through introduction of the Lagrange multiplier are the same as
in the straightforward calculation in which the constraint is solved
explicitly through construction of $P$. It is clear then that the
normalization of these functions should be the standard normalization
Eq.(\ref{norma}).

\section{The BFKL limit of the general evolution equation.}

In this section we will show in some detail how our general 
expressions for $\chi^{ab}$ and $\sigma^{a}$ give the BFKL
kernel. To do so, 
we have to take the limit of small $\rho$ in the evolution equation
Eq.(\ref{finalrg}).
In fact it is more convenient to consider directly the equation
for the density correlation function Eq.(\ref{prop}). As was
shown in \cite{linear}, using Eqs.(\ref{tree}) and (\ref{resa})
the unintegrated gluon density $\phi(k_\perp )$ (which is the quantity
which evolves according to the BFKL equation) to leading order in $\rho$
is just the charge density two point function $<\rho(k_\perp )\rho(-k_\perp )>$.
The BFKL equation should therefore be just the weak field limit of 
Eq.(\ref{prop}).

To verify this 
we will need the expressions for $\chi^{ab}$
and $\sigma^{a}$ expended to first order in $\rho$.
Let us consider the contributions of the real diagrams given by
$\delta \rho_{real}$ in equation~(ref{eq:totalrhoreal}).
Using the expressions for $v_{+}^{i}$ and $\gamma_{+}$ and $1/D^2$ 
expanded to
first order in $\alpha$
\begin{eqnarray}
\gamma^{i}_{+} &=& 2 {\partial^i \over \partial^2_\perp } 
\bigg[ \alpha v_{-} + (\partial \alpha ){\partial v_{-} \over 
\partial^2_\perp }\bigg]\nonumber\\
v^{i}_{+} &=& 2 \alpha^{i} {\partial v_{-} \over \partial^2_\perp } + v^{i}_{-} 
- 2 {\partial^i \over \partial^2_\perp } 
\bigg[ \alpha v_{-} + (\partial \alpha ){\partial v_{-} \over 
\partial^2_\perp }\bigg] \nonumber\\
{1 \over D^2_\perp }& =& {1 \over \partial^2_\perp } - 
{1 \over \partial^2_\perp } (\partial \alpha + \alpha \partial ) 
{1 \over \partial^2_\perp }
\label{eq:expandedagamma}
\end{eqnarray}
leads to
\begin{eqnarray}
\delta \rho^{a}_1(x_\perp ) = -2 f^{abc} \bigg[\alpha^{b} v^{c}_{-}
- \rho^{b} {\partial v^{c}_{-} \over \partial^{2}_{t} } \bigg] 
\label{eq:expandedrhoreal}
\end{eqnarray}
In the momentum space representation  
\begin{eqnarray}
{\partial^{i} \over \partial^{2}_{t}} v^{i}_{-} = i 
\int {d^2 p_\perp  \over (2\pi)^2} d^2 y_\perp  {p^i \over p^2_\perp } 
e^{i p_\perp  (x_\perp  -y_\perp )} v^{i}_{-}(y_\perp )
\end{eqnarray}
With these expressions we have
\begin{eqnarray}
\delta \rho^{a}_1(x_\perp )\delta \rho^{b}_1(y_\perp ) = 4 g^2
f^{acd}f^{bef} \bigg[\alpha^c v^{d}_{-} -  \rho^{c} 
{\partial v^{d}_{-} \over \partial^2_\perp }\bigg]_{x_\perp }
 \bigg[\alpha^e v^{f}_{-} - \rho^{e} 
{\partial v^{f}_{-} \over \partial^2_\perp }\bigg]_{y_\perp }
\end{eqnarray}
To this order the normalization of $v^i_i$ is 
\begin{eqnarray}
<v^{id}_{-}(x_\perp )v^{\ast jf}_{-}(y_\perp )> = \delta^{ij}\delta^{df} 
\delta^{2}(x_\perp  -y_\perp ) {1 \over 2\pi}{1 \over 2}
\end{eqnarray}
We then have
\begin{eqnarray}
\lefteqn{<\delta \rho^{a}_1(x_\perp )\delta \rho^{b}_1(y_\perp )> = 
g^2f^{acd}f^{bed} {1 \over \pi} \ln 1/x  \bigg\{ \alpha^{c}(x_\perp )
\alpha^{e}(y_\perp ) \delta^{2}(x_\perp  - y_\perp )}
\nonumber\\ &+& 
i\bigg[ \alpha^{ic}(x_\perp )\rho^{e}(y_\perp ) - 
\rho^{c}(x_\perp ) \alpha^{ie}(y_\perp )\bigg]\int  {d^2 p_\perp  \over (2\pi)^2}
{p^i \over p^2_\perp } e^{i p_\perp  (x_\perp  - y_\perp )}\nonumber\\
&+& \rho^{c}(x_\perp )\rho^{e}(y_\perp )\int  {d^2 p_\perp  \over (2\pi)^2}
{1 \over p^2_\perp } e^{i p_\perp  (x_\perp  - y_\perp )}\bigg\}
\end{eqnarray}
This expression coincides with Eqns.(53,54,55) in 
\cite{linear}. Using $\alpha^i=-{\partial^i\rho\over \partial^2}$ we obtain
\begin{eqnarray}
\delta \rho^{a}_1 (k_\perp ) \delta \rho^{a}_1 (-k_\perp )  &=& 
{{2g^2 N_c}\over{(2\pi )^3}}
\int d^2 p_\perp  \rho^{a} (p_\perp )\rho^{a} (-p_\perp ) {{k_\perp ^2}
\over{p_\perp ^2 (p_\perp  - k_\perp )^2}}
\label{totreal}
\end{eqnarray}
This is precisely the real part of the BFKL kernel.

It is straightforward to repeat the above procedure for the
contribution of virtual diagrams $<\delta \rho^{a}_2(x_\perp
)>$~(\ref{eq:totalvir}).  In this case, the term $R^{ab}$ vanishes to
order $\alpha$. Using our expanded expressions for $a_{+}$ and
$\gamma_{+}$ and noticing that $\gamma_{+}$ starts at order $\alpha$,
the first line in (\ref{eq:totalvir}) gives
\begin{equation}
<\delta\rho^{a}_2(x_\perp )>_{(1)} \!= \!- {g^2  \over 2\pi}
f^{dbc}f^{abc} \int \!d^2 y_\perp  \rho^{d}(y_\perp ) 
\int \!{d^2p_\perp  \over (2\pi)^2 } {d^2q_\perp  \over (2\pi)^2 }
{ p_\perp \cdot q_\perp  \over p^2_\perp  q^2_\perp } 
e^{i(p_\perp  +q_\perp )(x_\perp -y_\perp )}
\end{equation}
This agrees with the corresponding term in \cite{linear}. In the second
line in the expression for $\delta \rho^{a}_2$~(\ref{eq:totalvir}),
we can take the order $\alpha^{0}$ inside the brackets since there is
already an explicit factor of $\rho$ present. This gives
\begin{eqnarray}
<\delta\rho^{a}_2(x_\perp )>_{(2)} = - {g^2 N_c \over (2\pi)^3} \ln 1/x
\rho^{a}(x_\perp ) \int {d^2p_\perp  \over p^2_\perp }
\end{eqnarray}
which is exactly equation (50) in \cite{linear}.
Collecting all the contributions, substituting them into Eq.(\ref{prop})
and identifying the density correlator with the unintegrated gluon
density $\phi$ we obtain
\begin{eqnarray}
{d\over d\ln{1\over x}}\phi(k_\perp )=- 
{{g^2 N_c}\over{(2\pi)^3}}\int d^2 p_\perp  
{{k_\perp ^2}\over{p_\perp ^2 (p_\perp  - k_\perp )^2}}
\Bigg[\phi(k_\perp ) - 2 \phi(p_\perp)\Bigg]
\end{eqnarray}

This is precisely the BFKL equation \cite{BFKL}.

{\bf Acknowledgements} We are grateful to Larry McLerran for numerous
discussions on a variety of topics related to the subject of this
paper. We have also benefited from interesting discussions with A. Leonidov,
E. Levin, A. White and M. W\"usthoff.
 The work of J.J.M. was supported by DOE contract DOE-Nuclear
DE-FG02-87ER-40328. 
A.K. was supported by DOE contract 
DOE High Energy DE-AC02-83ER40105 and the PPARC Advanced
Fellowship. 
HW was supported by the EC Program ``Training and
Mobility of Researchers", Network ``Hadronic Physics with High Energy
Electromagnetic Probes", contract ERB FMRX-CT96-0008.

\end{document}